\documentclass[
nofootinbib,
 amsmath,amssymb,
 aps,
]{revtex4-2}
\pdfoutput=1

\usepackage{graphicx}% Include figure files
\usepackage{dcolumn}% Align table columns on decimal point
\usepackage{bm}% bold math
\usepackage{tikz-cd}
\usetikzlibrary{decorations.markings}
\tikzset{mid vert/.style={/utils/exec=\tikzset{every node/.append style={outer sep=0.8ex}},
postaction=decorate,decoration={markings,
mark=at position 0.5 with {\draw[-] (0,#1) -- (0,-#1);}}},
mid vert/.default=0.75ex}
\tikzcdset{every label/.append style = {font = \normalsize}}

\usepackage{array}
\newcolumntype{H}{>{\setbox0=\hbox\bgroup}c<{\egroup}@{}}

\usepackage{xtab}

\usepackage{diagbox}
\usepackage{wasysym}
\usepackage{multirow}

\usepackage{xcolor}
\usepackage{subcaption}

\definecolor{darkGreen}{rgb}{0,0.5,0}
\definecolor{darkRed}{rgb}{0.7,0,0.1}
\usepackage{soul}
\usepackage{tikz}

\makeatletter
\newcommand{\doublewidetilde}[1]{{%
  \mathpalette\double@widetilde{#1}%
}}
\newcommand{\double@widetilde}[2]{%
  \sbox\z@{$\m@th#1\widetilde{#2}$}%
  \ht\z@=.9\ht\z@
  \widetilde{\box\z@}%
}
\makeatother
\usepackage{placeins}
\usepackage{mathtools}
\newcommand\red[1]{\textcolor{red}{#1}}

\usepackage{empheq}
\usepackage{cases}
\usepackage{makecell}

\begin{document}

\preprint{APS/123-QED}
\title{Effects of particle size and background rotation on the settling of particle clouds}% Force line breaks with \\

\author{Quentin Kriaa}
\email{Corresponding author: quentin.kriaa@univ-amu.fr}

\author{Eliot Subra}
\author{Benjamin Favier}
\email{benjamin.favier@cnrs.fr}
\author{Michael Le Bars}
\email{michael.le-bars@univ-amu.fr}

\affiliation{%
 Aix Marseille Univ, CNRS, Centrale Marseille, IRPHE, Marseille, France%\\
}%

\date{\today}

\begin{abstract}
We experimentally investigate the behaviour of instantaneous localised releases of heavy particles falling as turbulent clouds in quiescent water, both with and without background rotation. We present the results of 514 systematic experiments for no rotation and for three rotation rates $\Omega = 5,10,20$rpm, and for the size of particles in the range $5\mu m$ to $1mm$, exploring four decades of the Rouse number $\mathcal{R} \in [6\times 10^{-4},4]$ which quantifies the inertia of particles. In the canonical framework of turbulent thermals described by Morton \textit{et al.}, [Proc. R. Soc. A: Math. Phys. Sci. \textbf{234}, 1 (1956)], we compare particle clouds with salt-water thermals to highlight specificities due to the particulate nature of the turbulence forcing. In the absence of rotation, particle clouds initially behave as salty thermals with a modulation of their entrainment capacity, which is optimally enhanced for a finite inertia $\mathcal{R}\simeq 0.3$ due to particulate effects. However this regime of turbulence is limited in time due to the inertial decoupling between turbulent eddies and particles. For the three values of $\Omega$ explored here, the particulate enhancement of entrainment is inhibited. Moreover the cloud's expansion is interrupted when the Coriolis force overcomes its inertia, forcing the cloud to transform into vortical columnar flows which considerably increase the residence time of particles.
\end{abstract}

\maketitle

%%%%%%%%%%%%%%%%%%%%%%%%%%%%%%%%%%%%%%%%%%%%%%%%
%%%%%%%%%%%%%%%%%%%%%%%%%%%%%%%%%%%%%%%%%%%%%%%%
\section{\label{sec:Introduction}Introduction}

Interactions between a fluid and solid particles can take many forms. In granular media of maximum packing fraction, particles constrain the fluid motion. Conversely, in configurations of vanishingly small concentrations, particles are seemingly isolated and the fluid tends to constrain the motion of the particles. The continuous transition between these end-members \citep{andreottiMilieuxGranulairesEntre2012} and the influence of other particle properties like their size and density offer a vast spectrum of interactions and phenomena (see \citep{brandtParticleLadenTurbulenceProgress2022} for a review), which notably manifest in landslides \citep{fritzLituyaBayLandslide2009b}, the transport of sediments in rivers \citep{abramianBoltzmannDistributionSediment2019}, the performance of agricultural sprays \citep{lakeEffectDropSize1977}, or the formation of planets in protoplanetary disks \citep{meheutDusttrappingRossbyVortices2012}.
\par
When the size of particles, their volume fraction and the particle-to-fluid density anomaly are low, the fluid governs the motion of particles whose feedback on the flow is negligible, a situation called \textit{one-way coupling} \citep{balachandarTurbulentDispersedMultiphase2010}. Many studies have considered the one-way coupling between particles and pre-established idealised isotropic turbulent flows whose properties are well controlled \citep{brandtParticleLadenTurbulenceProgress2022}, evidencing that turbulence is a source of non-uniformities in the field of particle concentration  (e.g., \citep{alisedaEffectPreferentialConcentration2002,yoshimotoSelfsimilarClusteringInertial2007,salazarExperimentalNumericalInvestigation2008}) and that turbulence alters the velocity and trajectory of settling particles (e.g., \citep{maxeyGravitationalSettlingAerosol1987,goodSettlingRegimesInertial2014}).
\par
Although these studies have brought substantial light on the interactions between the fluid and particles, in several situations the feedback of particles on the fluid has a non-negligible impact on the flow \citep{monchauxSettlingVelocityPreferential2017} -- a situation referred to as \textit{two-way coupling} \citep{balachandarTurbulentDispersedMultiphase2010}. This is especially true when the flow is nourished by the particles, as in downdrafts which can be accompanied by intense rainfalls and are accelerated by the evaporation of droplets \citep{krugerDynamicsDowndraughtsCold2020}, or when the flow is produced by the particles themselves e.g., in turbidity currents \citep{ouillonTurbidityCurrentsPropagating2019a,neckerHighresolutionSimulationsParticledriven2002a}. The initial motivation of the present study fits in this latter framework: past studies suggest that the magnetic field of small telluric planets and moons like Mercury or Ganymede results from the fluid motions generated by iron snow flakes settling into liquid metal -- a phenomenon called \textit{iron snow} \cite{ruckriemenFeSnowRegime2015a}. As a first step to understand this phenomenon, this study focuses on the instantaneous release of heavy particles falling as a cloud in water.
\par
The motion of buoyant clouds has been widely studied by releasing a finite volume of denser miscible fluid (often salt water) in fresh water, which almost immediately becomes turbulent. A decisive aspect of the dynamics of these clouds is the efficiency of turbulence to entrain ambient fluid at the cloud interface. Entrainment is actually key to modelling numerous structures like gravity currents \citep{ouillonTurbidityCurrentsPropagating2019a,neckerHighresolutionSimulationsParticledriven2002a}, wildfire plumes \citep{paugamReviewApproachesEstimate2016}, moist convection cells \citep{yanoBasicConvectiveElement2014} or heat plumes in ventilated spaces \citep{lindenFluidMechanicsNatural1999}. By a simple modelling of entrainment through a single scalar coefficient, Morton \textit{et al.} \citep{mortonTurbulentGravitationalConvection1956} developed in 1956 the \textit{turbulent thermal} model, which has proved a successful model of finite releases of buoyant fluid in a multitude of contexts \citep{turnerTurbulentEntrainmentDevelopment1986}, even for finite buoyant releases made of immiscible fluid \citep{landeauExperimentsFragmentationBuoyant2014} and bubbles \citep{penasBubbleladenThermalsSupersaturated2021} generated from different initial conditions. Similarly, past experiments on finite releases of heavy particles have shown that after an initial regime of acceleration, the dynamics of such clouds can be described with the turbulent thermal model \citep{rahimipourDynamicBehaviourParticle1992,bushParticleCloudsHomogeneous2003a,laiTwophaseModelingSediment2013a,laiModelingExperimentsPolydisperse2016}. To the best of our knowledge, no specific influence of the particulate nature of the turbulence forcing has been observed on this initial dynamics. Yet, particles have proved capable of modulating turbulence in controlled turbulent flows (see \citep{balachandarTurbulentDispersedMultiphase2010} for a review) as well as in recent experiments on particle-laden plumes (i.e. continuous injections of buoyancy), whose entrainment efficiency was altered by particles when they crossed the plume interface \citep{mcconnochieEntrainmentParticleladenTurbulent2021a}. The absence of any such observation for instantaneous releases calls for systematic experiments to determine whether the particulate nature of the turbulence forcing can alter the entrainment efficiency of particle clouds with respect to miscible turbulent thermals. To do so, we investigate the role of particle inertia by covering a large range of particles' sizes from a regime dominated by the fluid motions (when particles have a low inertia) to a regime dominated by the inertia of particles.
\par
Motivated by the influence of planetary rotation during iron snow, we also investigate the influence of background rotation on the clouds' dynamics. Various experimental studies exist on plumes in a rotating ambient \cite{fernandoDevelopmentPointPlume1998a,goodmanHydrothermalPlumeDynamics2004,frankAnticyclonicPrecessionPlume2017,frankEffectsBackgroundRotation2021,sutherlandPlumesRotatingFluid2021}. Some recent studies brought substantial light on their particular behaviour due to the solid body rotation, both for bubble-laden \citep{frankEffectsBackgroundRotation2021} and miscible plumes \citep{sutherlandPlumesRotatingFluid2021}. These studies can qualitatively guide the analysis of thermals in a rotating ambient, but while plumes tend to reach a permanent regime, thermals are inherently transient so that scaling laws inevitably differ. A few experimental studies exist for instantaneous buoyant releases that are miscible with water \citep{ayotteMotionTurbulentThermal1994,helfrichThermalsBackgroundRotation1994}, but to the best of our knowledge, no such experiments have been conducted with particle clouds.
\par
In this work, we present the results of 514 experiments performed by systematically varying the size of particles and the angular velocity of the background rotation. The experimental apparatus and governing dimensionless numbers are introduced in section \ref{sec:ExperimentalProcedures}. Section \ref{sec:StillEnvironment} analyses the behaviour of particle clouds in a still environment, starting from the reference salt water thermals in the canonical framework of Morton \textit{et al.} \citep{mortonTurbulentGravitationalConvection1956} in section \ref{subsec:OnePhaseThermals}. The evolution of clouds is analysed by distinguishing two dynamical regimes, and we observe specific effects resulting from the particulate nature of the turbulence forcing when clouds behave as turbulent thermals. Section \ref{sec:RotatingEnvironment} analyses the additional influence of background rotation. Again distinguishing two regimes, we observe that rotation inhibits most of the particulate effects observed without rotation, and considerably increases the residence time of particles when they fall. In section \ref{sec:Conclusion}, main results are summed up and final remarks are made.
Lists of all experiments (table \ref{tab:NumbersExperiments}) and all notations (table \ref{tab:ListVariables}) can be found in appendixes.

%%%%%%%%%%%%%%%%%%%%%%%%%%%%%%%%%%%%%%%%%%%%%%%%
%%%%%%%%%%%%%%%%%%%%%%%%%%%%%%%%%%%%%%%%%%%%%%%%
\section{\label{sec:ExperimentalProcedures}Experimental apparatus}
\subsection{\label{subsec:ExperimentalSetup}Experimental setup}

\begin{figure}[htb]
\centering
    \begin{subfigure}[b]{.65\textwidth}
      \caption{}
      \centering
      \includegraphics[height=8cm]{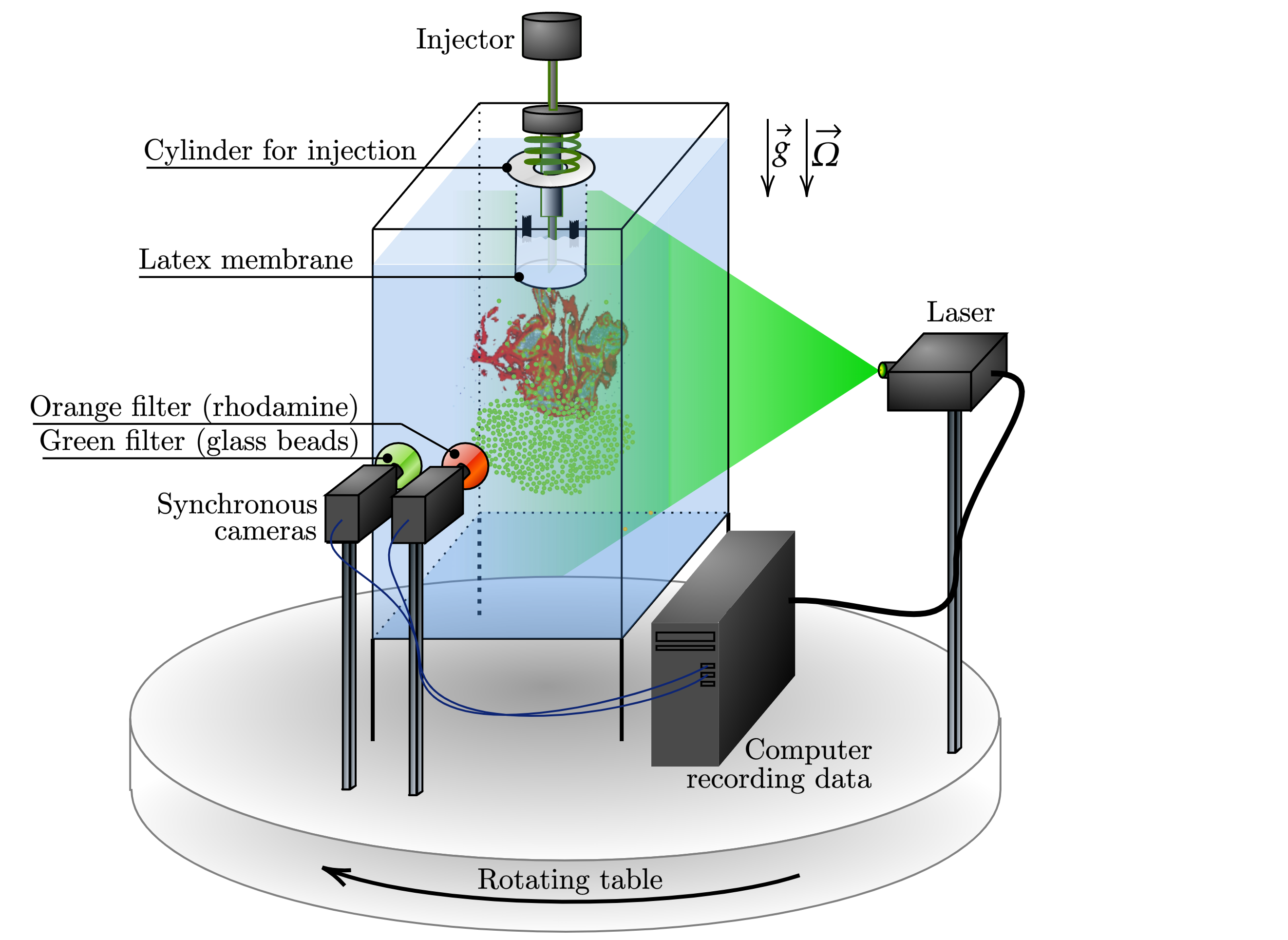}
      \label{subfig:Apparatus}
    \end{subfigure}
    \begin{subfigure}[b]{.34\textwidth}
    \caption{}
      \centering
      \includegraphics[height=8cm]{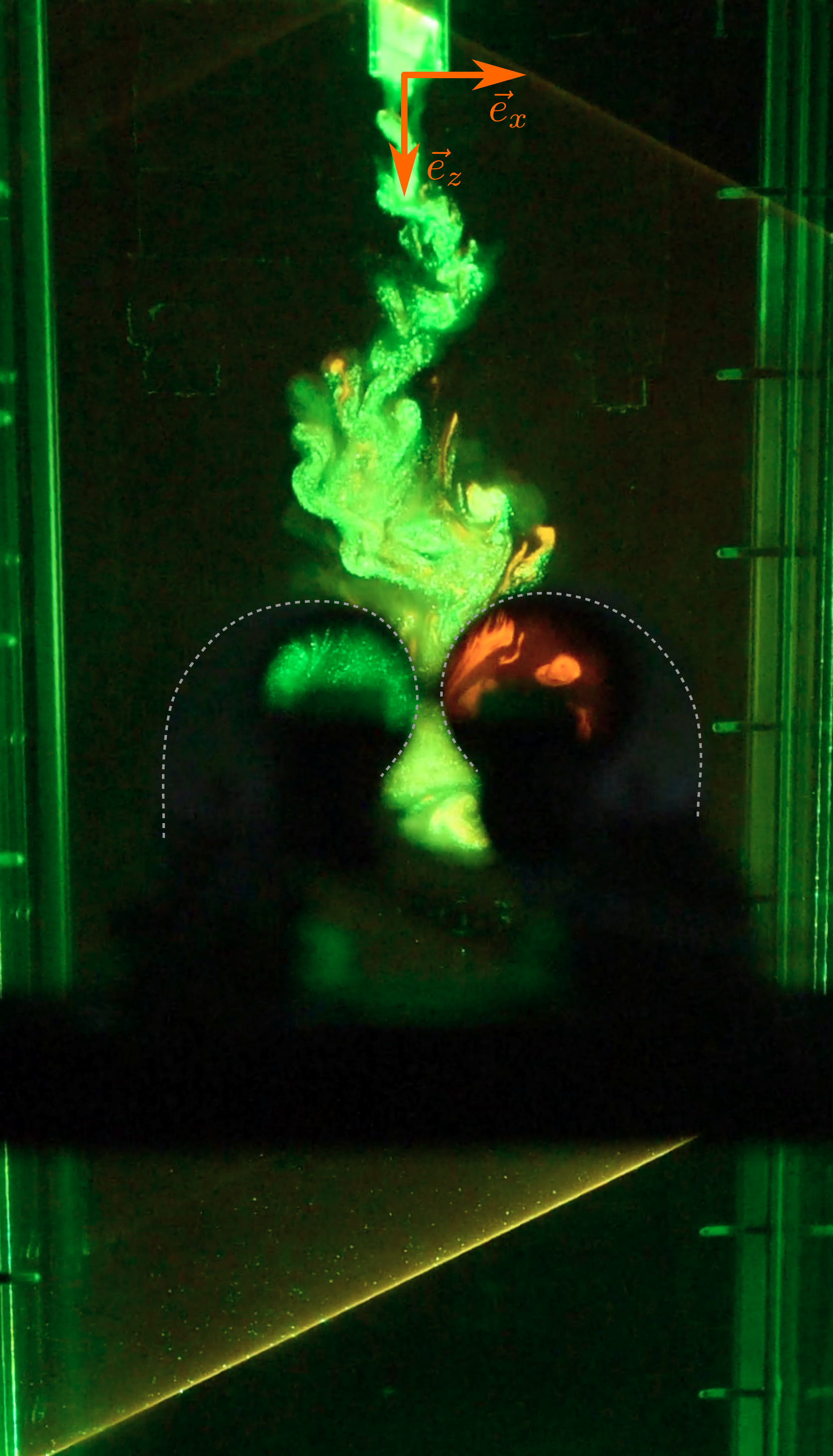}
      \label{subfig:CamerasLaserSheet}
    \end{subfigure}
    \caption{(a) Schematic of the experimental apparatus; see main text for additional details. (b) Turbulent cloud falling with rhodamine. Dotted gray lines highlight the edges of the two filter holders. Particles (respectively, rhodamine) are visible through the filter located on the left-hand side (respectively, right-hand side).}
  \label{fig:Apparatus}
\end{figure}

The apparatus is illustrated in figure \ref{fig:Apparatus}. The experiments are performed in a Plexiglas tank of height $100$cm and cross-section area $42\times42$cm$^2$ containing approximately 160L of fresh water ($\rho_f=998$kg.m$^{-3}$, $\nu=9.57\times 10^{-5}$m$^2$.s$^{-1}$). The tank is filled long before the experiments to ensure that water is at room temperature, i.e. 22$^\circ$C on average. The tank is fixed in the middle of a rotating table whose angular velocity $\Omega$ varies from 0 (no rotation) to 20 rotations per minute (rpm). A lid is placed on top of the tank to limit air motions above the free surface during the experiments. A hole at the centre of the lid enables to insert a cylinder of inner diameter $D_{\textrm{cyl}}=3.2$cm to release the buoyant material.
\par
Clouds are either made of salt water, or a mixture of 26.1mL of fresh water and a fixed mass $m_0=1.0$g of spherical glass beads of density $\rho_p = 2500$kg.m$^{-3}$ and a mean radius $r_p$ ranging from $2.5\mu$m to $500\mu$m (see discussion in section \ref{subsec:ParticleDistributions}). In all cases, the buoyancy introduced into the system is the same. Adding fresh water to particles has two motivations. First, in air, small particles tend to cluster because of electrostatic interactions which are easily removed by placing particles in water with a small amount of soap (typically one drop for 20cl of water). Secondly, if particles fall from air into water, we observe that they entrain some air with them, form clusters, and expell air only gradually as ascending bubbles which are detrimental for the detection of particles and might affect the particle dynamics of interest here. These effects are therefore avoided; the reader is referred to reference \citep{zhaoFormationParticleClouds2014a} for more information about clouds containing such initial clumps of particles. After the cylinder's bottom nozzle has been sealed by a latex membrane which is stretched and taped onto the cylinder itself, the buoyant material is poured into the cylinder. Water in the cylinder always occupies a volume rising at a height $H_0 = 3.3$cm above the latex membrane.\\

\begin{figure}[htb]
    \centering
    \includegraphics[width=\textwidth]{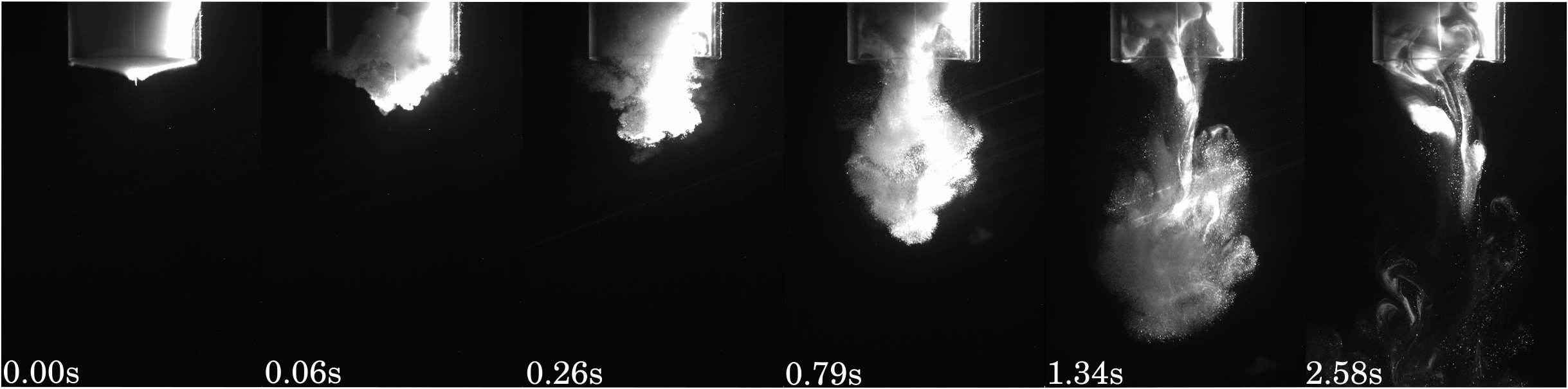}
    \caption{Rupturing (first frame) of the latex membrane, releasing rhodamine and glass beads whose diameter is in the range 90-150$\mu$m. Photographs are recorded by a single monochrome camera with no filter. Particles are sufficiently large to be distinguished from the continuous field of rhodamine. The cylinder being transparent, the needle and recirculation of rhodamine are visible on the last two photographs.}
    \label{fig:RuptureMembrane}
\end{figure}

Initially there is no relative motion between water and the tank (either water and the tank are motionless, or they are in solid body rotation with the rotating table). At $t=0$ the experiment starts by rupturing the latex membrane with a needle, releasing the content of the cylinder; see figure \ref{fig:RuptureMembrane} for an illustration. The 1-mm-thick tip of the needle is sharpened to ensure an efficient and fast rupturing of the latex membrane in less than $0.02$s according to the videos recorded. Once the membrane retracts, particles are observed to fall out of the cylinder because of their weight. For most particle sizes, the downward acceleration of particles quickly transmits to the fluid, the buoyant material rolls up and the cloud almost immediately becomes turbulent.
\par
Visualisations are performed in a vertical laser sheet with half-angle of divergence $30^\circ$, using a Powell lens and a laser of wavelength $532$nm with a power of $1$W or $1.5$W depending on the series of experiments (Laser Quantum 532nm CW laser 2 W). Since particles and water have different motions, two identical PointGrey cameras are synchronised and record the same experiment with two different filters. The first camera has a green filter (band-pass filter from Edmund Optics, CWL 532nm, FWHM 10nm) to record the motion of glass beads which reflect and refract the laser beam, while the second camera has an orange filter (high-pass filter above 570nm). By colouring the fluid inside the cylinder with rhodamine, the second camera records the motion of the turbulent eddies which appear in orange in the laser sheet because of the fluorescence of rhodamine. In doing so, there is no overlap of information between the two cameras. Both of them save images in a format $960\times600$ pixels, corresponding to a field of view whose size is $45$cm in depth and $28$cm in width. Cameras are synchronised, recording images at 50 frames per second (fps). To minimise effects of parallax, they are placed as close as possible to one another, and as far as possible from the laser sheet. A calibration grid was placed in the plane of the laser sheet and photographed by both cameras. This enables to define a coordinate system for each of the two cameras, and to dewarp images so that synchronous photographs can be sumperimposed using Python's library OpenCV.

\subsection{\label{subsec:InitialCondRelease}Initial conditions of release}

The initial release of particles is notably controlled by the effective density of the fluid within the cylinder, which depends on the volume fraction in particles. Two initial volume fractions can be imposed. Either particles are stirred in the cylinder to form a suspension in water before rupturing the membrane, and in that case the initial release is said to be \textit{dilute}. Otherwise, particles are left to settle before the experiment so that they form a compact layer at the bottom of the cylinder, directly lying on the latex membrane, and in that case the initial release is said to be \textit{compact}.
\par
Since larger particles settle faster, only particles with $r_p \leq 30 \mu$m could be maintained in suspension prior to puncturing the membrane. This is why to compare particle clouds of different particle sizes, experiments have been performed with a compact release for all particle ranges. Then, additionally, experiments have been performed with a dilute release when feasible. The motivation for these dilute releases is twofold: it enables to assess the effect of this initial condition on the cloud development, and it enables to compare the behaviour of particles with respect to salt water clouds, for which salt is always dilute in the whole volume of water within the cylinder.
\par
Then, we want to consider a quasi-instantaneous release of particles, but the smallest ones can take up to 7min to settle over the depth of the cylinder. Hence, the mass excess $m_0=1$g needs to be expelled from the cylinder. Yet, we are interested in the flow generated only by the glass beads settling. A compromise consists in immersing the cylinder on the last $2.2$cm of the $3.3$cm-high volume of water. Once the membrane is ruptured, the resulting hydrostatic imbalance with the ambient fluid leads to a downward acceleration of the buoyant mixture which is released on a much shorter time scale, at first order independent from the beads' mean radius and the dilute or compact initial conditions.
\par
In order to assess the robustness of our protocole, let us determine the depth beyond which this initial momentum has a negligible influence on the cloud evolution. When the buoyant material comes out of the cylinder, friction on the inner walls of the cylinder as well as shear on the sides of the buoyant fluid both generate circulation, forcing the buoyant fluid to roll up. The initial acceleration of all the fluid in the cylinder due to the hydrostatic imbalance actually adds some more circulation, which levels the difference between compact and dilute releases: in fact, we observe that for both dilute and compact releases in the range $r_p \leq 30 \mu$m, all clouds initially roll up and end up approximately spherical with typical size $D_{\textrm{cyl}}$ at a depth of order 1-2$D_{\textrm{cyl}}$. For a buoyant cloud with initial momentum, the Morton length \citep{mortonForcedPlumes1959,turnerTurbulentEntrainmentDevelopment1986} quantifies the distance beyond which the cloud buoyancy predominates over the initial momentum. For our experiments, the Morton length can be expressed [see equation (A.15) in \citep{deguenExperimentsTurbulentMetalsilicate2011}]
\begin{equation}
\label{eq:MortonLength}
    l_M = \left [ \frac{(\rho_0/\rho_f)^2 U_{\textrm{ref}}^2 D_{\textrm{cyl}}^3}{g(\rho_0/\rho_f-1)} \right ]^{1/4},
\end{equation}
with $g=9.81$m.s$^{-2}$. The initial cloud density $\rho_0$ relevant for its dynamics is computed once the cloud has rolled up as a sphere of radius $D_{\textrm{cyl}}$ hence $\rho_0 = \rho_f + (1-\rho_f/\rho_p) m_0/(4\pi D_{\textrm{cyl}}^3/3)$. Then, the typical cloud velocity $U_{\textrm{ref}}$ is prescribed by the cloud size and reduced gravity as 
\begin{equation}
\label{eq:Uref}
    U_{\textrm{ref}} = \sqrt{g \left (1-\frac{\rho_f}{\rho_0} \right)D_{\textrm{cyl}}},
\end{equation}
which is the same for all experiments. We find that $l_M \simeq D_{\textrm{cyl}}$. Hence, after the cloud has rolled up and travelled a distance $l_{M}$, corresponding to a total distance of order 2-3$D_{\textrm{cyl}}$, the influence of the initial momentum can be neglected.

\subsection{\label{subsec:ParticleDistributions}Particle distributions}
Particles are provided by Sigmund Lindner (SiLibeads) in sets of polydisperse distributions. For every set of particles, a Gaussian is fitted on the histogram of the distribution of particles' radii. This provides the average radius $r_p$ of the set of particles, as well as the standard deviation $\sigma_p$ with respect to $r_p$. The ratio $\mathcal{S} = \sigma_p/r_p$ is a measure of the polydispersity of a given set of particles. The average radius $r_p$ and the ratio $\mathcal{S}$ are listed in table \ref{tab:NumbersAllExp} for polydisperse, monodisperse and bidisperse particle sets.
\par
Monodisperse sets are obtained due to sieves, each of them corresponding to a given size which truncates the Gaussian. The distribution obtained is integrated to compute the resulting mean radius. The characteristic deviation $\sigma_p$ is computed simply as the difference in radius between $r_p$ and the cutoff value of the sieve.
\par
Bidisperse sets are obtained by mixing together a percentage $p$ of the total mass $m_0=1$g from the monodisperse set $90-100\mu$m, and a percentage $1-p$ of $m_0$ from the monodisperse set $140-150\mu$m. The characteristic deviation $\sigma_p$ is computed as the difference between the resulting average radius $r_p$, and the average radius of the monodisperse set whose percentage ($p$ or $1-p$) is the largest.

\begin{table}[htb]
\renewcommand{\arraystretch}{1.1}
\begin{tabular}{ccc||c|c|Hc|c|c|cHH|c}
&&Range of diameters ($\mu$m) & Symbol & $r_p$ ($\mu$m) & $\Omega$ (rpm) & $\mathcal{R}$   & $Re_p$          & $N_p$    & $\mathcal{S}$ & $\Pi_\rho$ & $m_0$ (g) & $\Pi$ ($\times 10^{-4}$) \\
\hline
\parbox[t]{2mm}{\multirow{8}{*}{\rotatebox[origin=c]{90}{Polydisp.}}} &&
0-20                            & $\Circle$ & 2.6 & 0 & 6.00$\times 10^{-4}$ & 1.16$\times 10^{-4}$ & 5.43$\times 10^{9}$ & 1.0 & 2.5 & 1 & 0.813 \\
&&40-70                           & $\Circle$ or $\CIRCLE$ & 29.9 & 0 & $7.57\times 10^{-2}$ & 0.167 & 3.57$\times 10^6$ & 0.35 & 2.5 & 1 & 9.34 \\
&&90-150                          & $\CIRCLE$ & 64.4  & 0 & 0.308 & 1.47 & 3.58$\times 10^5$ & 0.3 & 2.5 & 1 & 20.1 \\
&&150-250                         & $\CIRCLE$ & 108.1 & 0 & 0.696 & 5.57 & $7.56\times 10^4$ & 0.23 & 2.5 & 1 & 33.8 \\
&&250-500                         & $\CIRCLE$ & 192.4 & 0 & 1.48 & 21.1 & $1.34\times 10^4$ & 0.22 & 2.5 & 1 & 60.1 \\
&&500-750                         & $\CIRCLE$ & 339.8 & 0 & 2.73 & 68.8 & $2.43\times 10^3$ & 0.24 & 2.5 & 1 & 106 \\
&&750-1000                        & $\CIRCLE$ & 465.9 & 0 & 3.70 & 128 & 944 & 0.18 & 2.5 & 1 & 146 \\
\hline
\parbox[t]{2mm}{\multirow{4}{*}{\rotatebox[origin=c]{90}{Monodisp.}}}
&&90-100                          & $\blacksquare$ & 45.7 & 0 & 0.168 & 0.567 & 1.00$\times 10^6$ & 0.029 & 2.5 & 1 & 14.3 \\
&&140-150                         & $\blacksquare$ & 76.1  & 0 & 0.406 & 2.29 & $2.17\times 10^5$ & 0.029 & 2.5 & 1 & 23.8 \\
&&224-250                         & $\blacksquare$ & 120.8 & 0 & 0.814 & 7.28 & $5.42\times 10^4$ & 0.07 & 2.5 & 1 & 37.8 \\
&&$\geq$ 1000                     & $\blacksquare$ & 524.5 & 0 & 4.13 & 160 & 662 & 0.093 & 2.5 & 1& 164 \\
\hline
\parbox[t]{2mm}{\multirow{4}{*}{\rotatebox[origin=c]{90}{Bidisp.}}}
&&$p=80\%$ & $\blacktriangle$ & 51.8  & 0 & 0.210 & 0.805 & $6.87\times 10^5$ & 0.118 & 2.5 & 1 & 16.2 \\
&&$p=60\%$ & $\blacktriangle$ & 57.8  & 0 & 0.255 & 1.09 & $4.95\times 10^5$ & 0.211 & 2.5 & 1 & 18.1 \\
&&$p=40\%$ & $\blacktriangle$ & 63.9  & 0 & 0.304 & 1.44 & $3.66\times 10^5$ & 0.191 & 2.5 & 1 & 20.0 \\
&&$p=20\%$ & $\blacktriangle$  & 70.0  & 0 & 0.354 & 1.84 & $2.78\times 10^5$ & 0.087 & 2.5 & 1 & 21.9 \\
\end{tabular}
\caption{Governing numbers of all series of experiments. The column `Range of diameters' describes the range of diameters of the glass spheres (in microns) as referenced by the trademark SiLiBeads. Percentages $p$ of bidisperse clouds are defined in the main text. The symbols are those used in figures throughout the paper. Shapes depend on the dispersity of particle sets. Filled symbols correspond to a compact release of particles, and empty symbols correspond to a dilute release, as defined in the text. Particle clouds are also compared to salt water clouds of Rouse number $\mathcal{R}=0$, also represented by an empty circle $\Circle$.}
\label{tab:NumbersAllExp}
\end{table}

\subsection{\label{subsec:DimlessNumbers}Dimensionless numbers}

In the present study, particles are considered to interact only through hydrodynamic interactions. This assumption is based on three arguments: particles of size $r_p \geq 2.5\mu$m are non-Brownian at temperatures around 290-300K \cite{andreottiMilieuxGranulairesEntre2012}; soap was used during experiments to avoid physico-chemical interactions; finally, collisions are expected to play little part in the cloud dynamics except during the phase of acceleration when particles are released from a compact layer in the cylinder. Hence, particles interact by inducing a velocity perturbation on other particles with a magnitude decreasing with distance, or by wake interactions.
\par
Thus, the following dimensional quantities describe the dynamics of the clouds and the particles they are laden with: gravity $g$, the fluid density $\rho_f$ and kinematic viscosity $\nu$, the particles' density $\rho_p$, the total mass of particles $m_0$, the average radius of particles $r_p$, the standard deviation $\sigma_p$ of the supposedly Gaussian distribution of particles' radii, and finally the diameter of the cylinder $D_{\textrm{cyl}}$. The tank angular velocity is not included yet; effects of rotation are discussed in section \ref{sec:RotatingEnvironment}. The tank dimensions are not considered, assuming they are large enough not to influence the dynamics.\\

These parameters enable us to define the terminal settling velocity $w_s$ of a single particle for the two following end members.
Very small particles have a negligible particulate Reynolds number $Re_p = 2r_p w_s/\nu \ll 1$ and therefore move in a Stokes flow. Assuming sphericity, the balance between their buoyancy and the linear Stokes drag leads to defining
\begin{equation}
    \label{eq:wsStokes}
    w_s^{\textrm{Stokes}} = \frac{2gr_p^2(\rho_p-\rho_f)}{9\nu \rho_f} \propto r_p^2.
\end{equation}
As for large particles characterised by a large particle Reynolds number $Re_p\gg 1$, the balance between their buoyancy and a quadratic drag law leads to
\begin{equation}
    \label{eq:wsNewton}
    w_s^{\textrm{Newton}} = \sqrt{\frac{8g(\rho_{p}-\rho_{f})}{3C_d \rho_{f}} r_p} \propto \sqrt{r_p},
\end{equation}
with $C_d$ the drag coefficient, approximately constant and equal to 0.445 for a sphere in the range $Re_p\in [750, 3.5\times 10^5]$ (see \cite{croweMultiphaseFlowsDroplets2011}). Several empirical expressions exist in the literature to capture the smooth transition from regime \eqref{eq:wsStokes} to regime \eqref{eq:wsNewton} when $r_p$ increases. For particulate Reynolds numbers lower than $\sim$800 as is the case here, a classical equation is provided by the Schiller-Naumann correction to the Stokes velocity \citep{croweMultiphaseFlowsDroplets2011}:
\begin{equation}
    \label{eq:wsSchillerNaumann}
    w_s = \frac{w_s^{\textrm{Stokes}}}{1+0.15Re_p^{0.687}}.
\end{equation}

Finally, the eight quantities listed above involve three dimensions, thus according to the Vaschy-Buckingham theorem, five dimensionless numbers are defined as listed below:
\begin{equation}
    \mathcal{S}=\frac{\sigma_p}{r_p};
    \qquad N_p = \frac{3m_0}{4\pi r_p^3 \rho_p};
    \qquad \Pi = \frac{r_p}{D_{\textrm{cyl}}};
    \qquad Re_p = \frac{2r_pw_s}{\nu};
    \qquad \mathcal{R} = \frac{w_s}{U_{\textrm{ref}}}.
\end{equation}
The ratio $\mathcal{S}$ quantifies the dispersity of particle distributions. The total number of particles $N_p$ plays an important part in the particles' interactions, and governs the initial fall. The ratio $\Pi$ is typically adequate to characterise the influence of the velocity perturbations induced by a single particle at initial times when the cloud size is of order $D_{\textrm{cyl}}$ \citep{subramanianEvolutionClustersSedimenting2008,pignatelFallingCloudParticles2011}. The particulate Reynolds number $Re_p$ compares advection and molecular diffusion in the flow produced around a particle as it settles. Finally, particles and water have different motions since particles have inertia, as quantified by their Rouse number $\mathcal{R}$ which compares the inertia of a particle (through its terminal velocity $w_s$) and that of the cloud which sustains that particle. Since the reference fluid velocity $U_\textrm{ref}$ is the same for all experiments (equation \eqref{eq:Uref}), in this study the Rouse number varies only with the particles' radius: the larger the particle, the larger its inertia, the larger the Rouse number. Dimensionless numbers for each set of beads are listed in table \ref{tab:NumbersAllExp}, and appendix \ref{sec:ListExperiments} additionally provides a complete list of the experiments performed.\\

%%%%%%%%%%%%%%%%%%%%%%%%%%%%%%%%%%%%%%%%%%%%%%%%
%%%%%%%%%%%%%%%%%%%%%%%%%%%%%%%%%%%%%%%%%%%%%%%%
\section{\label{sec:StillEnvironment}Particle clouds in a still environment}

\subsection{\label{subsec:OnePhaseThermals}The turbulent thermal as a one-phase reference}
In this section, we focus on the release of salt water clouds in fresh water, and interpret our results in the framework of Morton \textit{et al.} \citep{mortonTurbulentGravitationalConvection1956} (the paper hereafter abbreviated as MTT56), whose equations have proved applicable in a multitude of contexts (see section \ref{sec:Introduction}), highlighting the role of attractor of the model of turbulent thermal. We then use this purely fluid case to highlight and understand any specificity due to the particulate nature of the turbulence forcing in the next sections.
\par

\begin{figure}[htb]
    \begin{subfigure}[b]{.82\textwidth}
    \caption{}
    \centering
    \includegraphics[height=4.65cm]{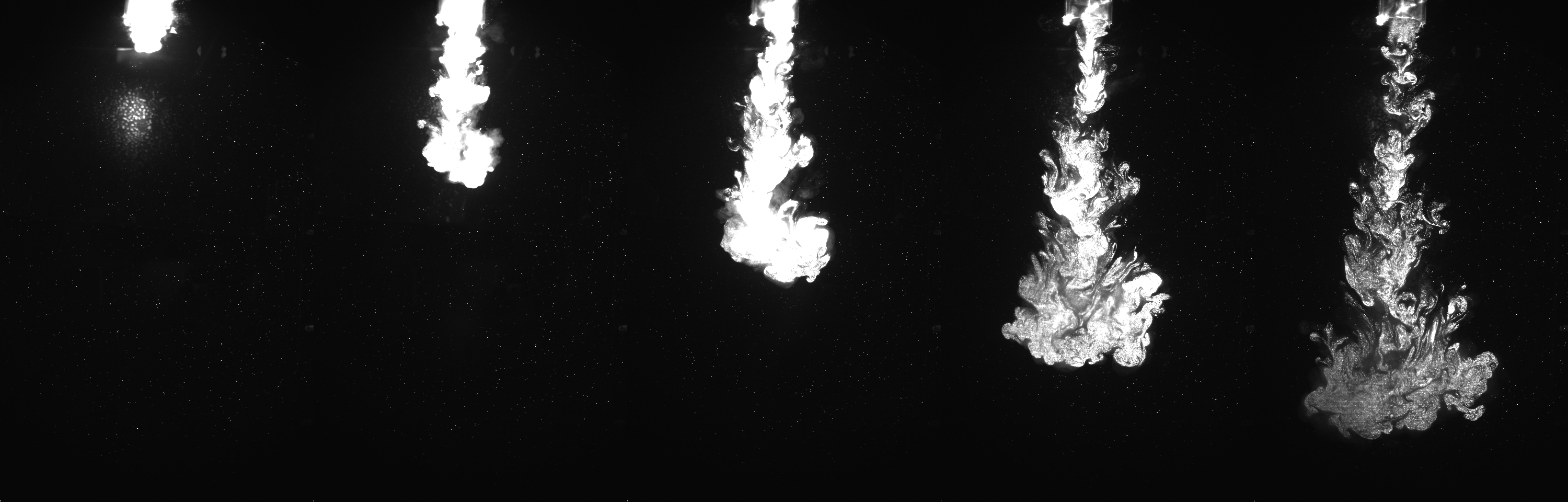}
    \label{subfig:SnapshotsSalt0rpm}
    \end{subfigure}
    \hfill
    \begin{subfigure}[b]{.17\textwidth}
    \caption{}
    \centering
    \includegraphics[height=4.65cm]{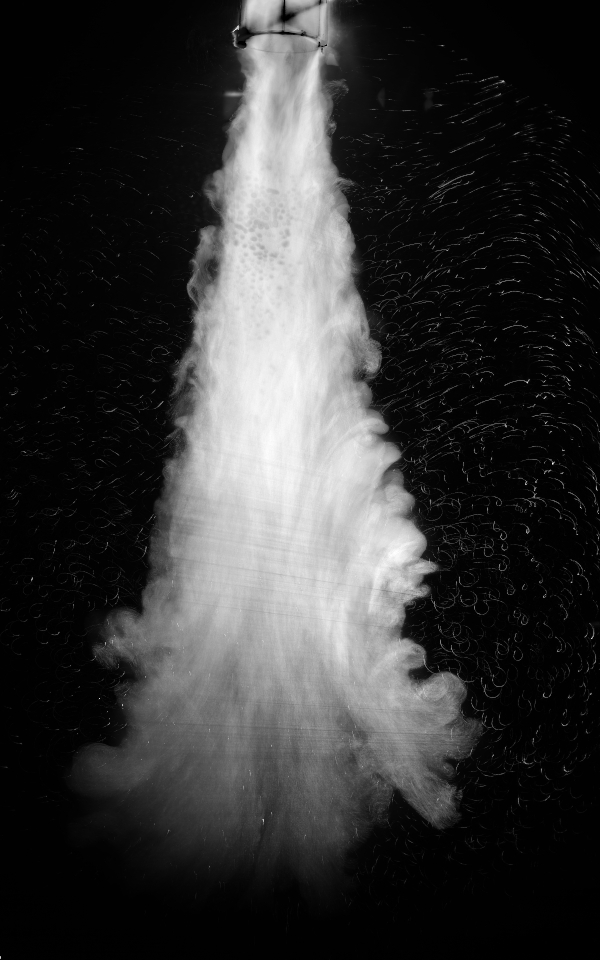}
    \label{subfig:AVGSalt0rpm}
    \end{subfigure}
    \caption{(a) Salt-water cloud falling in still, fresh water. The time lapse between frames is 1.6s, and all photographs are 45cm high. (b) Pixel-by-pixel standard deviation of light intensity during the same experiment, over 8.0s.}
    \label{fig:saltWaterReference}
\end{figure}

The robustness of MTT56 comes from its integral description of the main physics of the thermal as well as the small number of hypotheses required. The shape of the thermal is expected to play little part on its dynamics, and it is approximated by a sphere of radius $r(t)$ and uniform density $\rho(t)$ so that describing the thermal evolution only requires three variables: $r(t)$, $\rho(t)$, and the vertical velocity $\dot{z}(t)=dz/dt$ whose integration leads to the thermal position in depth $z(t)$. Knowing the thermal velocity, these quantities can alternatively be computed as functions of $z$: $r(z)$, $\rho(z)$, $\dot{z}(z)$.
\par
Figure \ref{fig:saltWaterReference} shows the evolution of a salt-water thermal of excess mass $m_0=1$g released in the same condition as any particle cloud. We observe that the turbulent cloud radius grows linearly with depth. This is captured by MTT56 through the entrainment hypothesis. The turbulence developing inside the cloud is considered responsible for a uniform inflow of ambient fluid into the spherical thermal -- a process called \textit{entrainment} -- with a characteristic inflow velocity $v_e$. The \textit{entrainment hypothesis} states that this velocity is proportional to the sole characteristic velocity scale of the system, the thermal vertical velocity $\dot{z}$, so that the entrainment velocity reads

\begin{equation}
\label{eq:EntrainmentVelocity}
    v_e=\alpha \dot{z},
\end{equation}
with $\alpha$ a positive constant called the \textit{coefficient of entrainment}. From this model, the mass of the thermal increases in time due to the uniform entrainment of ambient fluid over the thermal's surface. Neglecting any \textit{detrainment}, i.e. any outflow of buoyant material which would be lost in the wake (or \textit{stem}) behind the thermal, the mass anomaly $m_0$ remains constant within the cloud. Finally the evolution of the cloud's momentum is mainly driven by the constant buoyancy force $m_0g$, and mitigated by a quadratic drag term. The resulting equations of evolution are

\begin{subnumcases}{}
	\frac{d}{dt} \left[ \frac{4}{3}\pi r^3 \rho \right ] & $= 4\pi r^2 v_e \rho_f = 4\pi r^2 \alpha \dot{z} \rho_f$ \label{eq:ThMassCons}\\
	\frac{d}{dt} \left [ \frac{4}{3}\pi r^3 (\rho-\rho_f) \right ] & $= \dfrac{dm_0}{dt} = 0$ \label{eq:ThExcessCons}\\
	\frac{d}{dt} \left [ \frac{4}{3} \pi r^3 \rho \dot{z} \right ] & $= m_0g - \frac{1}{2} \rho_f \dot{z}^2 \pi r^2 C_D$ \label{eq:ThMomCons}
\end{subnumcases}
with $C_D$ a drag coefficient. Combining the equations of conservation of mass \eqref{eq:ThMassCons} and mass excess \eqref{eq:ThExcessCons} immediately yields $\dot{r}=\alpha \dot{z}$ which shows that $\alpha$ directly quantifies the constant growth rate of the thermal in depth with $\alpha=dr/dz$. Figure \ref{fig:saltWaterReference} confirms the linear growth of the turbulent thermal in depth with snapshots [figure \ref{subfig:SnapshotsSalt0rpm}] and with the pixel-by-pixel standard deviation of light intensity captured by the camera over 8.0s of the cloud fall [figure \ref{subfig:AVGSalt0rpm}]. Knowing the couple $(r,\rho)$ at a given depth $z_0$ enables to compute $(r,\rho)$ at any depth $z$ from the sole knowledge of $\alpha$ and conservation of $m_0$. Finally, integration of the whole model shows that the cloud decelerates as it entrains ambient fluid. With no specific assumption on $C_D$, self-similar solutions can be found at large times, which scale like (see \citep{escudierMotionTurbulentThermals1973})

\begin{subnumcases}{}
	 r\sim z \sim t^{1/2}, \label{eq:similarR}\\
	 \dot{z}\sim z^{-1}\sim t^{-1/2}, \label{eq:similarVelocity}
\end{subnumcases}
In the present context, particle clouds follow this regime only after an initial phase of acceleration and beyond the Morton length [equation \eqref{eq:MortonLength}].

\subsection{\label{subsec:ThermalRegime}Thermal regime: specificities of particle-induced entrainment}

\begin{figure}[htbp]
\centering
    \begin{subfigure}[b]{0.82\textwidth}
    \caption{$\mathcal{R}=6.00\times 10^{-4}$, $\Delta t=2.5s$}
      \centering
      \includegraphics[height=4.52cm]{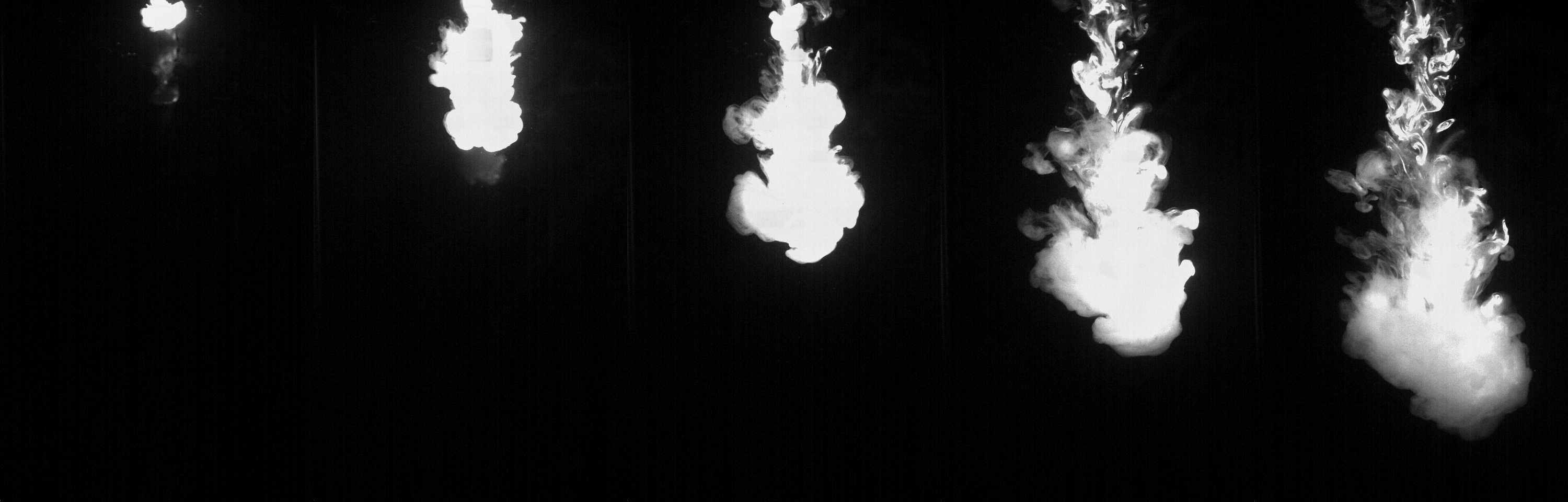}
      \label{subfig:Sizes-0rpm-0-20}
    \end{subfigure}
    \begin{subfigure}[b]{0.17\textwidth}
    \caption{}
      \centering
      \includegraphics[height=4.52cm]{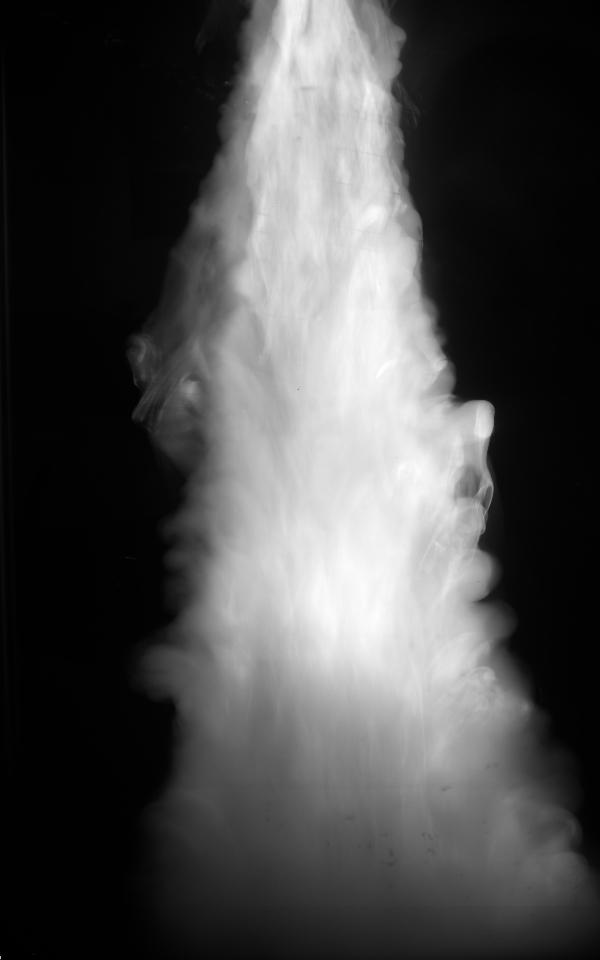}
      \label{subfig:SizesSTD-0rpm-0-20}
    \end{subfigure}

    \begin{subfigure}[b]{0.82\textwidth}
      \caption{$\mathcal{R}=7.57\times 10^{-2}$, $\Delta t=3s$}
      \centering
      \includegraphics[height=4.52cm]{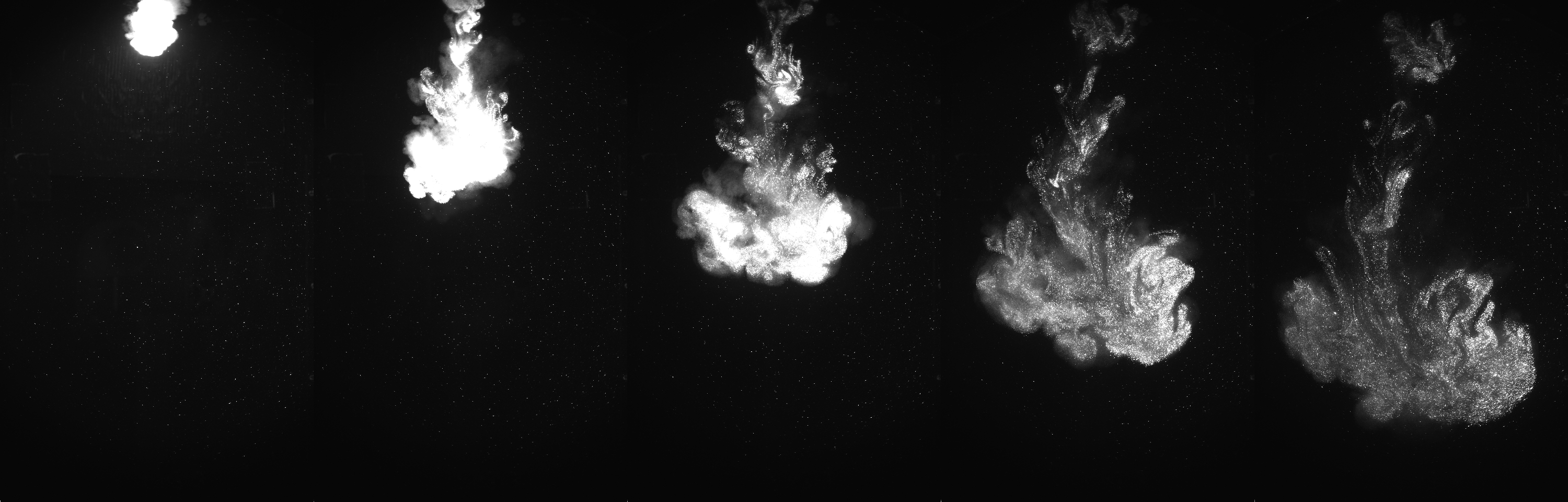}
      \label{subfig:Sizes-0rpm-40-70}
    \end{subfigure}
    \begin{subfigure}[b]{0.17\textwidth}
      \caption{}
      \centering
      \includegraphics[height=4.52cm]{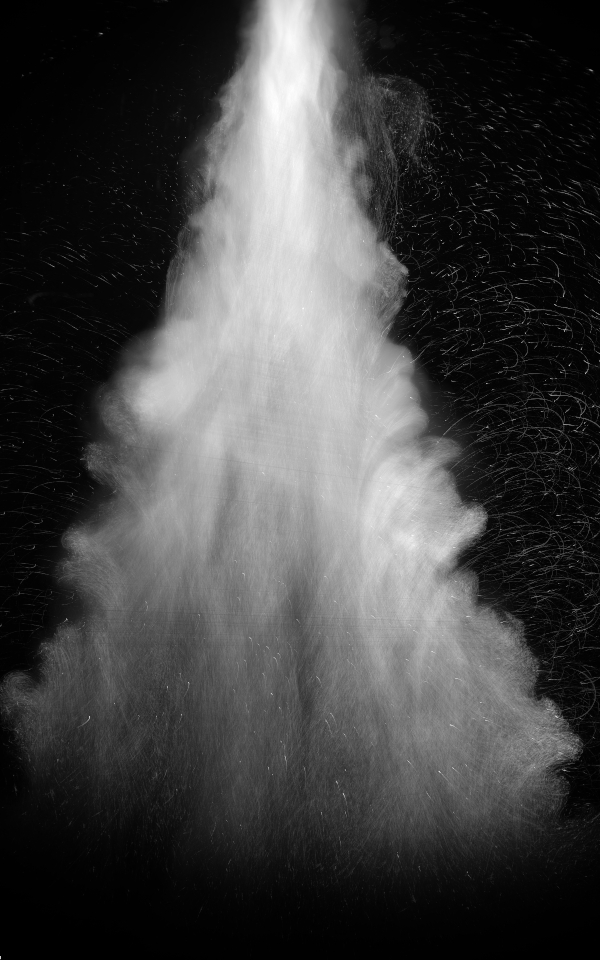}
      \label{subfig:SizesAVG-0rpm-40-70}
    \end{subfigure}

    \begin{subfigure}[b]{0.82\textwidth}
      \caption{$\mathcal{R}=0.406$, $\Delta t=2s$}
      \centering
      \includegraphics[height=4.52cm]{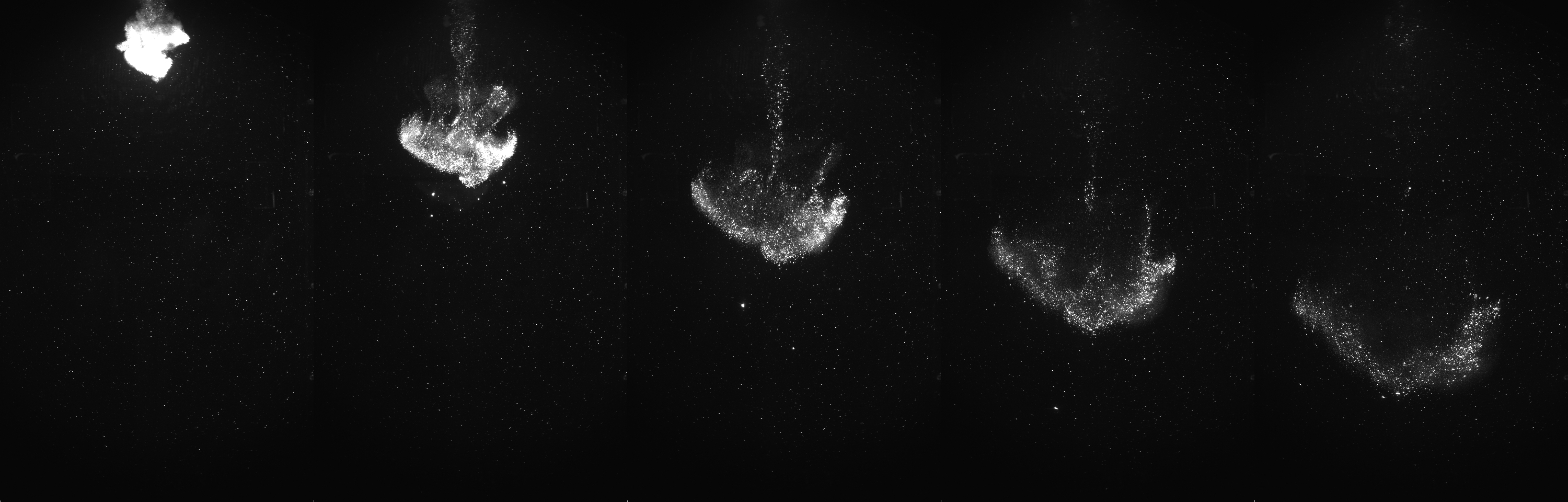}
      \label{subfig:Sizes-0rpm-140-150}
    \end{subfigure}
    \begin{subfigure}[b]{0.17\textwidth}
      \caption{}
      \centering
      \includegraphics[height=4.52cm]{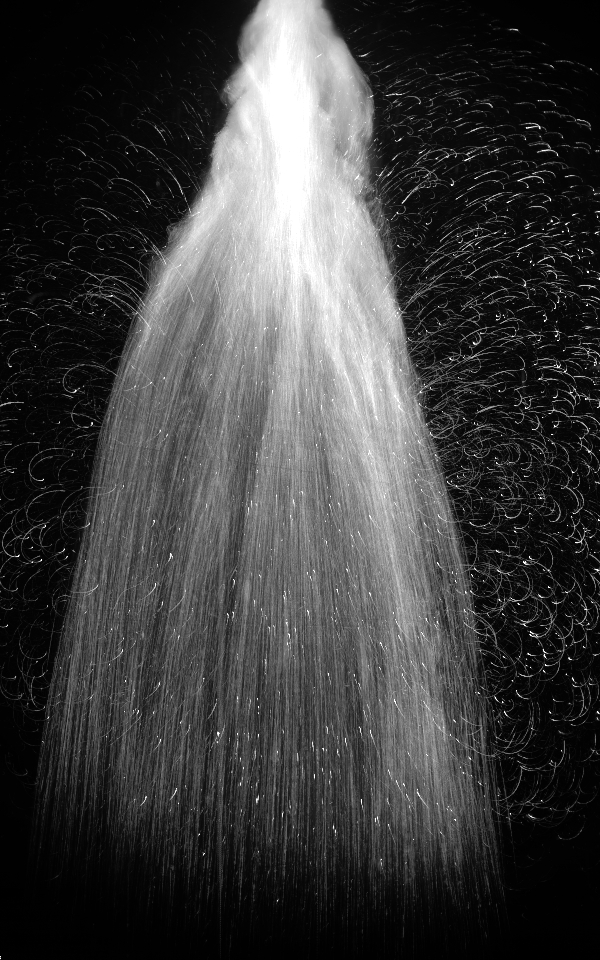}
      \label{subfig:SizesAVG-0rpm-140-150}
    \end{subfigure}

    \begin{subfigure}[b]{0.82\textwidth}
      \caption{$\mathcal{R}=3.70$, $\Delta t=0.5s$}
      \centering
      \includegraphics[height=4.52cm]{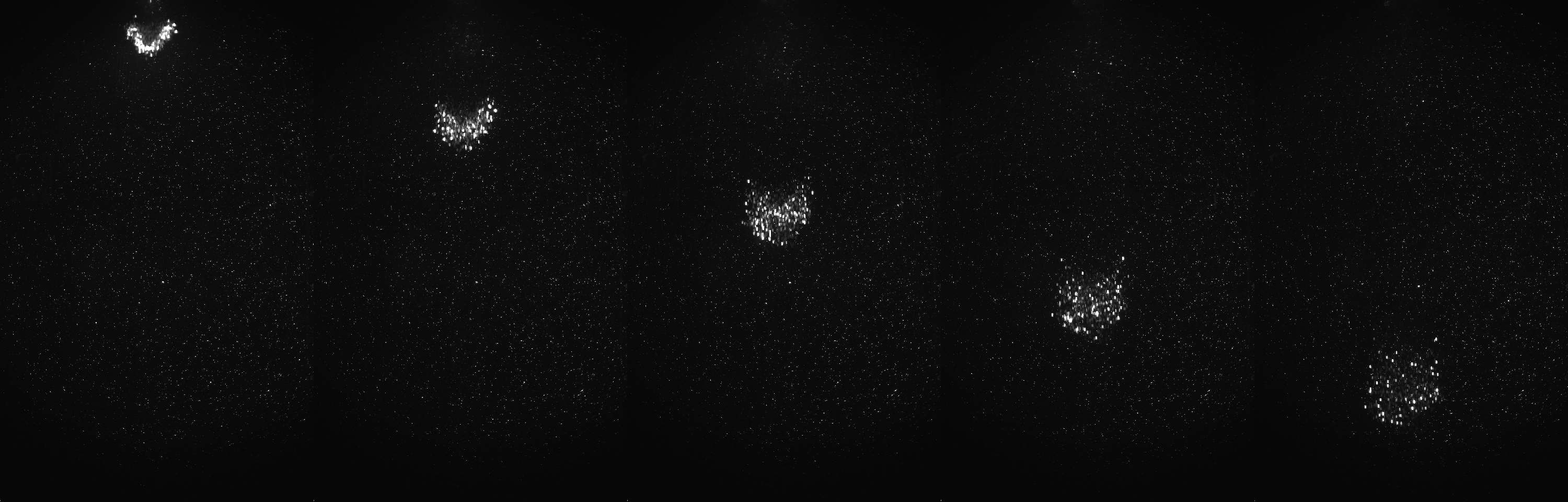}
      \label{subfig:Sizes-0rpm-750-1000}
    \end{subfigure}
    \begin{subfigure}[b]{0.17\textwidth}
      \caption{}
      \centering
      \includegraphics[height=4.52cm]{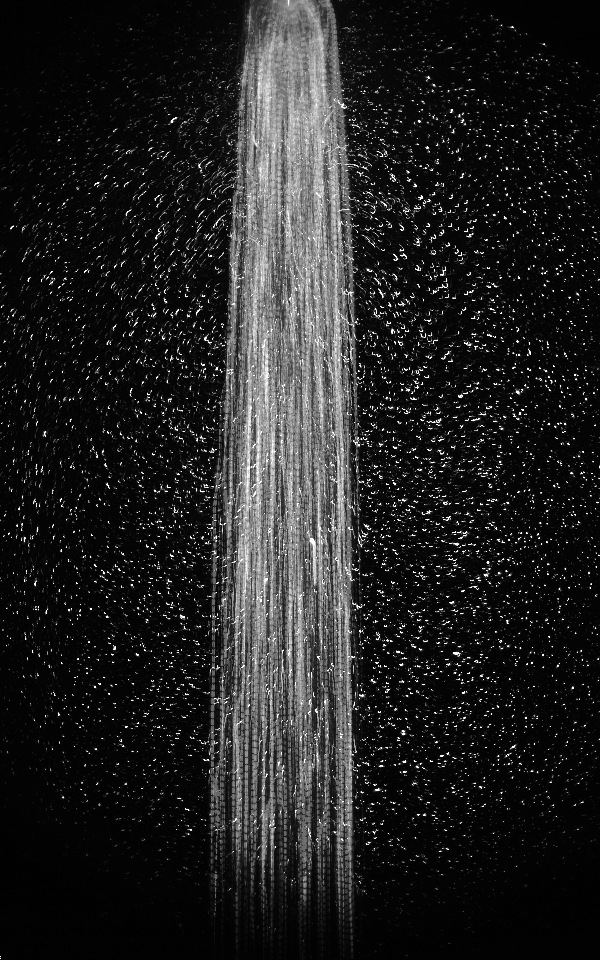}
      \label{subfig:SizesAVG-0rpm-750-1000}
    \end{subfigure}
    \caption{(a,c,e,g): Particle clouds falling for different Rouse numbers $\mathcal{R}$ in the absence of background rotation. The height of all snapshots is 45cm and the time lapse between two successive snapshots $\Delta t$ is indicated in each subtitle. (b,d,f,h): Pixel-by-pixel standard deviation of all photographs taken during the experiment on the same row, of respective durations (b) 20s, (d) 28s, (f) 16s, (h) 3.4s. Bright dots in the background correspond to remaining particles from previous experiments which have no influence during experiments.}
  \label{fig:sizeEffect}
\end{figure}

The broad behaviour of particle clouds is now compared to the reference one-phase turbulent thermals. Figure \ref{fig:sizeEffect} presents snapshots of particle clouds for four different Rouse numbers. At low Rouse numbers, clouds clearly appear to decelerate and linearly grow in depth, in much the same way as salt-water thermals. Yet, for large Rouse numbers, defining a coefficient of entrainment due to the cloud growth hardly seems feasible since the particle cloud spreads little during its fall. This observation is confirmed by the pixel-by-pixel standard deviation of light intensity computed over the fall of a cloud of Rouse number $\mathcal{R}=3.70$ (see figure \ref{subfig:SizesAVG-0rpm-750-1000}), as opposed to similar visualisations for $\mathcal{R} \leq 0.406$ (see figure \ref{subfig:SizesSTD-0rpm-0-20}). Note as well that for $\mathcal{R}=3.70$, particles are observed to mostly fall vertically without swirling in eddies, as visible due to the particle trajectories in figure \ref{subfig:SizesAVG-0rpm-750-1000}.

\begin{figure}[htb]
\centering
    \includegraphics[height=5cm]{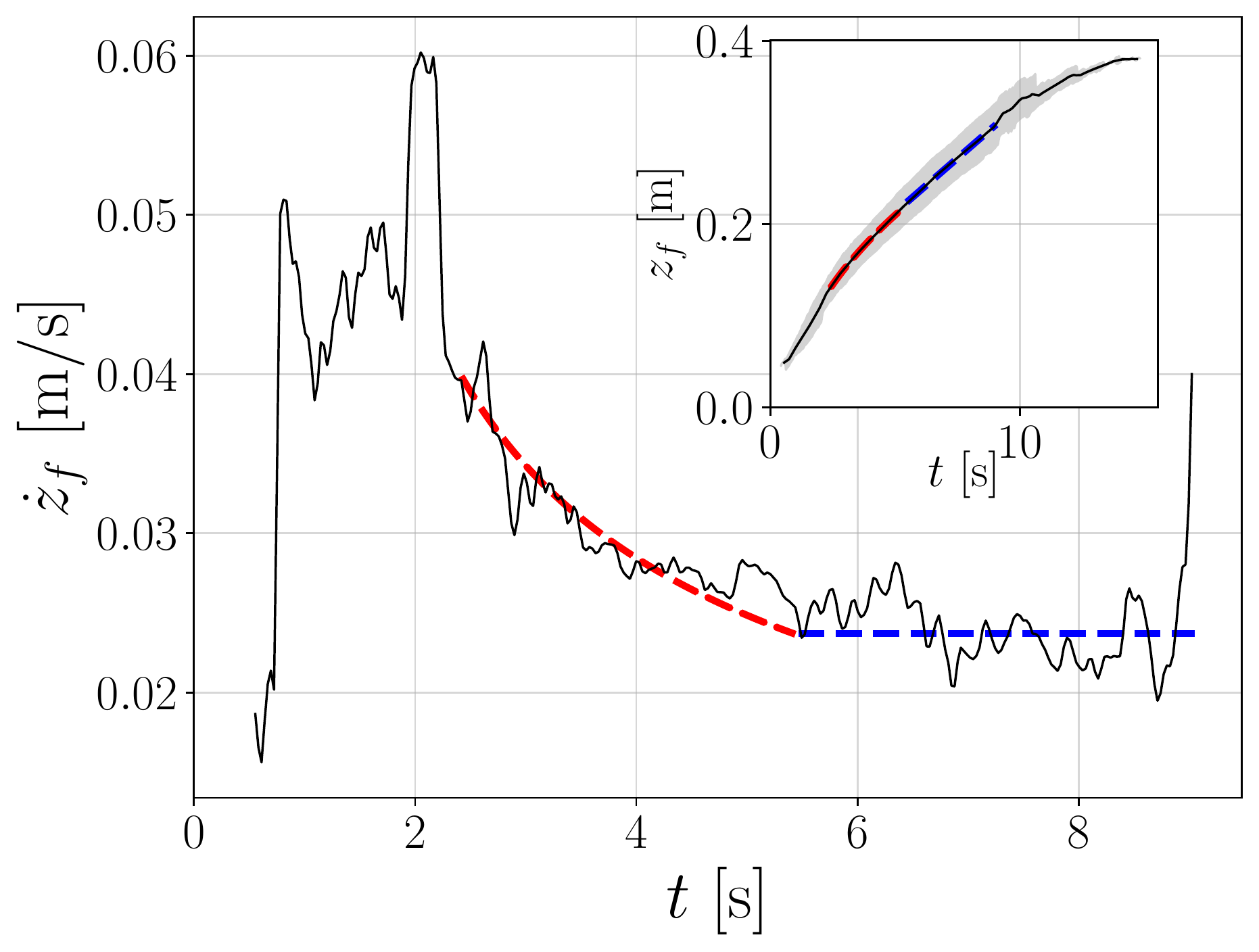}
    \caption{For a cloud of Rouse number $\mathcal{R}=0.308$, the main graph shows in dark solid line (\protect\tikz[baseline=-0.25ex] \protect\draw [color={rgb, 255:red, 1; green, 1; blue, 1},thick] (0,0) -- (0.6,0);) the average cloud front velocity up to $t \simeq 9$s after which the noise is too large. This velocity is computed from second order finite differences of the cloud front position $z_f(t)$, itself shown in inset (\protect\tikz[baseline=-0.25ex] \protect\draw [color={rgb, 255:red, 1; green, 1; blue, 1},thick] (0,0) -- (0.6,0);). The gray shaded area corresponds to the standard deviation due to averaging of $z_f(t)$ from different realisations of the same experiment. On both curves are represented the model for the thermal regime (\protect\tikz[baseline=-0.25ex] \protect\draw [color={rgb, 255:red, 255; green, 1; blue, 0},thick,dashed] (0,0) -- (0.6,0);) and the model of constant front velocity (\protect\tikz[baseline=-0.25ex] \protect\draw [color={rgb, 255:red, 1; green, 1; blue, 255},thick,dashed] (0,0) -- (0.6,0);). }
  \label{fig:ThermalKinematicsStill}
\end{figure}

These last observations emphasise a transition in the cloud dynamics which is due to a separation between particles and turbulent eddies, hereafter simply referred to as \textit{separation}, which interrupts the turbulence forcing. This transition is confirmed by analysing the kinematics of the cloud front whose depth is denoted $z_f$ (measuring techniques are described in Appendix \ref{sec:cloudTracking}). To do so, for a given Rouse number, the curves $z_f(t)$ from different realisations of the same experiment are averaged. Then, the resulting curve is filtered with a moving average to reduce noise, and the velocity $\dot{z}_f$ is computed. Finally, two laws can be fitted on $z_f(t)$: either we fit $z_f=C\sqrt{t-t_0}$ for the thermal regime where $t_0$ and $C$ are arbitrary constants, or we fit an affine law on $z_f(t)$ which is adequate after separation has happened. Figure \ref{fig:ThermalKinematicsStill} shows that after an initial duration of acceleration of the cloud front up to 2.5s, the latter decelerates in the thermal regime (see red dashed curves in the range 2.5s-5.2s). This highlights the relevance of the analogy between particle clouds and turbulent thermals described by MTT56, as previously verified in the literature (e.g., \citep{rahimipourDynamicBehaviourParticle1992,bushParticleCloudsHomogeneous2003a,deguenExperimentsTurbulentMetalsilicate2011}). However, after some time, the cloud front velocity becomes constant; see blue dashed curves in figure \ref{fig:ThermalKinematicsStill} after 5.2s. This transition is due to separation, which is further discussed in section \ref{subsec:SeparationSwarm}. We now exclusively focus on the cloud evolution before separation.\\

As mentioned in section \ref{subsec:OnePhaseThermals}, one of the main features of turbulent thermals is their ability to efficiently entrain surrounding fluid. According to the literature, thermals typically entrain with an average entrainment coefficient $\alpha=0.25 \pm 0.10$ \citep{deguenExperimentsTurbulentMetalsilicate2011,landeauExperimentsFragmentationBuoyant2014}, whose variability is due to the sensitivity of turbulence to initial conditions, and also to the dependence of $\alpha$ on the experimental configuration.
The coefficient of entrainment of all clouds is measured following the procedure described in Appendix \ref{subsec:CoefficientEntrainment}. Figure \ref{fig:EntrainmentStill} shows the entrainment capacity of particle clouds for all Rouse numbers, with respect to the reference of salty thermals whose measured entrainment coefficient is in our case $\alpha_{\textrm{salt}}=0.18\pm 0.02$, which lies in the usual range. Note the drastic decrease of the entrainment coefficient for Rouse numbers above $\simeq$0.3. This drop is directly due to particles of large Rouse number quickly separating from turbulent eddies, resulting in a short-lived turbulence forcing and therefore a low entrainment rate (see section \ref{subsec:SeparationSwarm}). Most importantly for the rest of this study, two regions are distinguished: clouds in the range $\mathcal{R}\leq 1$ go through the self-similar thermal regime, unlike those in the range $\mathcal{R} > 1$ (see the gray shaded area in figure \ref{fig:EntrainmentStill}) in which case $\alpha$ is still measured but does not have the same meaning (see Appendix \ref{subsec:CoefficientEntrainment}).
\par
Then, considering clouds in the range $\mathcal{R}\leq 1$, two main observations can be made. The first observation is that, in the thermal regime, particle clouds entrain more than salt water clouds using the same experimental apparatus, and for the same mass excess. The second observation is that this enhanced entrainment capacity operates most efficiently in a range of Rouse numbers centered on $\mathcal{R}\simeq 0.3$, since $\alpha/\alpha_{\textrm{salt}} \rightarrow 1$ when the Rouse number vanishes to zero, and $\alpha/\alpha_{\textrm{salt}} \rightarrow 0$ when the Rouse number largely exceeds 0.3.\par

\begin{figure}[htb]
    \centering
    \includegraphics[height=5cm]{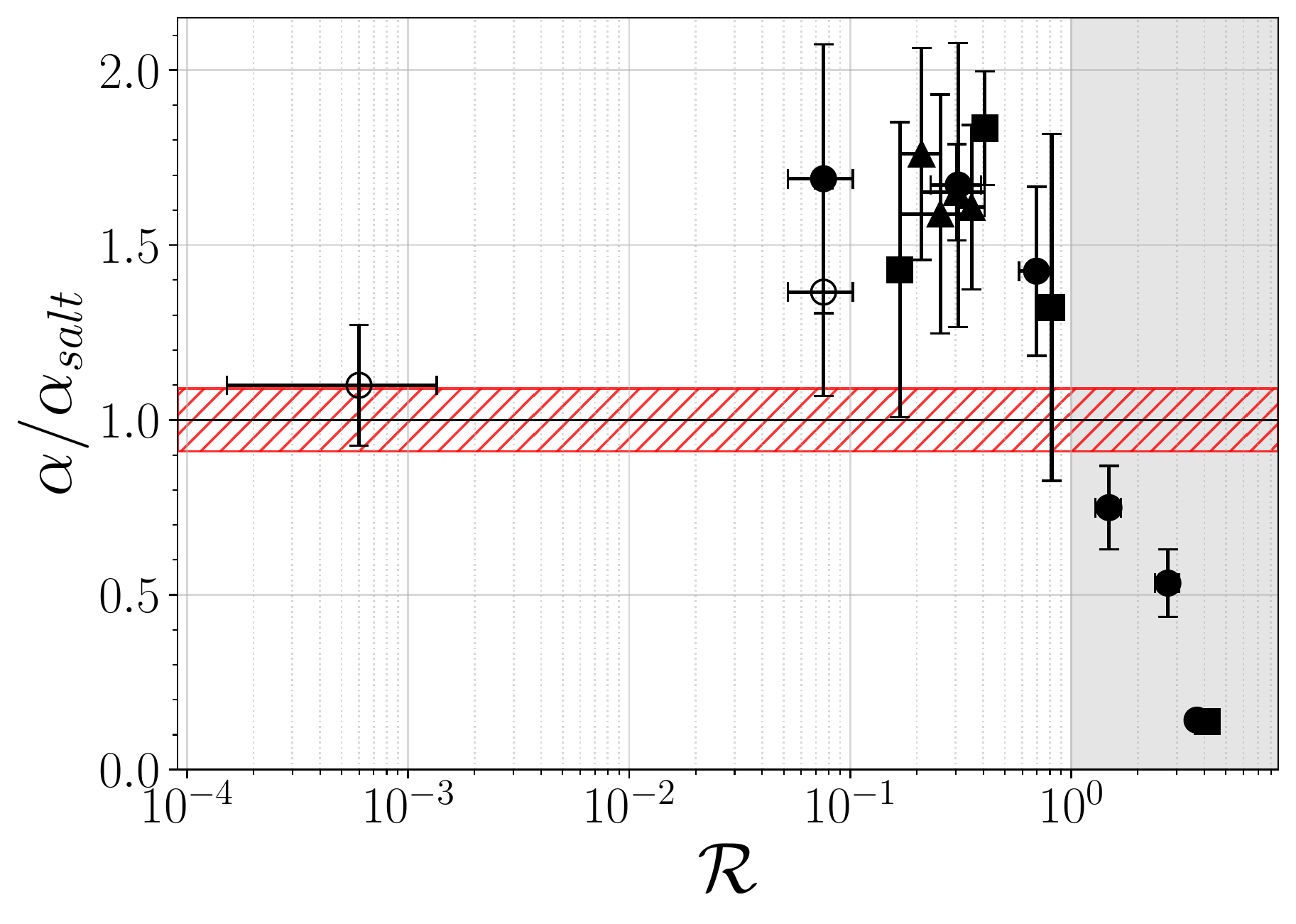}
    \caption{Entrainment capacity of particle clouds with respect to the reference case $\alpha_{\textrm{salt}}=0.18$. Symbols are listed in table \ref{tab:NumbersAllExp}. Red hatchings correspond to the errorbar for salt water whose Rouse number virtually corresponds to $\mathcal{R} = 0$. The gray shaded area highlights clouds which do not go through the thermal regime.}
    \label{fig:EntrainmentStill}
\end{figure}

This enhanced entrainment of particle-laden thermals compared to salt-water thermals is now interpreted in the light of past results on inertial particles. Particles of very large $\mathcal{R}$ hardly respond to modifications of the flow around them because of their large inertia \cite{yoshimotoSelfsimilarClusteringInertial2007}, hence they tend to settle vertically, whatever the flow direction around them. On the opposite, particles of vanishingly small $\mathcal{R}$ act as passive tracers in the fluid -- salt water corresponding to the asymptotic case $\mathcal{R} = 0$. In the intermediate range where we observe that $\alpha>\alpha_\textrm{salt}$, particles of finite inertia follow the fluid, yet they also gravitationally slip with respect to the fluid in their vicinity \cite{nitscheBreakupFallingDrop1997}. This gravitational slip is typically modelled by a vertical terminal velocity $w_s$ which adds up to the local fluid velocity for particles of low inertia \citep{balachandarTurbulentDispersedMultiphase2010,nasabPreferentialConcentrationMechanically2021}. Consequently, through drag, such particles can accelerate the fluid and modify its streamlines so that the fluid follows the particles due to \textit{two-way coupling} \cite{balachandarTurbulentDispersedMultiphase2010,monchauxSettlingVelocityPreferential2017}. A single particle may be too small to efficiently modify the flow around it, however the more concentrated the particles, the more efficient their forcing \cite{balachandarTurbulentDispersedMultiphase2010,monchauxSettlingVelocityPreferential2017,nasabPreferentialConcentrationMechanically2021}.
\par
In our experiments two effects are observed which modify the concentration of particles inside the clouds. First, particles preferentially concentrate on the edges of eddies because their density is larger than that of the ambient. This effect of \textit{preferential concentration} or \textit{inertial clustering} is notably due to pressure effects centrifuging dense particles outside of eddies (see \cite{brandtParticleLadenTurbulenceProgress2022} for further details). It can be derived from theory (see \citep{balachandarTurbulentDispersedMultiphase2010,mcconnochieEntrainmentParticleladenTurbulent2021a}) and has been evidenced in numerous experimental \citep{alisedaEffectPreferentialConcentration2002} and numerical \citep{nasabPreferentialConcentrationMechanically2021,yoshimotoSelfsimilarClusteringInertial2007} past studies. Second, particles preferentially settle on the side of eddies that moves with a downward velocity (parallel to $+\vec{g}$). This effect of \textit{fast-tracking} or \textit{preferential sweeping} has been evidenced in numerical simulations \citep{maxeyGravitationalSettlingAerosol1987,wangSettlingVelocityConcentration1993}, experiments \citep{alisedaEffectPreferentialConcentration2002}, and recently atmospheric flows \citep{liEvidencePreferentialSweeping2021}.

\begin{figure}[htb]
\centering
    \begin{subfigure}[b]{.33\textwidth}
    \caption{}
      \centering
      \includegraphics[height=5cm]{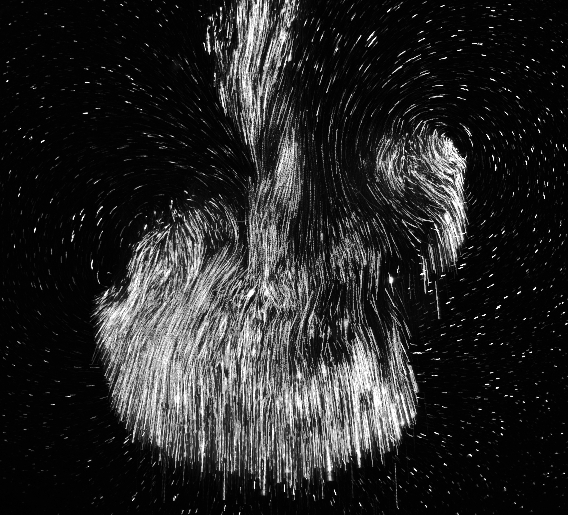}
      \label{subfig:OutOfEddy}
    \end{subfigure}
    \begin{subfigure}[b]{.33\textwidth}
      \caption{}
      \centering
      \includegraphics[height=5cm]{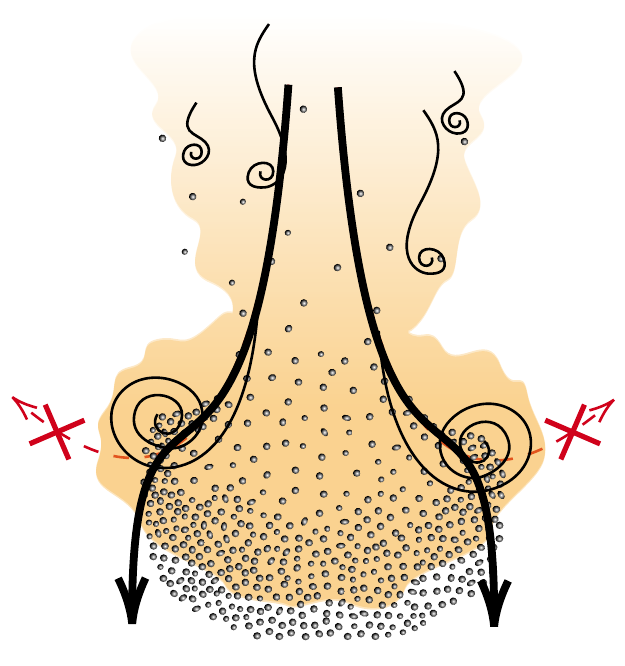}
      \label{subfig:medium}
    \end{subfigure}
    \par
     \begin{subfigure}[b]{\textwidth}
      \caption{}
      \centering
      \includegraphics[width=\textwidth]{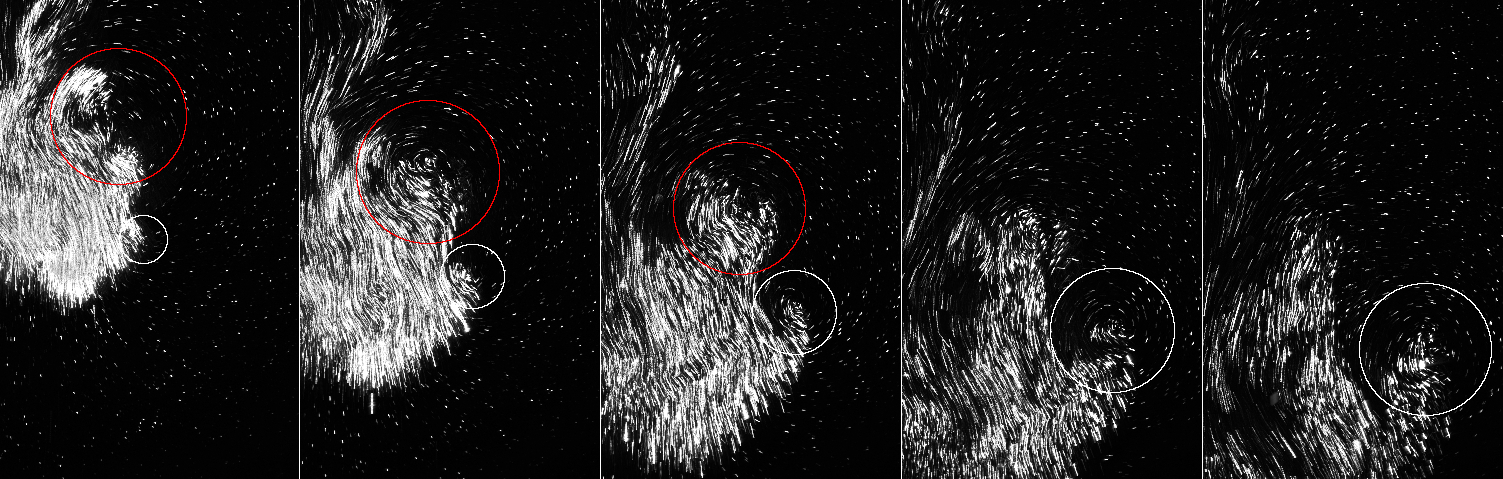}
      \label{subfig:SuccessiveEddies}
    \end{subfigure}
    \caption{(a) Particles ($\mathcal{R}=0.304$) preferentially fall along the down-going sides of eddies (image height: 24.0cm). (b) Schematic of the situation photographed in (a), with particles going downward (see dark arrows) without rolling upward on the up-going sides of eddies (see red dashed arrows which are crossed out). (c) Illustration of the ability of particles ($\mathcal{R}=0.308$) to force a new eddy (centred on the white circle) after they separate from a former one which vanishes (centred on the red circle). Each image is a pixel-by-pixel moving standard deviation of light intensity over 0.3s. The time interval between frames is $\Delta t=1.4$s and their height is 20.7cm. See the supplemental material \textit{Rouse0p308\_0RPM.avi} for an animation showing a particle cloud of Rouse number $\mathcal{R}$ falling in a still ambient.}
  \label{fig:ParticulateEffectsOutEddies}
\end{figure}

The combination of both phenomena is shown in figure \ref{subfig:OutOfEddy}. In this snapshot, the particle cloud has led to the formation of a vortex ring, whose cross-section evidences two vortical structures, centered on the toroidal vortex core. Many particles settle on the edges of these two vortices, and particles almost exclusively sweep on the downgoing side. Together, preferential concentration and preferential sweeping lead particles to fall on the inner sides of these eddies without rolling upwards with the fluid on the other side. Hence, the work of the buoyancy force exerted on particles is always positive, and no turbulent kinetic energy needs to be converted into potential energy to lift particles on the outer ascending side of those eddies. This enhances the efficiency of the particulate forcing and consequently the entrainment rate of particle clouds compared to salt-water thermals. Additionally, both phenomena increase the local effective density of the fluid on the downgoing side of eddies where eddies are being forced by particles, and they also enhance the gravitational slip of particle clusters with respect to the fluid; see in particular \citep{alisedaEffectPreferentialConcentration2002}. This enhances the capacity of concentrated particles to impose their own trajectory to the fluid and force the turbulent flow through two-way coupling, as we initially observe during the cloud acceleration for all $\mathcal{R}$. All in all, figure \ref{subfig:SuccessiveEddies} illustrates the ability of particles to modify the flow and force a new eddy through two-way coupling. On the first photograph particles aim at separating from a vortex, yet one observes the formation of an additional eddy on the following frames. This is likely due to the large concentration and fast downward velocity of the particles sweeping along the former vortex. This enhances entrainment and makes it longer-lasting.
\par
Our observations of the enhancement of entrainment by particles contrast with the study of McConnochie \textit{et al.} \citep{mcconnochieEntrainmentParticleladenTurbulent2021a} who studied particle-laden turbulent plumes whose buoyancy was due to two agents: particles as well as a difference in salinity. They observed that particles modify the entrainment coefficient \textit{only} when they settle in the direction opposite to the plume fluid buoyancy flux. This modification was interpreted as being due to particle clusters crossing the plume interface and interacting with it, which requires particles to settle opposing the fluid buoyancy flux.  The fact that, in the present study, particles modify the entrainment coefficient as they fall down with the turbulent thermal, might be due to the very distinct nature of a plume and a thermal: once the former reaches a steady state, particles evolve in a pre-established stream, while the particles of a thermal evolve in a fundamentally unsteady flow which develops because of their forcing. This distinction between plumes and thermals in the context of particle-laden flows remains to be fully explored.

\subsection{\label{subsec:SeparationSwarm}Separation}

\begin{figure}[htb]
\centering
    \begin{subfigure}[b]{.325\textwidth}
    \caption{$\mathcal{R}=7.57\times 10^{-2}$}
      \centering
      \includegraphics[width=\textwidth]{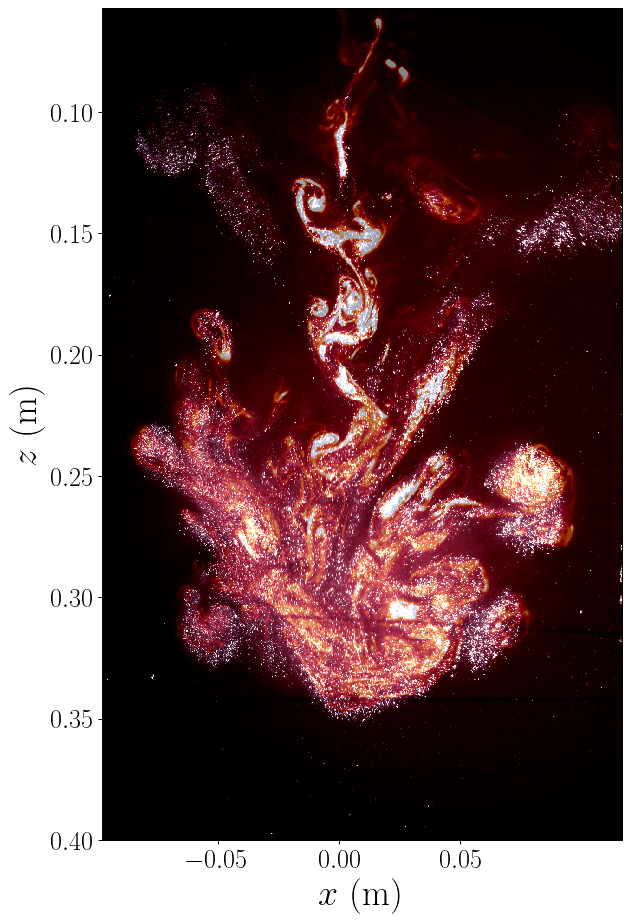}
      \label{subfig:RhodaLowest}
    \end{subfigure}
    \begin{subfigure}[b]{.325\textwidth}
      \caption{$\mathcal{R}=0.308$}
      \centering
      \includegraphics[width=\textwidth]{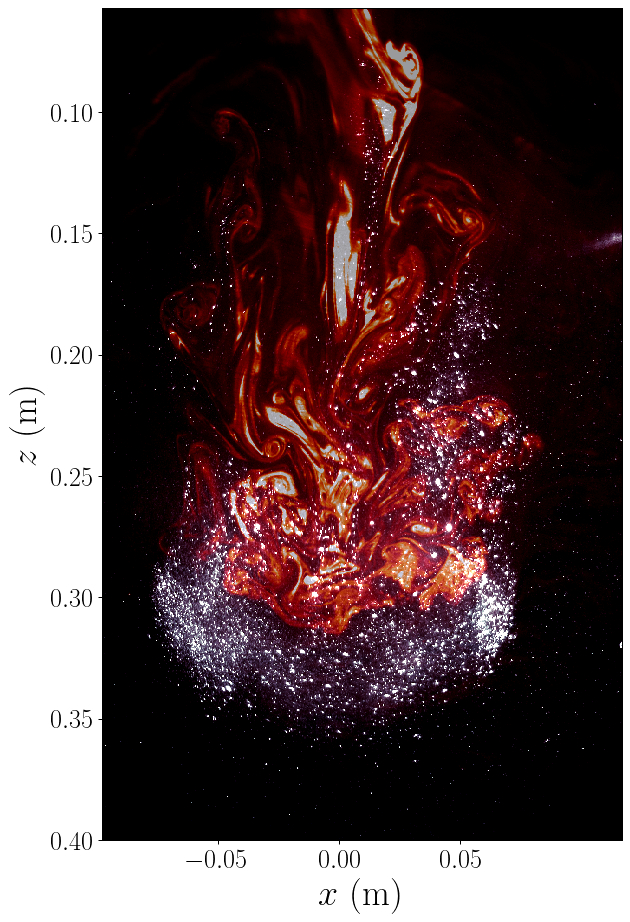}
      \label{subfig:RhodaIntermediate}
    \end{subfigure}
    \begin{subfigure}[b]{.325\textwidth}
      \caption{$\mathcal{R}=1.48$}
      \centering
      \includegraphics[width=\textwidth]{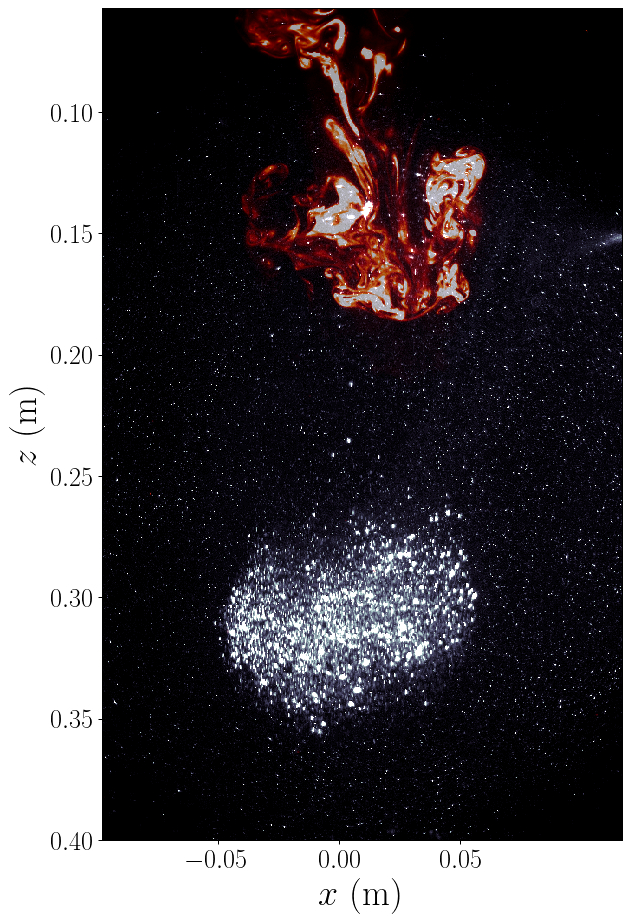}
      \label{subfig:RhodaLargest}
    \end{subfigure}
    \caption{Separation between glass beads (in gray/white) and rhodamine (in orange) for three different Rouse numbers, for the same front depth.}
  \label{fig:SepRhoda}
\end{figure}

To visualise separation, rhodamine was added to water in the cylinder of injection for a few experiments, and snapshots can be visualised in figure \ref{fig:SepRhoda}. For $\mathcal{R}=7.57\times 10^{-2}$ no separation is visible between particles (in gray/white) and rhodamine (in orange) over the depth of our set-up. On the opposite for large particles of Rouse number $\mathcal{R}=1.48$, separation quickly takes place even above $z\simeq 10cm$, and the rapid increase of the distance between rhodamine and particles emphasizes this process. Separation is a gradual phenomenon though, as evidenced by particles of intermediate Rouse number $\mathcal{R}=0.308$ which are still separating in the snapshot in figure \ref{subfig:RhodaIntermediate}. This process can be understood by the fact that when the thermal velocity becomes lower than the particles' settling velocity $w_s$, the rms velocity in the cloud (which scales like $v_e=\alpha \dot{z}$) becomes unable to sustain particles in eddies (e.g., \citep{rahimipourDynamicBehaviourParticle1992,bushParticleCloudsHomogeneous2003a}). Hence they separate at a depth $z_{\textrm{sep}}$ which is defined by $\dot{z}(z_{\textrm{sep}})=w_s$. From this definition of separation, and from the self-similar behaviour $\dot{z}\sim z^{-1}$ of MTT56 [equation \eqref{eq:similarVelocity}], the distance of separation is expected to scale like $\mathcal{R}^{-1}$, and we now attempt to verify this scaling experimentally.
\par
\begin{figure}[htb]
\centering
      \includegraphics[height=5cm]{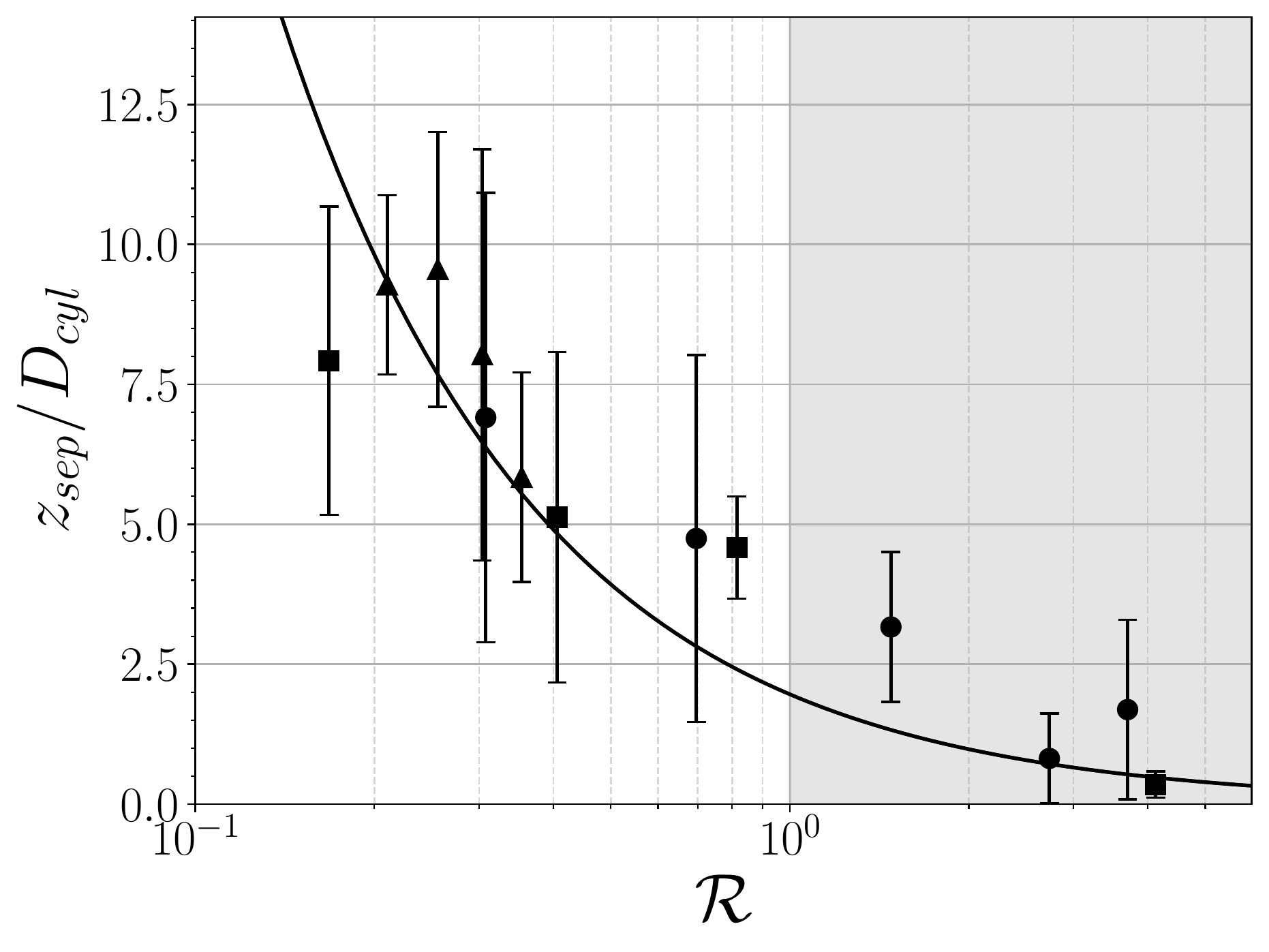}
    \caption{Depth of separation of particles from eddies. The gray shaded area corresponds to the range $\mathcal{R} > 1$ (see figure \ref{fig:EntrainmentStill}) containing clouds which never behave as turbulent thermals. The dark solid line corresponds to the scaling $z_\textrm{sep}\propto \mathcal{R}^{-1}$ obtained from the equations of Morton.}
  \label{fig:zSepExp}
\end{figure}

Appendix \ref{subsec:AppendixDepthSeparation} presents the protocol to measure the depth of separation $z_{\textrm{sep}}$ between turbulent eddies and particles. Essentially, when particles separate from eddies, the patterns made by their trajectories transition from curved and randomly oriented (due to particles whirling inside eddies) to mostly straight and vertical; this enables to measure $z_{\textrm{sep}}$. Figure \ref{fig:zSepExp} shows the depth of separation as measured in experiments, and the associated forced-fit curve following a law $z_{\textrm{sep}} \propto \mathcal{R}^{-1}$. The curve confirms the decrease of $z_{\textrm{sep}}$ with a larger particle size under the trend $z_{\textrm{sep}} \propto \mathcal{R}^{-1}$ provided by MTT56 in the thermal regime, which is also consistent with previous results in the literature (see the scaling laws in reference \citep{wangLargeEddySimulationSettling2014a}).\\

As soon as separation starts, the turbulent thermal has a decreasing mass excess which eventually vanishes to zero. After that, particles fall in a quiescent fluid. Hence, particle clouds can rigorously be considered as turbulent thermals only as long as particles have not separated yet. This has led previous authors \citep{rahimipourDynamicBehaviourParticle1992,bushParticleCloudsHomogeneous2003a,laiModelingExperimentsPolydisperse2016} to distinguish mainly three regimes in the cloud's evolution: (i) the \textit{acceleration regime}, (ii) the self-similar turbulent \textit{thermal regime} of MTT56, and (iii) the \textit{swarm regime} in which particles settle (mostly) vertically in a quiescent fluid, while the turbulent eddies are left behind particles and viscously decay. This latter regime is analysed in section \ref{subsec:SwarmRegime}. For now, we focus on the formation of the swarm \textit{during} separation.
\par
\begin{figure}[htb]
\centering
    \begin{subfigure}[b]{0.3\textwidth}
      \caption{$\mathcal{R}=0.814$, $\mathcal{S}=0.07$}
      \centering
      \includegraphics[height=5.5cm]{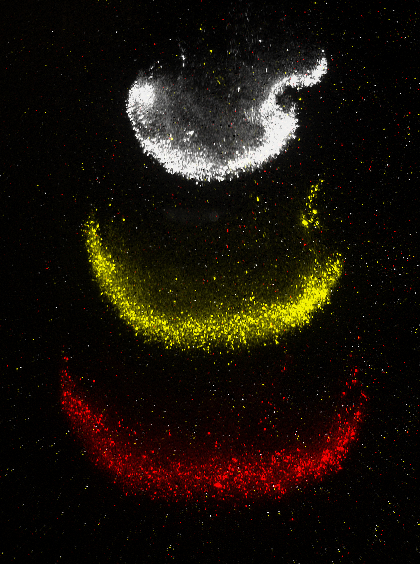}
      \label{subfig:MorphSwarm224250}
    \end{subfigure}
    \begin{subfigure}[b]{0.3\textwidth}
      \caption{$\mathcal{R}=0.696$, $\mathcal{S}=0.23$}
      \centering
      \includegraphics[height=5.5cm]{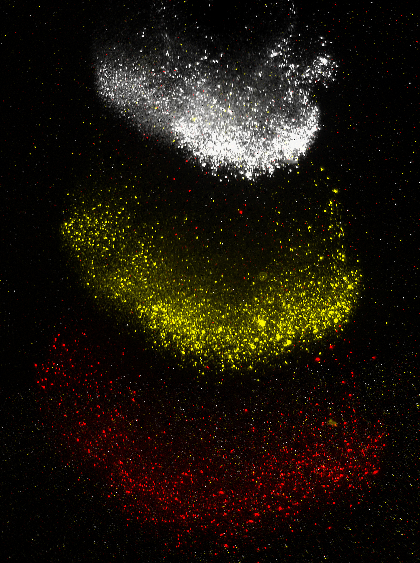}
      \label{subfig:MorphSwarmPolydisp}
    \end{subfigure}
    \caption{Morphology of swarms for a monodisperse (a) or polydisperse (b) distribution of particles. Colours are used to superimpose snapshots of a swarm at different times. Both photographs are 24.4cm-high, and the time delay between snapshots is (a) 1.6s., (b) 2.0s.}
  \label{fig:MorphologySwarms}
\end{figure}

All swarms change morphology during their fall. The swarms which originate from a turbulent thermal are shaped both by particles and by turbulent eddies over the course of separation. This results in a variety of possible swarm morphologies. Monodisperse (figure \ref{subfig:MorphSwarm224250}) and polydisperse (figure \ref{subfig:MorphSwarmPolydisp}) swarms are illustrated in figure \ref{fig:MorphologySwarms}. We observe that monodisperse clouds produce thin swarms which are concentrated in particles and therefore very bright on the camera. Conversely, polydisperse clouds produce swarms of larger height $\sigma_z$, which keeps increasing in time because of the difference in settling velocities between the smallest and largest particles. Consequently, the brightness of these polydisperse swarms decreases in time (compare the brightness of the three snapshots in figure \ref{subfig:MorphSwarmPolydisp}). Finally bidisperse swarms evidence a gradual splitting between two particle fronts (not shown here), due to the bimodal distribution of particle sizes.

The role of turbulent eddies in shaping the swarm during separation can be visualised in the Supplemental Material [\textit{Still-ambient-particles-and-rhodamine.avi} (see the case $\mathcal{R}=0.31$) and \textit{Rouse0p406\_STD\_0RPM.avi}]. Particles initially swirl within the turbulent thermal, until separation starts, as can be assessed from the vertical motion of particles at the cloud front. Then, two regions can be distinguished. On one hand, particles ahead of eddies have separated and fall vertically at a constant velocity $\dot{z}_\textrm{f,s}$ which is the (front) velocity of the swarm (hence the subscript \textit{s}), of order $w_s$. On the other hand, particles in the eddies keep swirling and fall faster than $\dot{z}_\textrm{f,s}$, say at a velocity
\begin{equation}
    \label{eq:PartVelWhenInsideEddies}
    \dot{z}_{p,e} = \dot{z}_\textrm{f,s}+u(t),
\end{equation}
with $u(t)>0$ the velocity excess. The width of the swarm is equal to that of the turbulent region which feeds it with particles. During separation the turbulent cloud keeps growing so that particles are gradually shed outwards in the swarm (see figures \ref{subfig:MorphSwarm224250}-\ref{subfig:MorphSwarmPolydisp}).\\

From these observations, we can propose a model of swarm formation in the framework of MTT56. The description hereafter is for a cloud transitioning from thermal to swarm regime ($\mathcal{R}\leq 1$), but it is also valid for a cloud which directly accelerates to the swarm regime ($\mathcal{R} > 1$) if one cancels the velocity $\dot{z}_e(t)$ of eddies in equations \eqref{eq:HorizontalExtSwarm} and \eqref{eq:VerticalExtAllPart}. New notations are introduced to rigorously distinguish the behaviours of eddies, particles within eddies, and particles which have separated from eddies.
\par
\begin{figure}[htb]
\centering
\includegraphics[height=6cm]{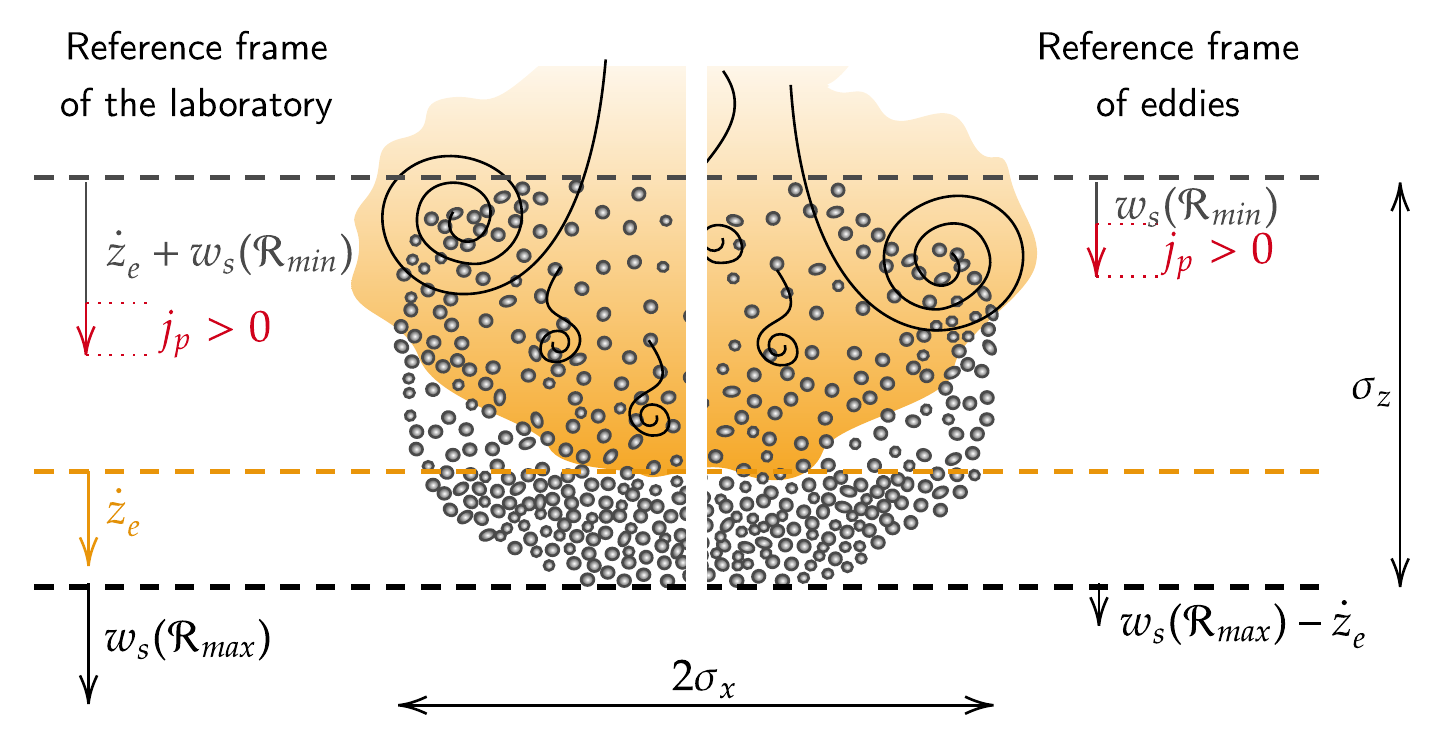}
\caption{Sketch of the process of cloud contraction and compaction of particles during separation for $\dot{z}_e(t) > \Delta w_s$.}
\label{fig:ContractionCompaction}
\end{figure}

During separation, consider that turbulent eddies fall with velocity $\dot{z}_e(t)$ (see the left column in figure \ref{fig:ContractionCompaction}) and, for simplicity, assume that they remain in the same self-similar regime despite the loss of particles. This notably implies that $\alpha$ remains constant during separation. Then, the radii of both the patch of turbulent eddies and the nascent swarm increase at a rate
\begin{equation}
    \label{eq:HorizontalExtSwarm}
    \dot{\sigma}_x(t) = \alpha\dot{z}_e(t).
\end{equation}

All particle sets are polydisperse to some extent. Hence, consider a cloud of minimum (respectively maximum) Rouse number $\mathcal{R}_{\textrm{min}}$ (respectively $\mathcal{R}_{\textrm{max}}$). The difference between the maximum and minimum settling velocities reads $\Delta w_s = w_s(\mathcal{R}_{\textrm{max}})-w_s(\mathcal{R}_{\textrm{min}})>0$. Since the depth of separation $z_{\textrm{sep}}\propto \mathcal{R}^{-1}$ decreases with $\mathcal{R}$, the cloud's largest particles separate first, with a velocity $w_s(\mathcal{R}_{\textrm{max}})$ in the reference frame of the laboratory (see the left column in figure \ref{fig:ContractionCompaction}). Meanwhile, smaller particles keep falling through turbulent eddies. Assuming that the drag and buoyancy of these smaller particles are balanced, their upper rear front falls with velocity $w_s(\mathcal{R}_{\textrm{min}})$ in the reference frame of eddies (right column in figure \ref{fig:ContractionCompaction}). Since eddies fall at velocity $\dot{z}_e(t)$ with respect to the laboratory, it means that the smaller particles fall at velocity $\dot{z}_e(t)+w_s(\mathcal{R}_{\textrm{min}})$ in the reference frame of the laboratory. Hence, the vertical extent of the particle cloud detected by the cloud-tracking algorithm varies with a rate
\begin{equation}
    \label{eq:VerticalExtAllPart}
    \dot{\sigma}_z(t) = w_s(\mathcal{R}_{\textrm{max}}) - (\dot{z}_e(t)  +w_s(\mathcal{R}_{\textrm{min}})) = \Delta w_s - \dot{z}_e(t).
\end{equation}

Let us finally turn to the transfer of particles from eddies to the swarm. We just concluded that in the reference frame of the laboratory, the velocity of the smallest particles inside the turbulent eddies is $\dot{z}_e(t)+w_s(\mathcal{R}_{\textrm{min}})$. For the sake of simplicity, we adopt the definition $\dot{z}_{p,e}=\dot{z}_e(t)+w_s(\mathcal{R}_{\textrm{min}})$. Since the largest particles separate first, we expect the swarm front to fall with a velocity $\dot{z}_\textrm{f,s} = w_s(\mathcal{R}_{\textrm{max}})$. Combining these expressions of $\dot{z}_{p,e}$ and $\dot{z}_\textrm{f,s}$ with equation \eqref{eq:PartVelWhenInsideEddies}, we find $u(t)=\dot{z}_e(t) - \Delta w_s$. Consequently the volume flux $j_p(t) = \phi(t)u(t)$ of particles shed in the emerging swarm is
\begin{equation}
    \label{eq:VolFluxPartSwarm}
    j_p(t) = \phi(t) [\dot{z}_e(t) - \Delta w_s] = -\phi(t)\dot{\sigma}_z(t).
\end{equation}
In equation \eqref{eq:VolFluxPartSwarm}, $\phi(t)=N_p[r_p/r(t)]^3$ is the supposedly uniform particle volume fraction within turbulent eddies, with $N_p$ the total number of particles in the cloud, $r_p$ their radius and $r(t)$ the radius of the supposedly spherical cloud in the framework of MTT56. The whole process of swarm formation is sketched in figure \ref{fig:ContractionCompaction}.

The expressions of $\dot{\sigma}_x(t)$, $\dot{\sigma}_z(t)$ and $j_p(t)$ are controlled by the velocity of eddies $\dot{z}_e(t)$ which cannot instantaneously vanish as soon as particles reach the depth $z_{\textrm{sep}}(\mathcal{R}_{\textrm{max}})$. Neglecting polydispersity for now ($\Delta w_s=0$), the fact that eddies keep falling explains why the particle cloud is necessarily contracted ($\dot{\sigma}_z<0$) and thus particles are compacted ($j_p>0$) from the thermal to the swarm: this explains why monodisperse swarms have a large aspect ratio $\sigma_x/\sigma_z$ even though they originate from an approximately spherical cloud. During separation, eddies keep widening ($\dot{\sigma}_x>0$) so that particles are gradually shed outward and behind former particles. As a result the swarm takes a bowl shape (see figure \ref{fig:MorphologySwarms}, and reference \citep{wangLargeEddySimulationSettling2014a} for a clear numerical illustration). However, polydispersity acts against the cloud contraction, and therefore against the compaction of particles. Once separation is over, in the swarm regime the volume flux $j_p$ loses meaning, the cloud keeps a constant lateral extension ($\dot{\sigma}_x=0$), and it grows vertically at a constant rate $\dot{\sigma}_z=\Delta w_s$ exclusively governed by polydispersity (see below, section \ref{subsec:SwarmRegime}).
\par
This simple description captures the evolution of large polydisperse clouds whose particles do seem to keep the entrainment active during separation; see the Supplemental Material \textit{Still-ambient-particles-and-rhodamine.avi} for $\mathcal{R}=0.308$. It also captures the evolution of particle clouds which never go through the thermal regime for $\mathcal{R}>1$ (see discussion in section \ref{subsec:SwarmRegime}). To sum up, the larger $\mathcal{R}$, the lower the depth of separation, hence the thermal at the start of contraction so the more efficient contraction; a lower depth of separation also implies that the thermal has a smaller size, and so does the swarm. Polydispersity, however, acts to increase the swarm height, hence monodisperse swarms are the thinnest (see figure \ref{fig:MorphologySwarms}).
\par
Note that the role of $\mathcal{R}$ in shaping clouds is not straightforward: the dependency of $\dot{\sigma}_z(t)$ and $j_p(t)$ with $\mathcal{R}$ is not explicit in equations \eqref{eq:VerticalExtAllPart}-\eqref{eq:VolFluxPartSwarm}. One can easily show that $\phi(t)$ is independent of $\mathcal{R}$ for a spherical cloud of constant $m_0$, and the velocity difference $\Delta w_s$ is not a function of the mean settling velocity $w_s$. The role of $\mathcal{R}$ is only implicit in the velocity of eddies $\dot{z}_e(t)$: since during separation the eddies are expected to fall at velocity $\dot{z}(z_\textrm{sep})=w_s(\mathcal{R})$, at first order we have $\dot{z}_e(t) \simeq \dot{z}(z_\textrm{sep})=w_s(\mathcal{R})$ during separation. Consequently, for a given velocity difference $\Delta w_s$, a cloud of larger particles experiences a larger $\dot{z}_e(t)$ during separation and compacts more than a cloud of smaller particles (see the expression of $\dot{\sigma}_z$ in equation \eqref{eq:VolFluxPartSwarm}). This trend is consistent with the observation of our clouds in experiments.

%%%%%%%%%%%%%%%%%%%%%%%%%%%%%%%%%%%%%%%%%%%%%%%%%%
\subsection{\label{subsec:SwarmRegime}Swarm regime}

\begin{figure}[htb]
\centering
    \begin{subfigure}[b]{0.495\textwidth}
      \caption{}
      \centering
      \includegraphics[height=6cm]{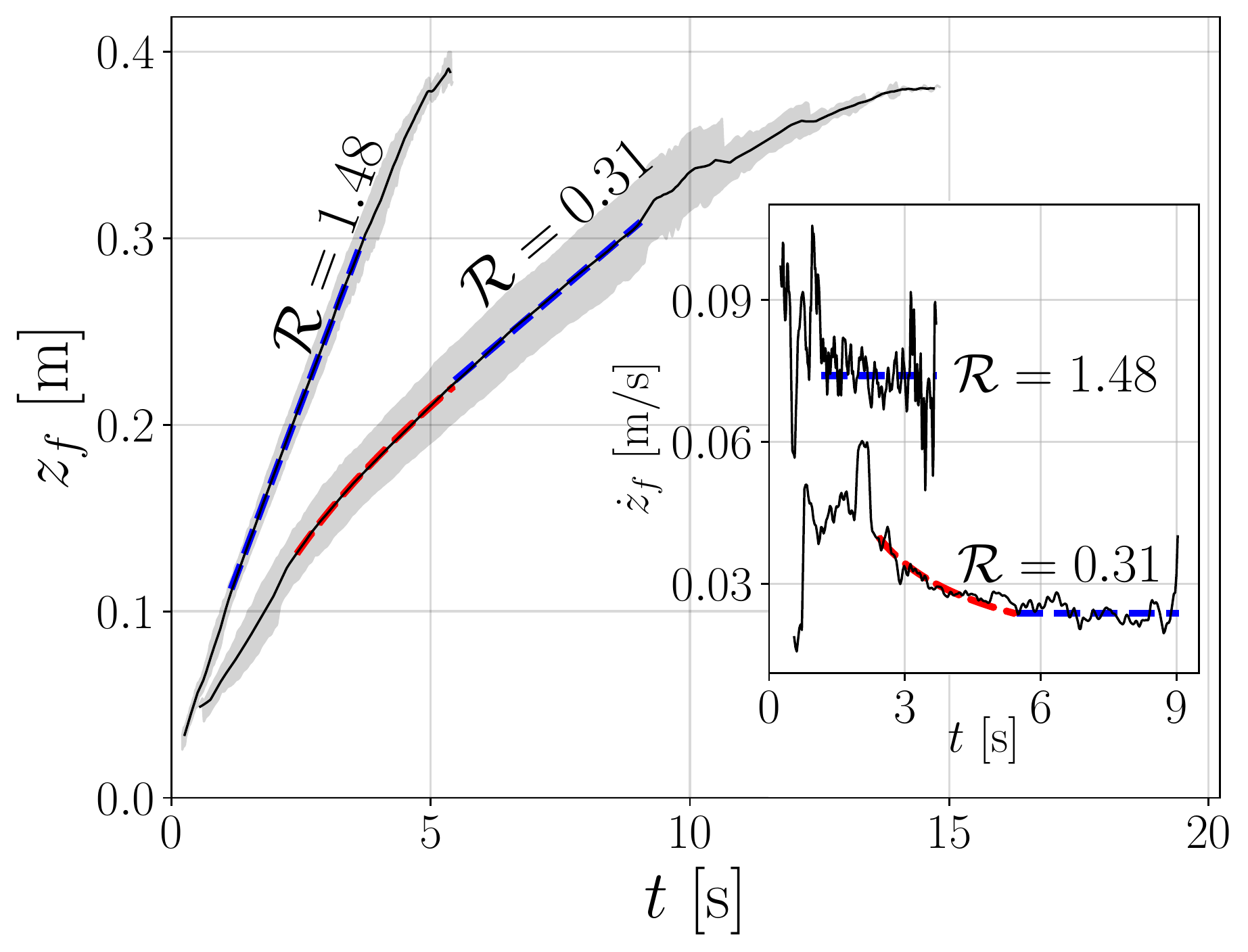}
      \label{subfig:fitThermalSwarmRegimes}
    \end{subfigure}
    \begin{subfigure}[b]{0.495\textwidth}
      \caption{}
      \centering
      \includegraphics[height=6cm]{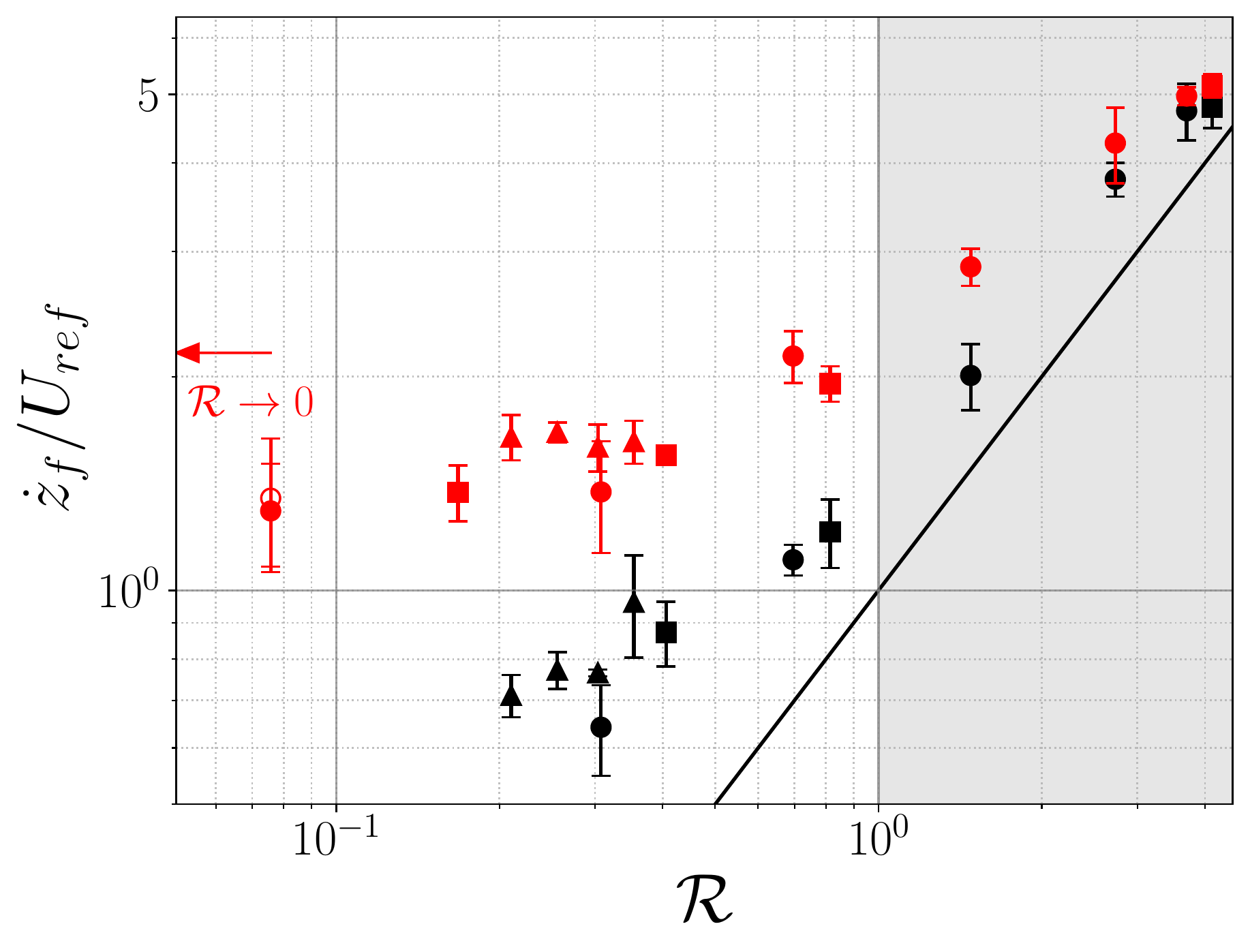}
      \label{subfig:swarmVelocityRouse}
    \end{subfigure}
    \caption{(a) Kinematics of two clouds of different Rouse numbers. (\protect\tikz[baseline=-0.25ex] \protect\draw [color={rgb, 255:red, 1; green, 1; blue, 1},thick] (0,0) -- (0.6,0);) Average cloud front position in time with the standard deviation due to averaging indicated in gray shaded area; (\protect\tikz[baseline=-0.25ex] \protect\draw [color={rgb, 255:red, 255; green, 1; blue, 0},thick,dashed] (0,0) -- (0.6,0);) fit of thermal regime; (\protect\tikz[baseline=-0.25ex] \protect\draw [color={rgb, 255:red, 1; green, 1; blue, 255},thick,dashed] (0,0) -- (0.6,0);) fit of swarm regime. 
    (b) $\CIRCLE$ Average cloud front velocity $\dot{z}_\textrm{f,s}$ during the swarm phase; \red{$\CIRCLE$} maximum cloud front velocity $\dot{z}_{\textrm{f,max}}$; (\protect\tikz[baseline=-0.25ex] \protect\draw [color={rgb, 255:red, 1; green, 1; blue, 1},thick] (0,0) -- (0.6,0);) line of equation $\dot{z}_f=w_s$. 
    The red arrow shows the measured value for the asymptotic limit $\mathcal{R} \rightarrow 0$ corresponding to salty thermals, and the gray shaded area corresponds to clouds which never behave as a turbulent thermal.}
  \label{fig:SwarmStill}
\end{figure}

Swarms are produced in different ways depending on the size of particles. Particles in the range $\mathcal{R}\leq 1$ initially accelerate, then transition to the thermal regime and finally separate as a swarm. On the opposite, particles in the range $\mathcal{R} > 1$ directly transition from their initial acceleration to the swarm regime, which is characterised by a constant fall velocity. This distinction introduced in figure \ref{fig:EntrainmentStill} is confirmed in figure \ref{subfig:fitThermalSwarmRegimes}, which complements the previous figure \ref{fig:ThermalKinematicsStill} with the kinematics of a cloud directly accelerating to the swarm regime; see blue dashed curves for $\mathcal{R}=1.48$.

Recall that particles are expected to separate from eddies at their terminal velocity so that the constant swarm velocity $\dot{z}_\textrm{f,s}$ should be of order $w_s$. Particles having $\mathcal{R}<0.2$ separate too deep in the tank for the swarm regime to be analysed. For other clouds, the constant velocity $\dot{z}_\textrm{f,s}$ is measured and visible in figure \ref{subfig:swarmVelocityRouse} in dark symbols. Up to a constant offset of order $O(1)$, measurements prove to be reasonably close to the reference value $w_s$ in dark solid line, all the more as $\mathcal{R}$ is larger. The fact that swarms of smaller particles fall faster than $w_s$ is interpreted as a consequence of the capacity of numerous smaller particles to drag fluid downward with them through their hydrodynamical interactions by displacing fluid \cite{yamamotoNumericalSimulationConcentration2015a,pignatelFallingCloudParticles2011}, whereas fewer larger particles behave individually and modify only the flow close to themselves in a small wake \citep{subramanianEvolutionClustersSedimenting2008}.
\par
The maximum cloud front velocity $\dot{z}_\textrm{f,max}$ measured during every experiment is also computed and shown with red symbols in figure \ref{subfig:swarmVelocityRouse}. As $\mathcal{R}$ increases, the velocities $\dot{z}_\textrm{f,s}$ and $\dot{z}_\textrm{f,max}$ get closer. This is due to the reduced duration of the thermal regime, which is ultimately shut off, leading clouds of large Rouse number to directly accelerate to reach their maximum velocity as a single particle would, hence $\dot{z}_\textrm{f,max}\simeq w_s$.

%%%%%%%%%%%%%%%%%%%%%%%%%%%%%%%%%%%%%%%%%%%%%%%%%%%%%%%%%%%%%%%%%%
%%%%%%%%%%%%%%%%%%%%%%%%%%%%%%%%%%%%%%%%%%%%%%%%%%%%%%%%%%%%%%%%%%
\FloatBarrier
\section{\label{sec:RotatingEnvironment}Particle clouds in a rotating environment}
\subsection{\label{subsec:OnePhaseThermals_rot}Formation of a columnar flow}

\begin{figure}[htbp]
\centering
    \begin{subfigure}[b]{0.82\textwidth}
    \caption{$\Omega=0$rpm}
      \centering%width=.8\textwidth
      \includegraphics[height=4.52cm]{AllFigures/figure3a.jpg}
      \label{subfig:saltWater0rpm}
    \end{subfigure}
    \begin{subfigure}[b]{0.17\textwidth}
    \caption{}
      \centering%width=.8\textwidth
      \includegraphics[height=4.52cm]{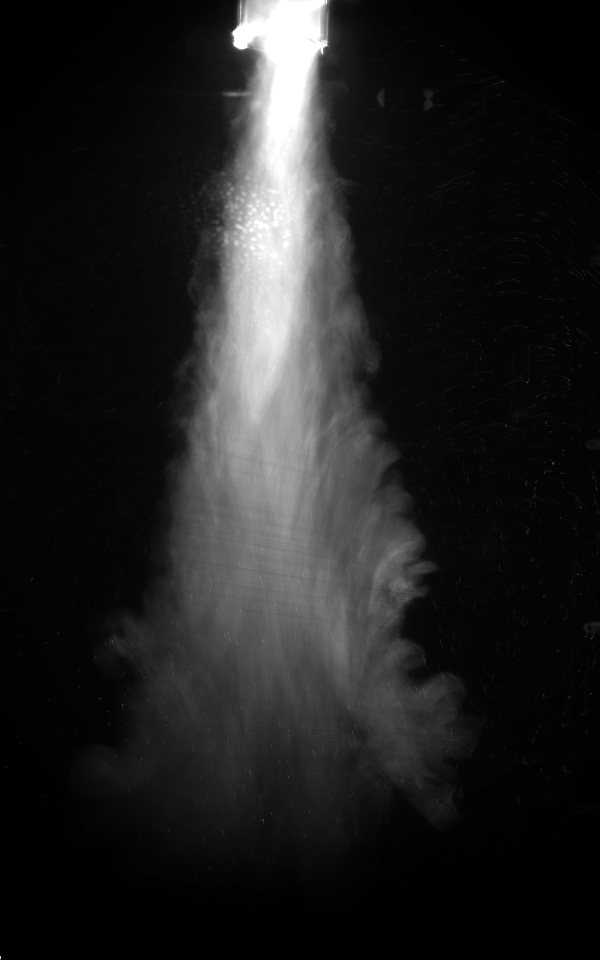}
    \end{subfigure}
    
    \begin{subfigure}[b]{0.82\textwidth}
      \caption{$\Omega=5$rpm}
      \centering
      \includegraphics[height=4.52cm]{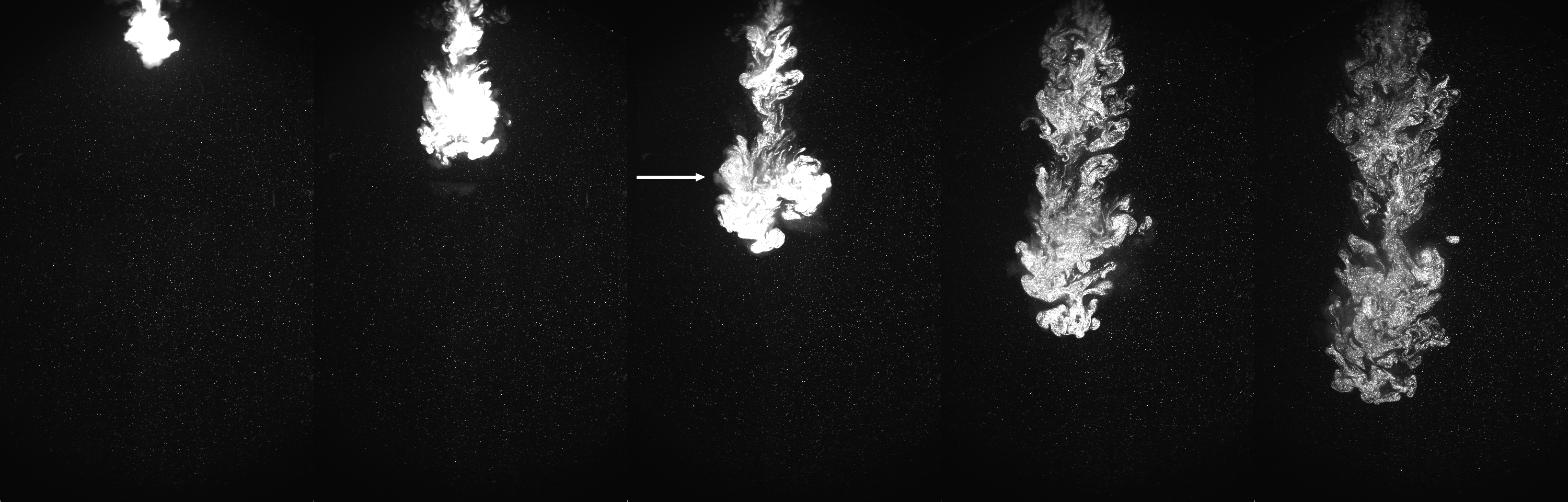}
      \label{subfig:saltWater5rpm}
    \end{subfigure}
    \begin{subfigure}[b]{0.17\textwidth}
      \caption{}
      \centering
      \includegraphics[height=4.52cm]{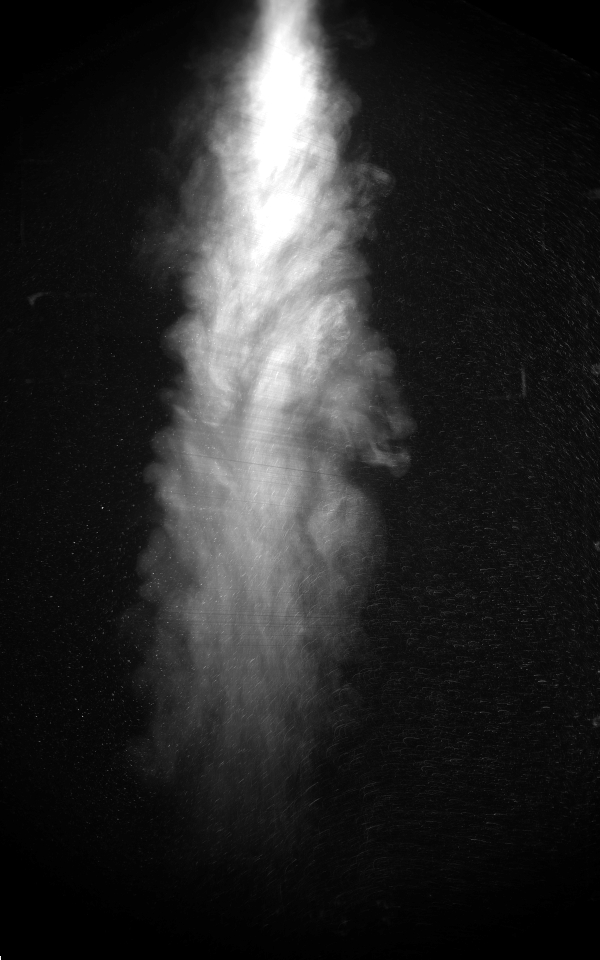}
      \label{subfig:saltWater5rpmAVG}
    \end{subfigure}
    
    \begin{subfigure}[b]{0.82\textwidth}
      \caption{$\Omega=10$rpm}
      \centering
      \includegraphics[height=4.52cm]{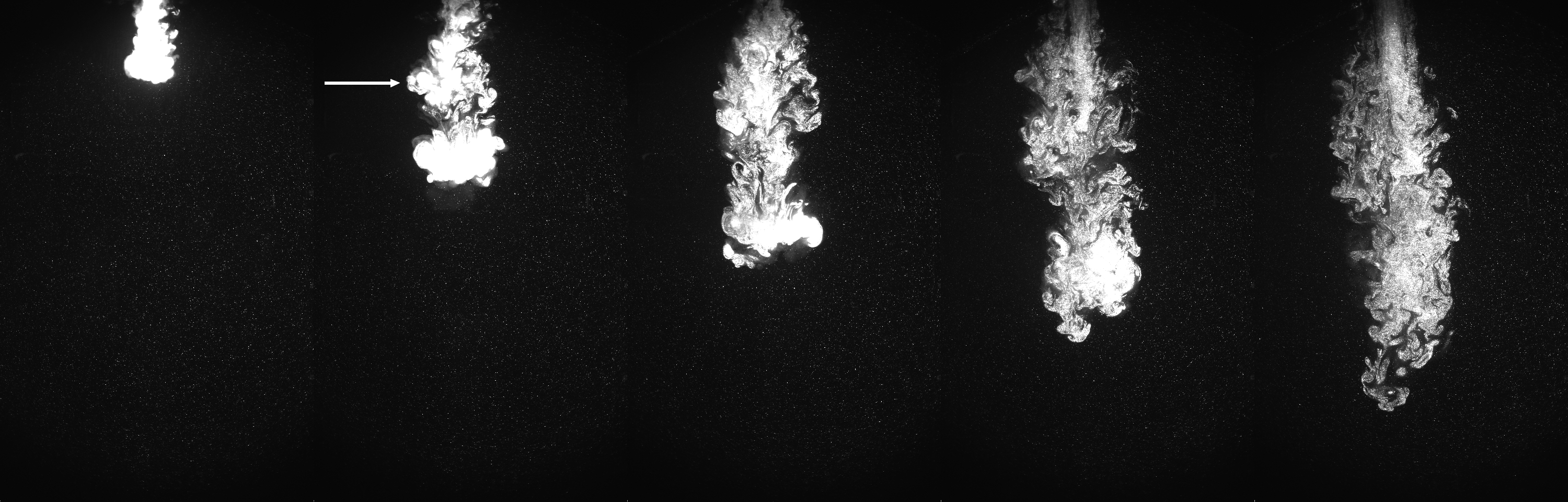}
      \label{subfig:saltWater10rpm}
    \end{subfigure}
    \begin{subfigure}[b]{0.17\textwidth}
      \caption{}
      \centering
      \includegraphics[height=4.52cm]{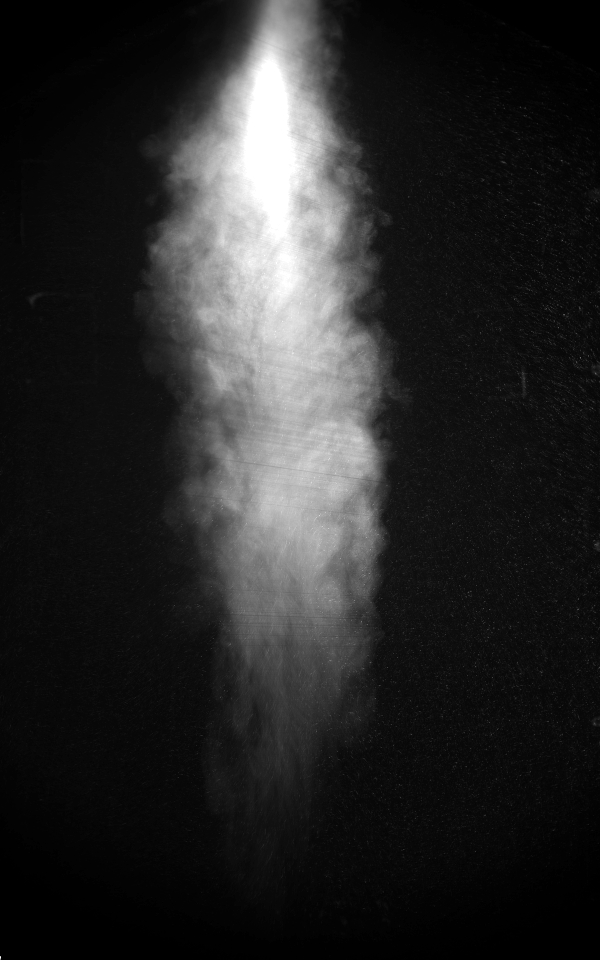}
      \label{subfig:saltWater10rpmAVG}
    \end{subfigure}
    
    \begin{subfigure}[b]{0.82\textwidth}
      \caption{$\Omega=20$rpm}
      \centering
      \includegraphics[height=4.52cm]{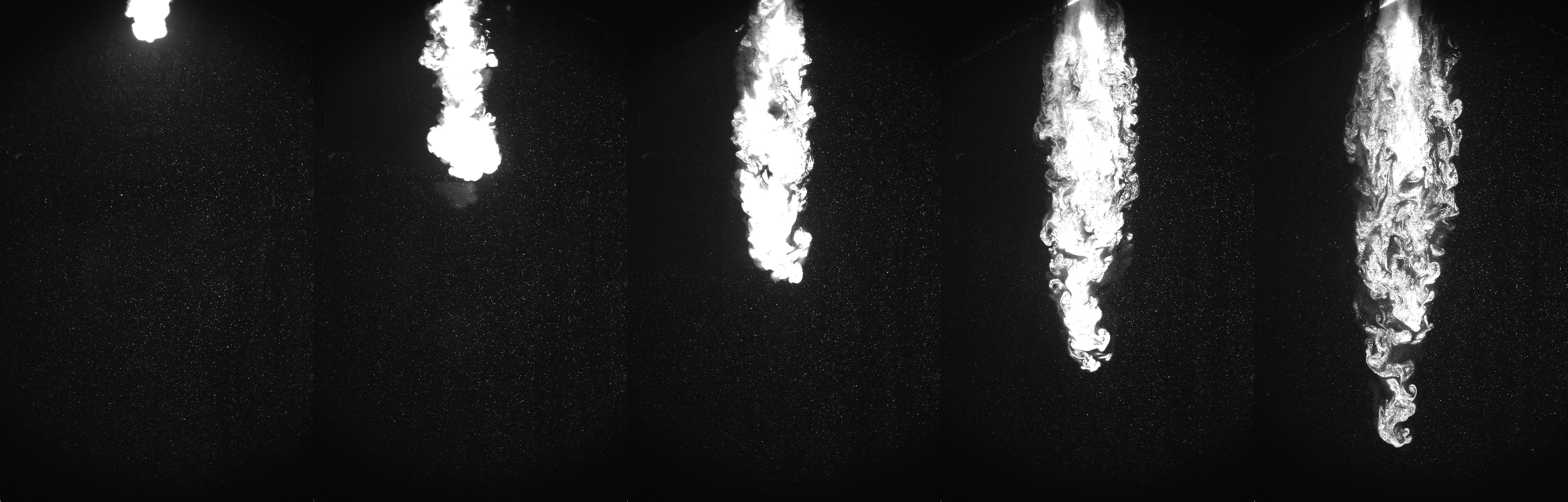}
      \label{subfig:saltWater20rpm}
    \end{subfigure}
    \begin{subfigure}[b]{0.17\textwidth}
      \caption{}
      \centering
      \includegraphics[height=4.52cm]{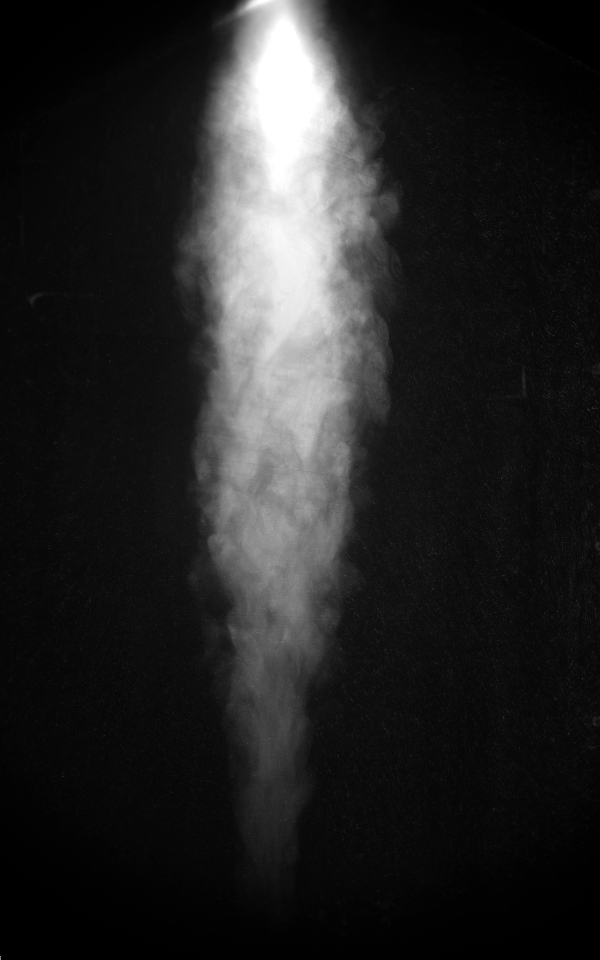}
      \label{subfig:saltWater20rpmAVG}
    \end{subfigure}
    \caption{(a,c,e,g): Salt water clouds for varying $\Omega$. The time lapse between snapshots is always 1.6s, and photographs are always 45cm-high. When observable, the depth $z_{f,\textrm{col}}$ where the cloud becomes columnar is indicated by a white arrow. (b,d,f,h): Pixel-by-pixel average of light intensity of all photographs taken during the experiment on the same row, of respective duration (b) 8.3s, (d) 9.1s, (f) 9.8s, (h) 7.8s.}
  \label{fig:saltWaterSeries}
\end{figure}

From the previous section, we can expect -- as the following section will confirm -- that clouds containing particles of vanishingly small Rouse number behave similarly as salt-water clouds, and the larger $\mathcal{R}$ the larger the discrepancy between their behaviours. Therefore, we start by analysing the influence of rotation on salt-water clouds.
\par
In figure \ref{fig:saltWaterSeries}, the most striking observation is that rotation interrupts entrainment at some depth $z_{f,\textrm{col}}$, marking a transition from a regime of expanding cloud to a vortical columnar flow of constant radial extension for $z\geq z_{f,\textrm{col}}$ (see white arrows in figures \ref{subfig:saltWater5rpm} and \ref{subfig:saltWater10rpm}; the transition is also visible in a video in the Supplemental Material [\textit{SaltWater\_10RPM.avi}]. Initially the cloud inertia is large, hence as long as the cloud is expanding, one may neglect the influence of rotation and consider that the cloud follows the model of MTT56: the cloud grows and decelerates due to the entrainment of ambient fluid, gradually reducing the cloud inertia. Simultaneously, entrainment enables the ambient fluid in solid body rotation to gradually penetrate through the turbulent thermal and increase its total kinetic momentum. The transition is expected to happen when the Coriolis force predominates over the cloud inertia \citep{ayotteMotionTurbulentThermal1994,helfrichThermalsBackgroundRotation1994,fernandoDevelopmentPointPlume1998a}. This can be quantified with a Rossby number based on the inflow of ambient fluid at the thermal interface. There, the entrained fluid is subject to the Coriolis force $2\Omega v_e$ with $v_e$ the entrainment velocity (equation \eqref{eq:EntrainmentVelocity}), and its inertia reads $(\dot{z}_f/r)v_e$, leading to
\begin{equation}
    \label{eq:Rossby}
    Ro(z) = \frac{\dot{z}_f}{2\Omega r}.
\end{equation}
The front velocity $\dot{z}_f$ is used rather than $\dot{z}$ because the front is easily traceable from videos, and more meaningful due to a modification of the residence time of particles when $\Omega>0$, as discussed in section \ref{sec:KinematicsResRotation}.

\definecolor{navy}{rgb}{0.05,0.1,0.6}
\definecolor{crimson}{rgb}{0.6,0.08,0.2}
\definecolor{lightblue}{rgb}{0.5,0.8,1}
\begin{figure}[htb]
        \centering
        \includegraphics[height=6.5cm]{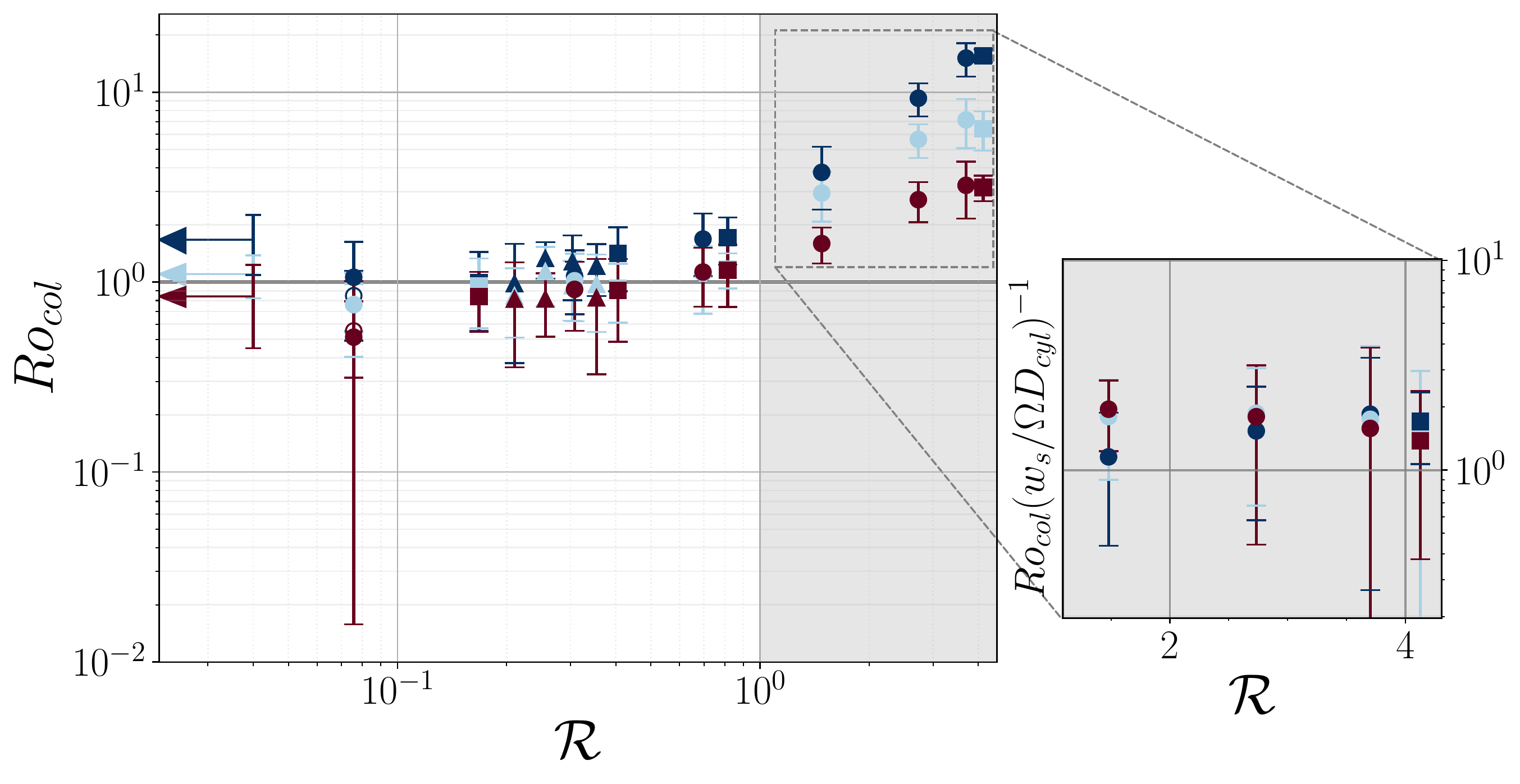}
    \caption{Transitional Rossby number $Ro_{\textrm{col}}$ computed at depth $z \geq z_{f,\textrm{col}}$. The inset validates the scaling $Ro_{\textrm{col}} \simeq w_s/\Omega D_\textrm{cyl}$ in the range $\mathcal{R}> 1$ (gray shaded area) corresponding to clouds which are columnar due to particles falling as swarms. Colour code: (\textcolor{navy}{$\CIRCLE$}) $\Omega=5$rpm, (\textcolor{lightblue}{$\CIRCLE$}) $\Omega=10$rpm, (\textcolor{crimson}{$\CIRCLE$}) $\Omega=20$rpm.}
  \label{fig:RossbyTransition}
\end{figure}

The transition is expected when the Rossby number is of order unity. For simplicity and up to a constant factor, we will consider $Ro_\textrm{col}=Ro(z_{f,\textrm{col}})=1$ as the condition of transition to a vortical columnar flow.
To test this condition, we compute the transitional Rossby number from the automatic cloud-tracking algorithm of Appendix \ref{sec:cloudTracking}. The depth of transition is defined due a change in the clouds' kinematics, from an initial phase of deceleration when $z<z_{f,\textrm{col}}$ to a phase of constant velocity when $z>z_{f,\textrm{col}}$ (see section \ref{sec:KinematicsResRotation}). Then, measurements of the Rossby number $\dot{z}_f(z)/2\Omega \sigma_x(z)$ are approximately constant for $z\geq z_{f,\textrm{col}}$, hence the values of this Rossby number are averaged for $z\geq z_{f,\textrm{col}}$ where noise is low enough. Results are shown in figure \ref{fig:RossbyTransition} in ordinate, as a function of $\mathcal{R}$. For any given $\Omega$, measurements confirm that $Ro_{\textrm{col}}\simeq 1$ within the uncertainty margin of experiments in the range $\mathcal{R} \leq 1$. Therefore, in a rotating environment, this range of Rouse numbers corresponds to clouds whose particles fall in a vortical columnar flow below $z_{\textrm{f,col}}$.
\par
Before transition $Ro(z)>1$ so inertia predominates over the Coriolis force, and if one neglects rotation, clouds are expected to behave as turbulent thermals. According to this simple picture, which is rigorously applicable when $Ro(z)\gg 1$, the condition $Ro_\textrm{col}=1$ yields $z_{\textrm{f,col}} \sim \Omega^{-1/2}$. Although we lack additional values of $\Omega$ to quantitatively confirm this scaling, it is compatible with our present measurements made for three different values of $\Omega>0$, and was already verified in past experiments with one-phase thermals (see references \citep{ayotteMotionTurbulentThermal1994,helfrichThermalsBackgroundRotation1994}). Thus, the faster the spinning, the lower the depth of transition, as observed in figure \ref{fig:saltWaterSeries}.
\par
In figure \ref{fig:RossbyTransition}, in the range $\mathcal{R}>1$ the transitional Rossby number $Ro_{\textrm{col}}$ is larger than unity. This is because such clouds are columnar not due to rotation, but because particles behave as swarms. Hence particles fall with a constant velocity of order $w_s$ and clouds hardly grow so their diameter is approximately equal to $D_{\textrm{cyl}}$. Consequently $Ro_{\textrm{col}} \simeq w_s/\Omega D_\textrm{cyl}$, as confirmed by the inset of figure \ref{fig:RossbyTransition}.

%%%%%%%%%%%%%%%%%%%%%%%%%%%%%%%%%%%%%%%%%%%%%%%%%%%%%%%%%%%%%%%%
\subsection{\label{sec:ThermalRotation}Thermal regime with background rotation: entrainment levelling}

\begin{figure}[htb]
\centering
    \begin{subfigure}[b]{.4\textwidth}
      \caption{}
      \centering
      \includegraphics[height=5cm]{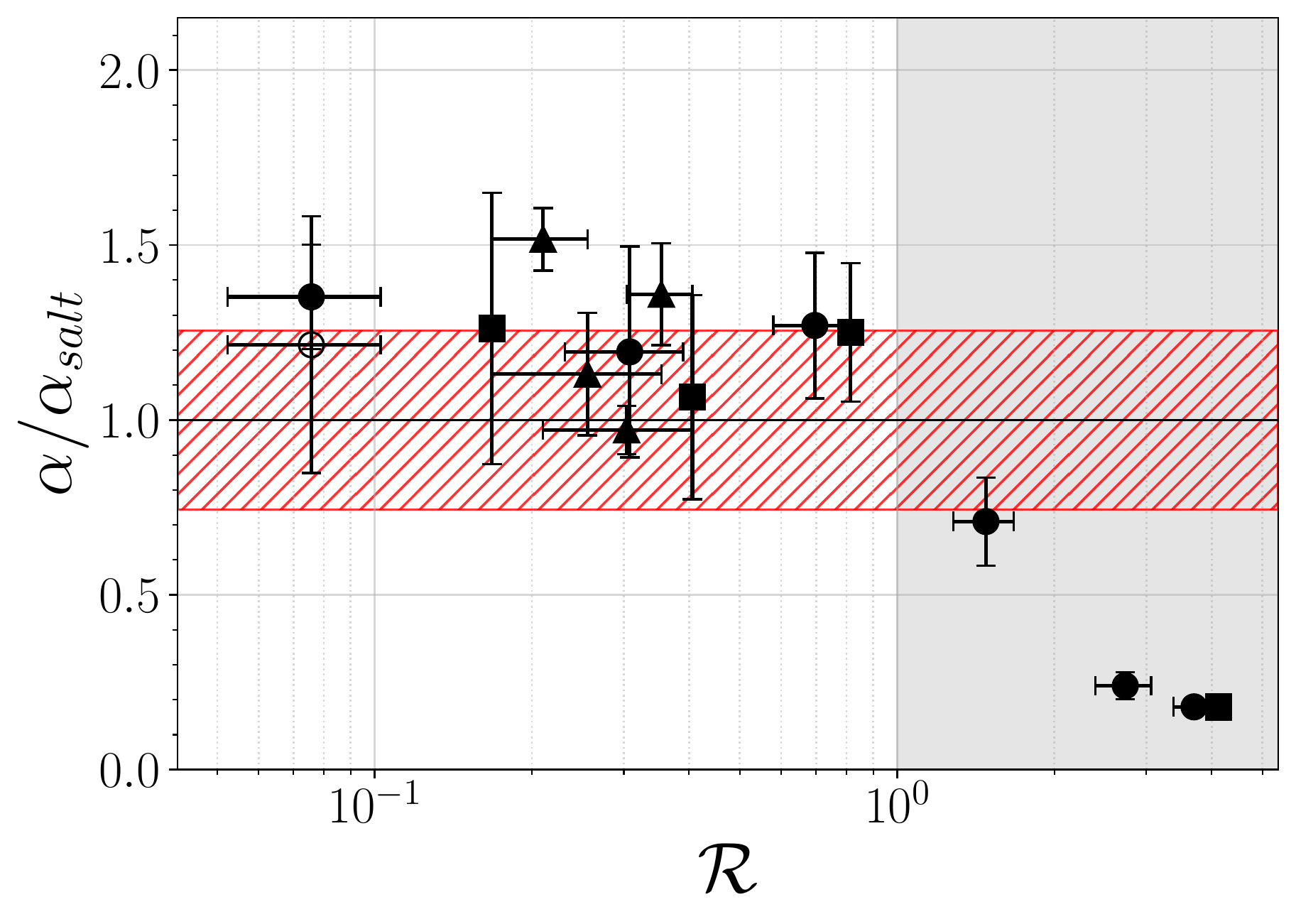}
      \label{subfig:alpha5rpm}
    \end{subfigure}
    \begin{subfigure}[b]{.4\textwidth}
      \caption{}
      \centering
      \includegraphics[height=5cm]{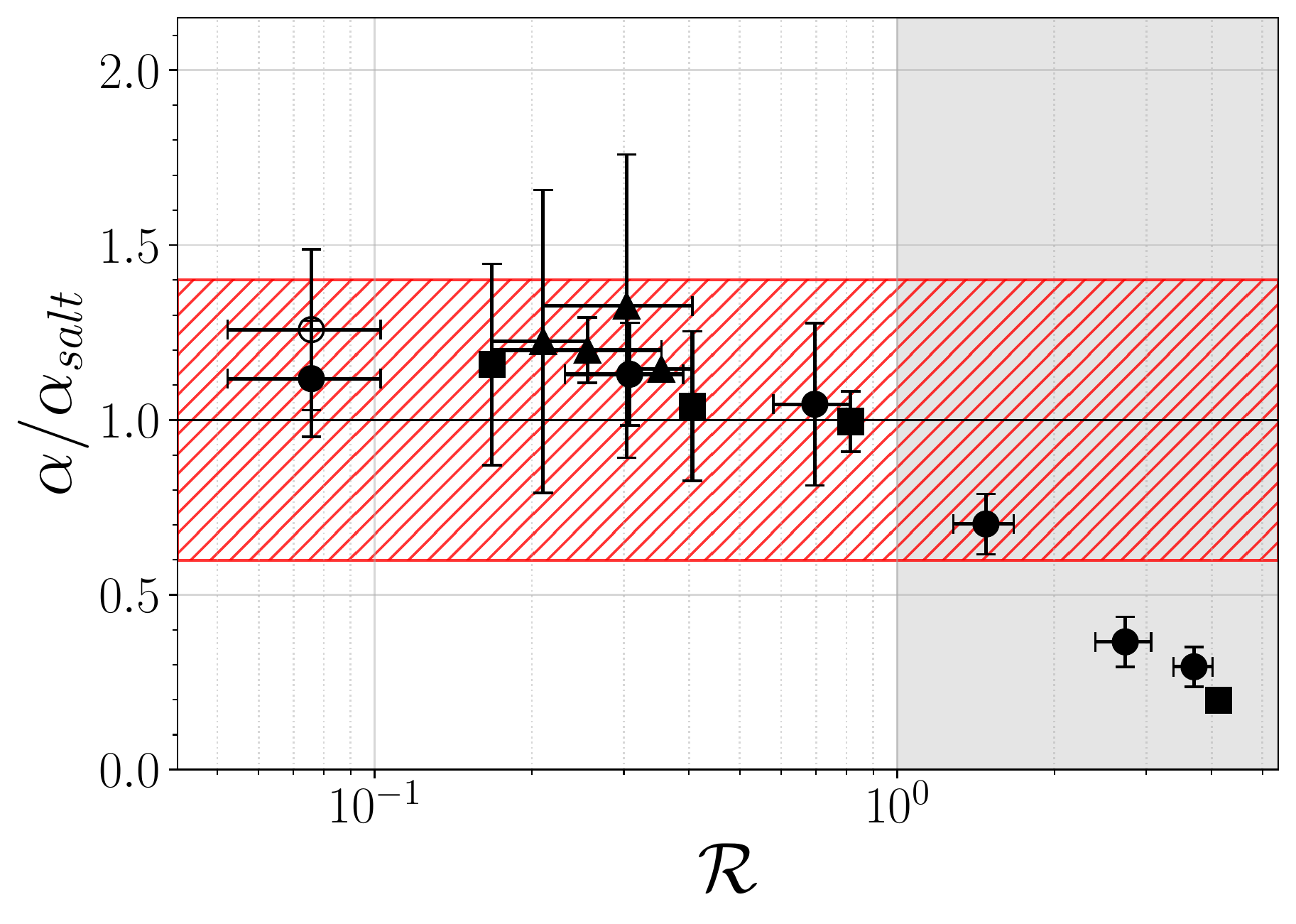}
      \label{subfig:alpha10rpm}
    \end{subfigure}
    \caption{Entrainment capacity of particle clouds for different Rouse numbers at (a) $\Omega=5$rpm and (b) $\Omega=10$rpm. For further information on colours and symbols, see figure \ref{fig:EntrainmentStill}.}
  \label{fig:entrainmentRotation}
\end{figure}

Let us focus on the evolution of clouds before the transition to a vortical columnar flow. As long as $Ro(z)>1$ and after a phase of acceleration, the cloud is considered to evolve as a turbulent thermal which entrains ambient fluid and grows linearly in depth. The entrainment capacity of all particle clouds is compared to the reference of salt water thermals in figure \ref{fig:entrainmentRotation}. This entrainment capacity is not presented for $\Omega=20rpm$ because at such rotation rate, the depth of transition $z_{\textrm{f,col}}$ is too low for the entrainment rate to be measurable above $z_{\textrm{f,col}}$.
\par
At both $\Omega=5rpm$ (figure \ref{subfig:alpha5rpm}) and $\Omega=10rpm$ (figure \ref{subfig:alpha10rpm}), the entrainment coefficient suddenly drops for $\mathcal{R} >1$ as previously seen for $\Omega=0rpm$. This is again due to particles decoupling from the fluid due to their inertia. Most interestingly, in the range $\mathcal{R} \leq 1$ figures \ref{subfig:alpha5rpm} and \ref{subfig:alpha10rpm} show that $\mathcal{R}$ hardly has any influence on $\alpha/\alpha_{\textrm{salt}}$, and the entrainment capacity $\alpha$ is levelled to a value close to $\alpha_{\textrm{salt}}$. This contrasts with observations at $\Omega=0$ in figure \ref{fig:EntrainmentStill}. In other words, because of rotation, particulate effects appear to be inoperative, hence particle clouds entrain approximately as much as their salt-water counterparts, as evidenced by the overlap of most data points with the errorbar of $\alpha_{\textrm{salt}}$ in red hatchings.

It may come as a surprise that rotation suppresses any local maximum value of $\alpha/\alpha_{\textrm{salt}}$ in the range $z< z_{f,\textrm{col}}$ even though $Ro(z< z_{f,\textrm{col}})>1$ i.e. inertia predominates over the Coriolis force. In a still environment, we observe an enhancement of the entrainment rate for $\mathcal{R}\simeq 0.3$ where particles concentrate, especially in the peripheral eddies of the particle cloud (see figure \ref{fig:ParticulateEffectsOutEddies}). At the particles' scale, the Coriolis force has a negligible dynamical influence (it is three to four orders of magnitude lower than the particles' inertia), so the disappearance of the enhancement of $\alpha$ is necessarily due to the influence of the Coriolis force on the fluid. This influence acts before $z=z_{f,\textrm{col}}$, because the condition $Ro(z_{f,\textrm{col}})=1$ holds for a transition of the \textit{entire} cloud, i.e. when rotation starts to fully govern the cloud dynamics. Conversely, the particulate effects at stake are rather \textit{localised} at the interface between the cloud and the ambient fluid, where the entrainment flux affects the flow as soon as the inflow starts. Consequently, rotation is expected to have a non-negligible influence on the outer shell of the cloud for a Rossby number $Ro(z)$ larger than unity (i.e. before $z_{f,\textrm{col}}$ is reached) by modifying the structure of the flow and gradually tilting vortices along $\vec{e}_z$. Such modifications likely prevent particles from concentrating and sweeping on the downward sides of eddies, thus switching particulate effects off. Because in our experiments at $\Omega\geq 5$rpm, the Rossby number never exceeds 6 when $\mathcal{R} \leq 1$, clouds are never sufficiently inertial for their interface to be unaffected by rotation (as pointed out by references \citep{frankAnticyclonicPrecessionPlume2017,fabregattomasNumericalSimulationsRotating2017a} in the case of continuous injections of buoyancy, rotation modifies a plume's dynamics even for modest Rossby numbers; furthermore, rotation eventually influences \textit{any} plume as long as its injection is long enough, due to the conservation of angular momentum \citep{frankAnticyclonicPrecessionPlume2017}). Although reference \cite{sutherlandPlumesRotatingFluid2021} bears upon the behaviour of plumes in background rotation, it inspired our reflections on the transition to vortical columnar flows and we refer the reader to this study for additional information on the flow that develops with a starting plume in background rotation. Further investigation on the behaviour of particles and the structure of the flow close to the interface requires additional visualisations of particles in horizontal planes and  quantification of the velocity field, which is beyond the scope of the present study.
\par
Another ingredient which may act to inhibit particulate effects is the centrifugal force. In fact, although this force is negligible compared to gravity (the ratio $r\Omega^2/g$ is only of order $10^{-1}$ for the largest rotation rate $\Omega=20$rpm and for $r=7D_{\textrm{cyl}}$ which corresponds to the maximum radial distance to the axis of rotation), depending on the value of $\Omega$ we observe that the centrifugal force due to the rotating table is comparable with the centrifugal force exerted on particles when they swirl in eddies at 0rpm. Hence, the centrifugal force due to $\Omega>0$ may switch off preferential concentration as well as preferential sweeping, so that particles cannot concentrate to force the flow through two-way coupling.

\subsection{\label{subsec:SwarmRotation}Swarm regime}

\begin{figure}[htb]
    \centering
    \begin{subfigure}[t]{.2\textwidth}
        \caption{}
        \centering
        \includegraphics[height=4.8cm]{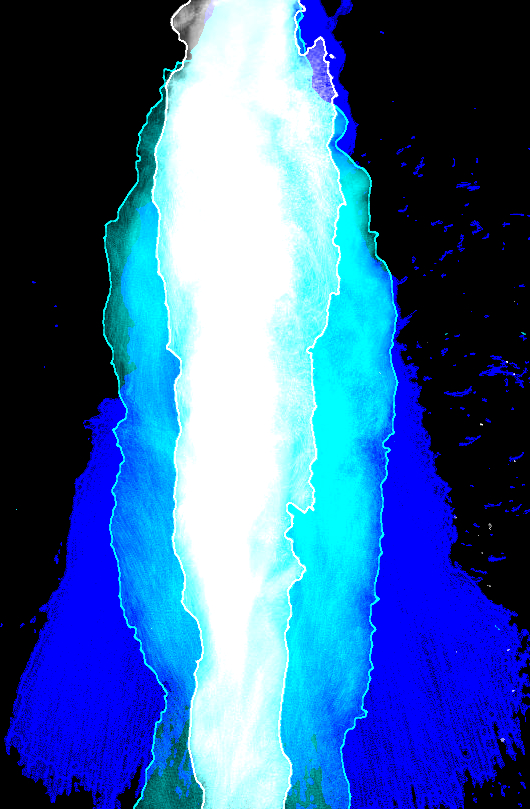}
        \label{subfig:overlayColumns}
    \end{subfigure}
    \begin{subfigure}[t]{.38\textwidth}
        \caption{}
        \centering
        \includegraphics[height=5cm]{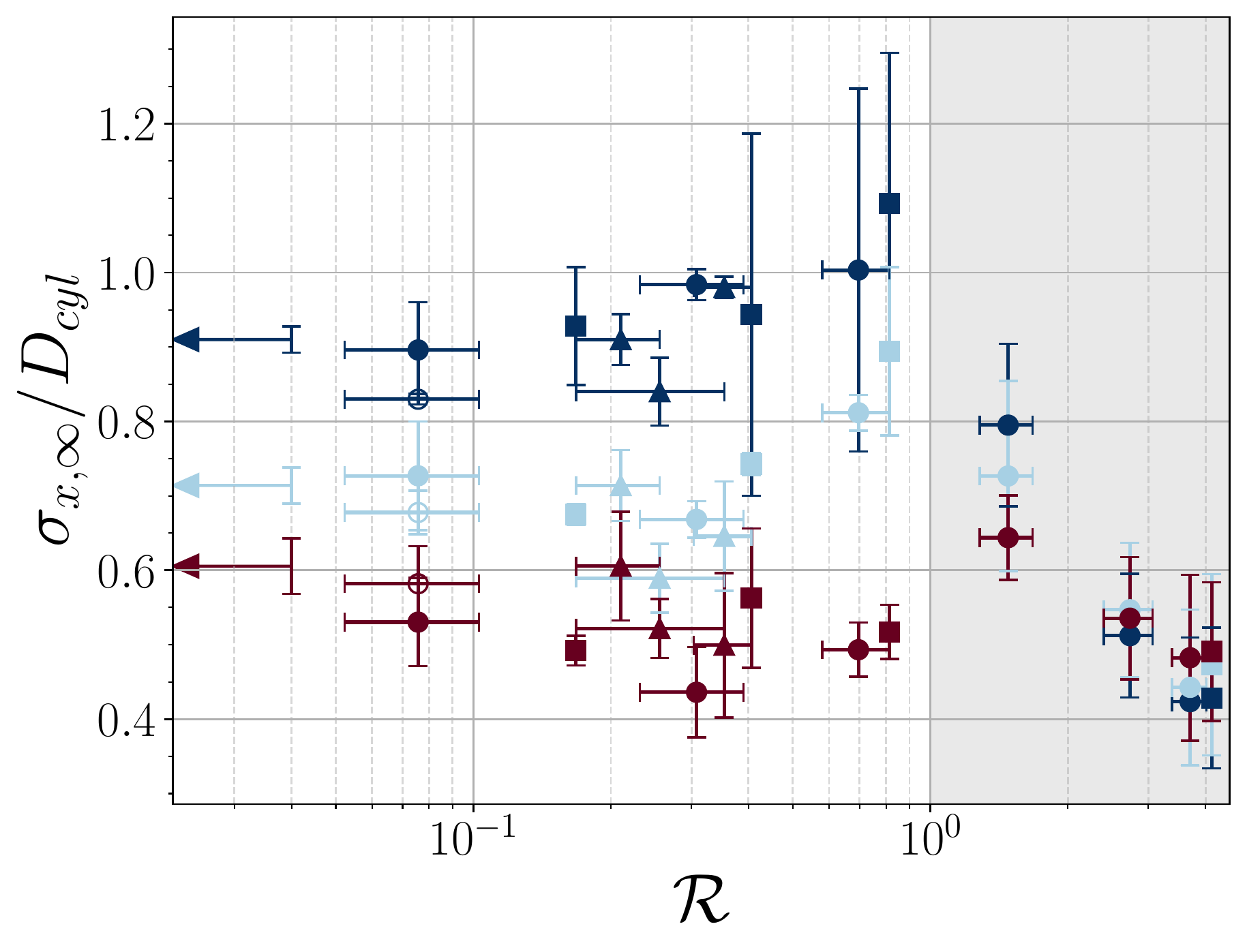}
        \label{subfig:sigmaXlimTC_severalRPM}
    \end{subfigure}
    \caption{(a) Overlay of colourised images of height 37.8cm, corresponding to pixel-by-pixel standard deviations of light intensity of three different experiments of Rouse number $\mathcal{R}=0.168$ at 0rpm (blue), 5rpm (cyan) and 20rpm (white). (b) Dimensionless width of the rotating columns in the range $z\geq z_{f,\textrm{col}}$ as a function of $\mathcal{R}$. Arrows and their errorbars correspond to values for salty thermals. Colour code: (\textcolor{navy}{$\CIRCLE$}) $\Omega=5$rpm, (\textcolor{lightblue}{$\CIRCLE$}) $\Omega=10$rpm, (\textcolor{crimson}{$\CIRCLE$}) $\Omega=20$rpm.}
  \label{fig:WidthsColumns}
\end{figure}

Figure \ref{subfig:overlayColumns} illustrates the influence of rotation due to the overlay of integral images for different rotation rates and identical Rouse numbers: the faster the rotation, the narrower the column (see also the Supplemental Material \textit{Rouse0p406\_0-5-10RPM.avi}), which is consistent with past observations for miscible thermals \citep{ayotteMotionTurbulentThermal1994,helfrichThermalsBackgroundRotation1994} and bubble-laden plumes \cite{frankEffectsBackgroundRotation2021}. To confirm this observation, we perform quantitative measurements of the width of columns through the quantity $\sigma_{x,\infty}$. For a given experiment, it corresponds to the average value of $\sigma_x(z)$ (the cloud width detected by the cloud-tracking algorithm; see appendix \ref{sec:cloudTracking}) after it becomes approximately constant. Figure \ref{subfig:sigmaXlimTC_severalRPM} confirms that the larger $\Omega$, the narrower the column. For $\Omega=20$rpm, the column half width is approximately $\sigma_{x,\infty}=D_{\textrm{cyl}}/2$ because $z_{f,\textrm{col}}$ is so low that the cloud has no time to grow through turbulent entrainment before becoming columnar, so it remains as large as the cylinder of injection. Similarly, all clouds in the range $\mathcal{R} >1$ never grow as a thermal and therefore have a width $\sigma_{x,\infty}=D_{\textrm{cyl}}/2$.
\par
The decreasing column width with increasing $\Omega$ can be explained from geometrical arguments. Before transition when $Ro(z)>1$, for simplicity one may assume that clouds behave as self-similar turbulent thermals. Then, from the scalings \eqref{eq:similarR}-\eqref{eq:similarVelocity} of MTT56, the width of a cloud should scale like $\sim \Omega^{-1/2}$ when $z_f=z_{f,\textrm{col}}$, as argued and verified in references \citep{ayotteMotionTurbulentThermal1994,helfrichThermalsBackgroundRotation1994}. Our experiments are perfomed for three different values of $\Omega>0$ which is insufficient to quantitatively confirm this scaling, yet we note that it is compatible with our present measurements, using $\sigma_{x,\infty}$ as a measurement of the cloud radius at $z=z_{f,\textrm{col}}$. In other words, as $\Omega$ increases, the condition $Ro(z)=1$ is reached at lower depths hence columns are narrower.

\begin{figure}[htb]
    \centering
    \includegraphics[height=5cm]{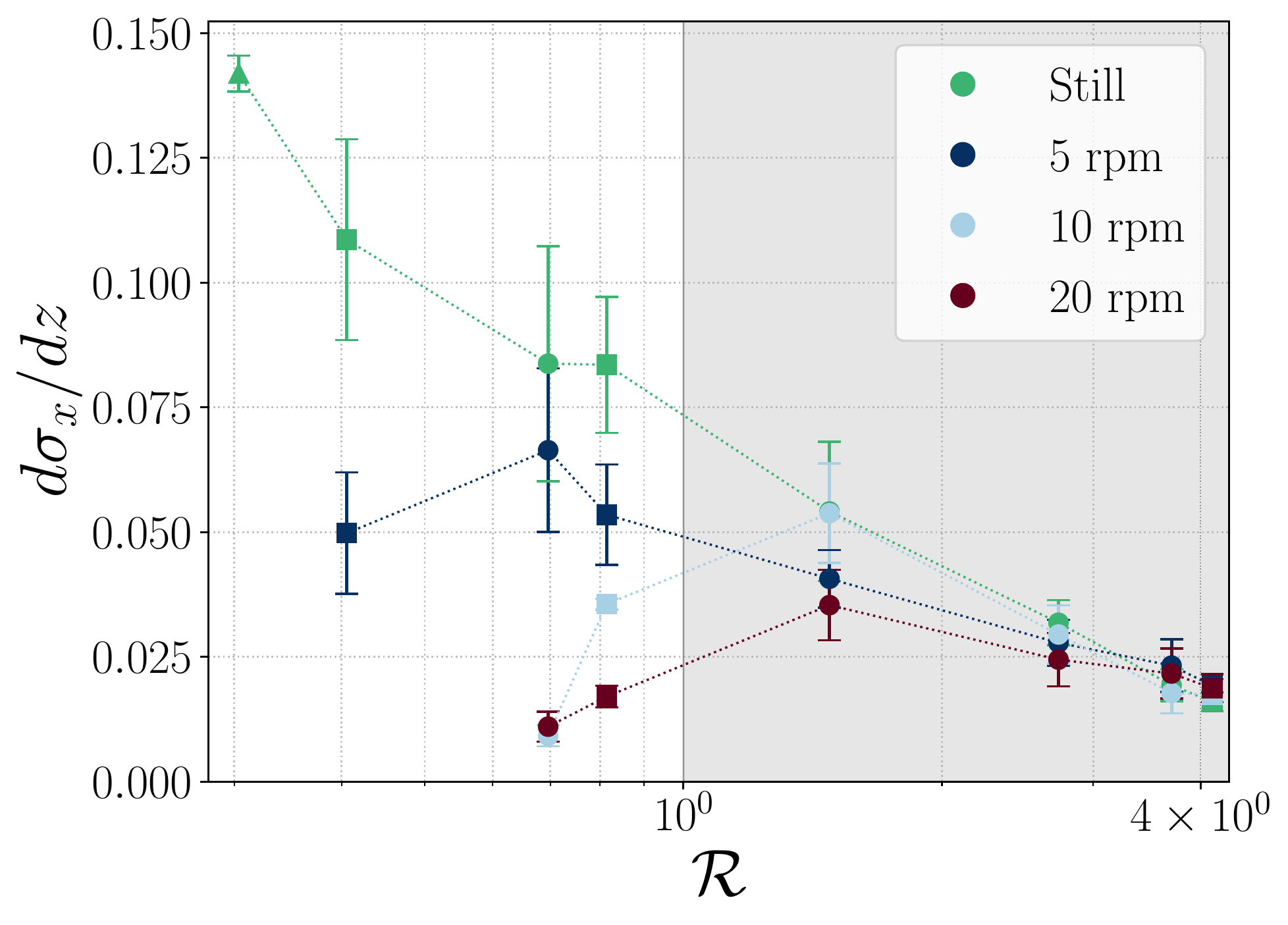}
    \caption{Lateral growth rate of clouds which behave as swarms and thus slightly grow in depth. Measurements are performed with least squares linear regressions on $\sigma_x(z)$, i.e. assuming that $d\sigma_x/dz$ is constant.}
    \label{fig:GrowthRateSwarmRotation}
\end{figure}

A remark should be made on the fact that the constant width of columns is a robust observation for low Rouse numbers, but for large Rouse numbers it is an approximation whose accuracy is quantified by the error bars in figure \ref{subfig:sigmaXlimTC_severalRPM}. In fact, for the largest Rouse numbers, particles ultimately fall as swarms. In the absence of background rotation, the latter slowly expand along their fall. In fact, Subramanian and Koch \citep{subramanianEvolutionClustersSedimenting2008} in the range $Re_p\ll 1$, and later Daniel \textit{et al.} \cite{danielClustersSedimentingHighReynoldsnumber2009} in the range $Re_p\in$[93-425], showed that hydrodynamical interactions between particles result in a growth of swarms as $\sigma_x\sim t^{1/3}$. Because of the variability between several realisations, and due to our limited field of view after separation happens, our measurements do not enable us to discriminate between a sublinear or linear evolution of $\sigma_x$ in depth for the swarm regime. Hence, we compute indicative constant values of the growth rate of swarms $d\sigma_x/dz$ obtained from linear regressions in the range $z>z_\textrm{sep}$ if $\Omega=0$, and we perform the same analysis in the range $z_f>z_{f,\textrm{col}}$ when $\Omega>0$. Results are shown in figure \ref{fig:GrowthRateSwarmRotation} for clouds which slightly grow indeed, behaving as swarms. In the range $\mathcal{R}>1$, we observe that the more inertial the particles, the more the curves collapse on the dark symbols for $\Omega=0$, indicating that particles are less and less sensitive to rotation due to their decoupling from the fluid. In the range $\mathcal{R}\leq 1$, the minimum Rouse number for which $d\sigma_x/dz>0$ is measurable tends to increase with $\Omega$; additionally, for a given Rouse number, the larger $\Omega$ the more $d\sigma_x/dz$ differs from the reference values in a still environment. Both of these observations are due to the larger azimuthal inertia of the fluid and the increasing stiffening of radial motions when $\Omega$ increases, which require a larger particle inertia to be overcome.

\subsection{\label{sec:KinematicsResRotation}Kinematics and residence time}
\begin{figure}[htb]
    \centering
    \begin{subfigure}[t]{.49\textwidth}
        \caption{}
        \centering
        \includegraphics[height=5.4cm]{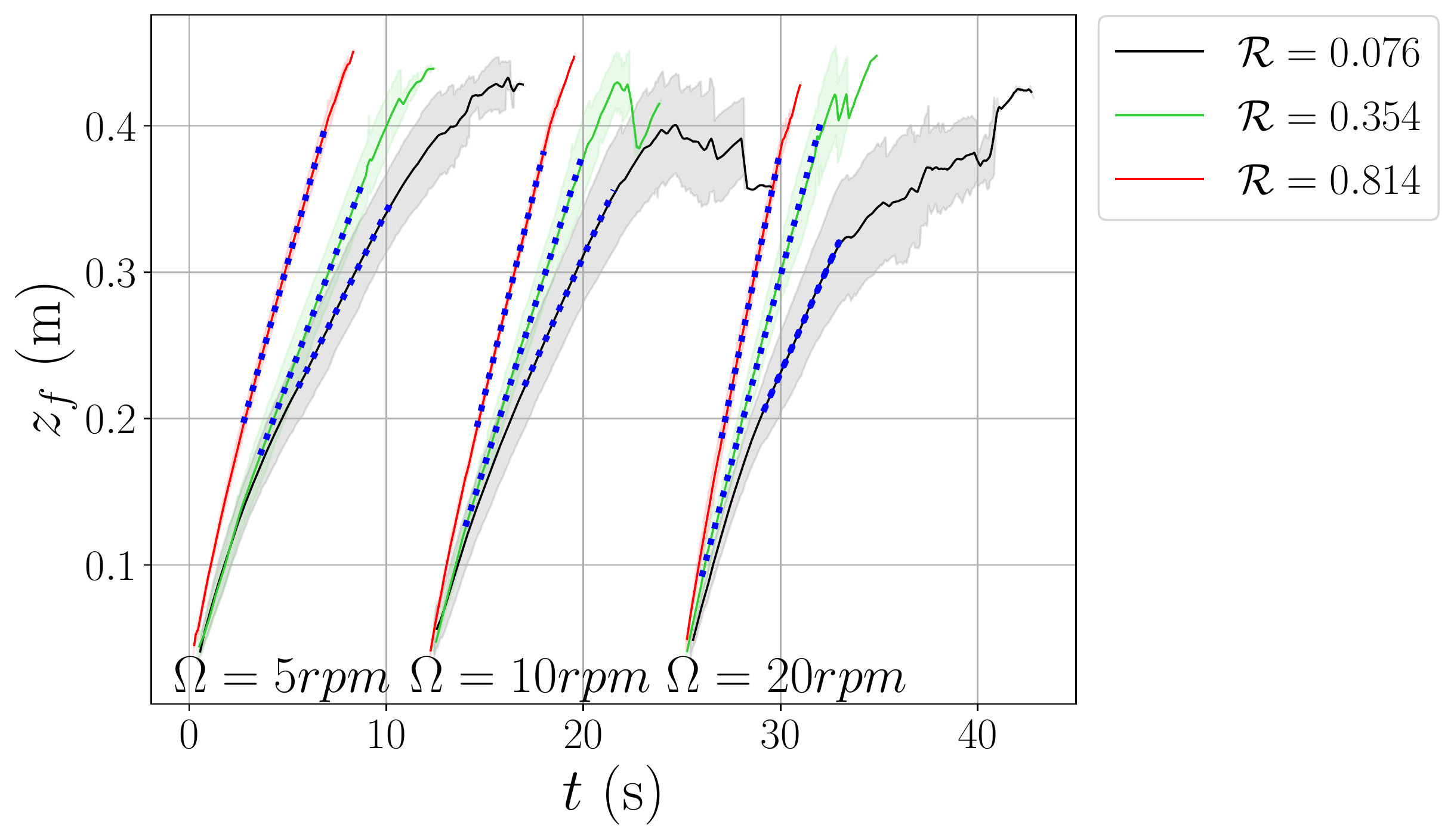}
        \label{subfig:SwarmFit_rotation}
    \end{subfigure}
    \begin{subfigure}[t]{.5\textwidth}
        \caption{}
        \centering
        \includegraphics[height=5.4cm]{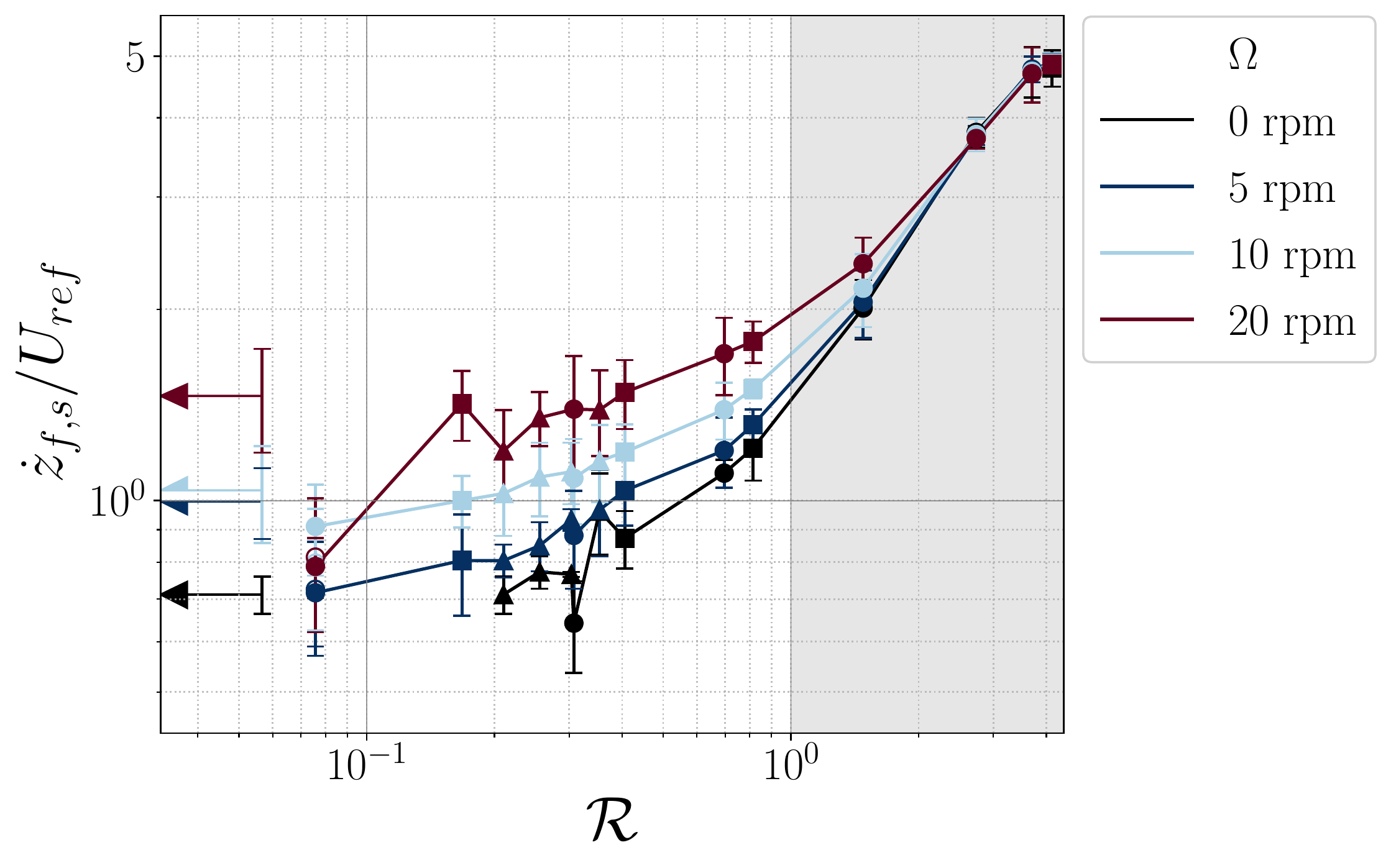}
        \label{subfig:SwarmVelocity_effectRotation}
    \end{subfigure}    
    \caption{(a) Examples of the determination of the constant front velocity during the swarm regime. Solid lines correspond to $z_f(t)$ averaged over all realisations, and shaded areas correspond to uncertainty margins, which are extremely small for large Rouse numbers. Colours vary with the Rouse number, and the curves are horizontally shifted for $\Omega=10rpm$ and $\Omega=20rpm$ for clarity. The linear fits are shown as dotted blue lines in the time ranges that minimise the error. (b) Evolution of the cloud front velocity in the swarm regime for various angular velocities of the rotating table. Arrows and their errorbars correspond to values for salty thermals.}
  \label{}
\end{figure}

The influence of rotation on a cloud front velocity is now assessed. The front velocity is observed to be approximately constant below the depth $z_{f,\textrm{col}}$. Since we can hardly distinguish between particles which fall in a vortical columnar flow, and swarms which are decoupled from the fluid, in both cases the constant velocity for $z>z_{f,\textrm{col}}$ is denoted $\dot{z}_\textrm{f,s}$ to be consistent with notations used for clouds in a still environment, and we talk about a swarm regime. This velocity is determined from a linear fit of $z_f(t)$ as illustrated in figure \ref{subfig:SwarmFit_rotation}. Results for the velocities are shown in figure \ref{subfig:SwarmVelocity_effectRotation}. In the range $\mathcal{R} >1$, the influence of $\Omega$ noticeably decreases with increasing $\mathcal{R}$ and curves collapse on the $\Omega=0$ curve. This is due to particles being more and more insensitive to the swirling flow, especially as $\Omega$ is lower. In the range $\mathcal{R} \leq 1$, the larger the rotation rate, the larger the front velocity (see also the Supplemental Material \textit{Rouse0p406\_0-5-10RPM.avi}). This trend is mainly interpreted through the reduction of the column radius $r\simeq \sigma_{x,\infty}$ (see figure \ref{subfig:sigmaXlimTC_severalRPM}) which, in turn, increases the particle volume fraction $\phi = 3m_0/4\pi \rho_p r^3$. From a macroscopic point of view, this results in a larger effective cloud density $\rho=\phi \rho_p + (1-\phi)\rho_f$, enhancing the cloud velocity through the reduced gravity $g(\rho/\rho_f-1)$.
\par
It should be mentioned that fitting an affine model on the average curve $z_f(t)$ is at odds with past measurements which seem in agreement with the scaling $z_f(t) \sim t^{1/2}$ proposed by the authors in references \citep{ayotteMotionTurbulentThermal1994,helfrichThermalsBackgroundRotation1994}. There seems to be no physical reason for columnar clouds to keep decelerating as turbulent thermals following the law $z_f(t) \sim t^{1/2}$ after the turbulent entrainment and lateral cloud growth are interrupted by rotation, thus cancelling the mixing drag. Conversely, the previous arguments are consistent with a constant cloud front velocity. Yet, we still note a subtle decrease of the clouds' velocity in time for the lowest Rouse numbers. This slight decrease is ultimately expected since a lot of the buoyant material remains in suspension behind the cloud front, to such an extent that the cloud front gradually thins out (see figure \ref{fig:saltWaterSeries}) thus reducing the front effective density and velocity. We expect the front deceleration to depend on the rate of detrainment of buoyant material in the columnar cloud wake, which cannot be measured from our planar visualisations.

\begin{figure}[h!]
    \centering
    \begin{subfigure}[t]{.75\textwidth}
        % \caption{}
        \centering
        \includegraphics[height=4.96cm]{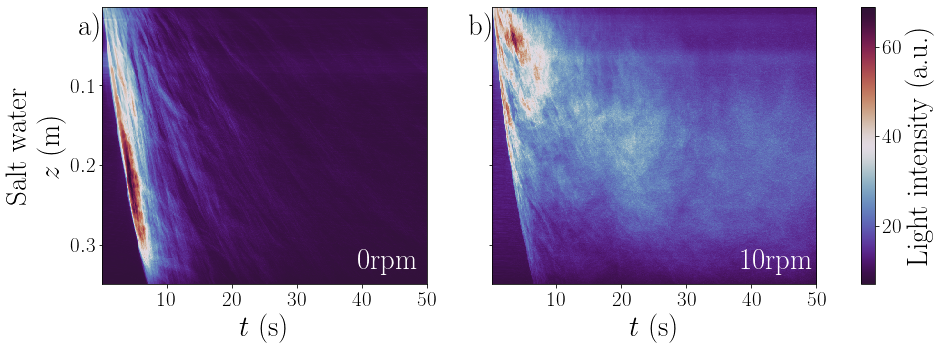}
        \label{subfig:profileSalt}
    \end{subfigure}
    \begin{subfigure}[t]{.24\textwidth}
        % \caption{}
        \centering
        \includegraphics[height=4.96cm]{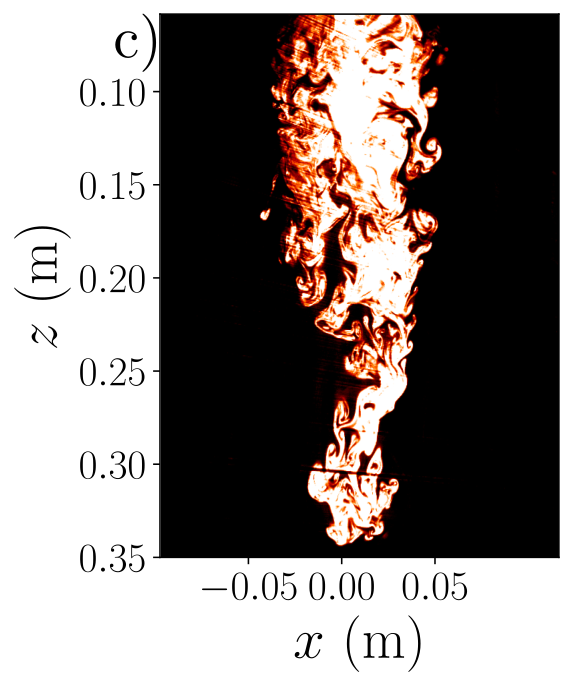}
        \label{subfig:rhodaSalt}
    \end{subfigure}
    \par
    \begin{subfigure}[t]{.75\textwidth}
        % \caption{}
        \centering
        \includegraphics[height=4.96cm]{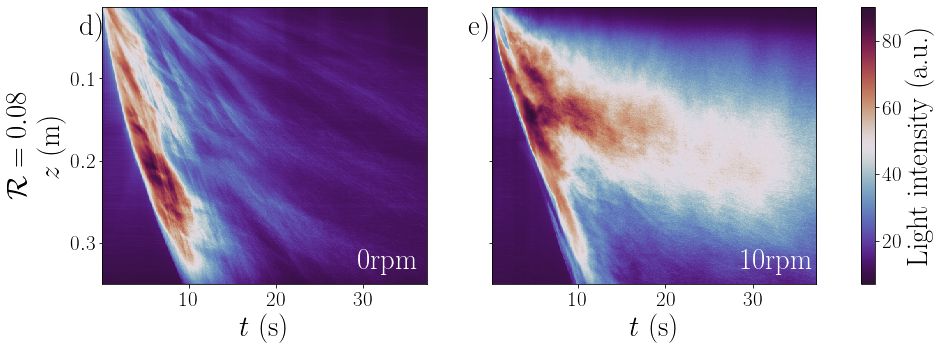}
        \label{subfig:profile4070}
    \end{subfigure}
    \begin{subfigure}[t]{.24\textwidth}
        % \caption{}
        \centering
        \includegraphics[height=4.96cm]{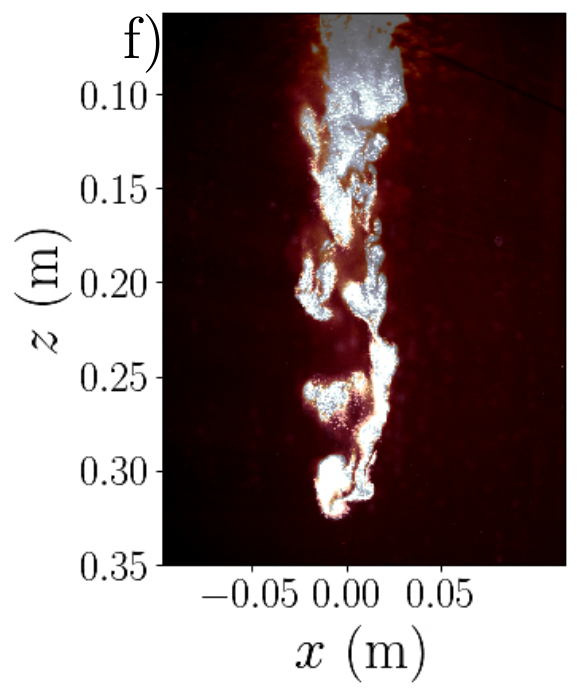}
        \label{subfig:rhoda4070}
    \end{subfigure}
    \par
    \begin{subfigure}[t]{.75\textwidth}
        % \caption{}
        \centering
        \includegraphics[height=4.96cm]{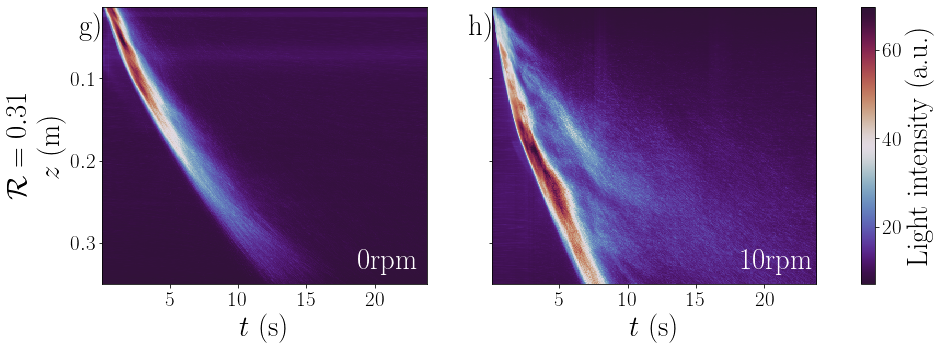}
        \label{subfig:profile90150}
    \end{subfigure}
    \begin{subfigure}[t]{.24\textwidth}
        % \caption{}
        \centering
        \includegraphics[height=4.96cm]{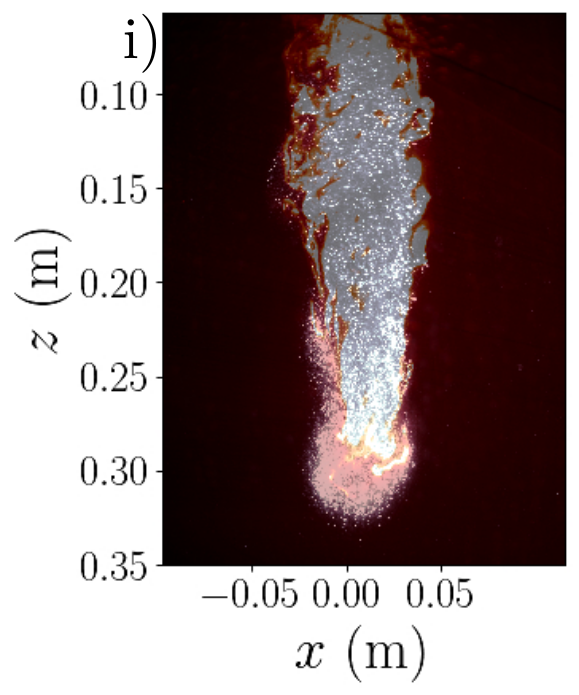}
        \label{subfig:rhoda90150}
    \end{subfigure}
    \par
    \begin{subfigure}[t]{.75\textwidth}
        % \caption{}
        \centering
        \includegraphics[height=4.96cm]{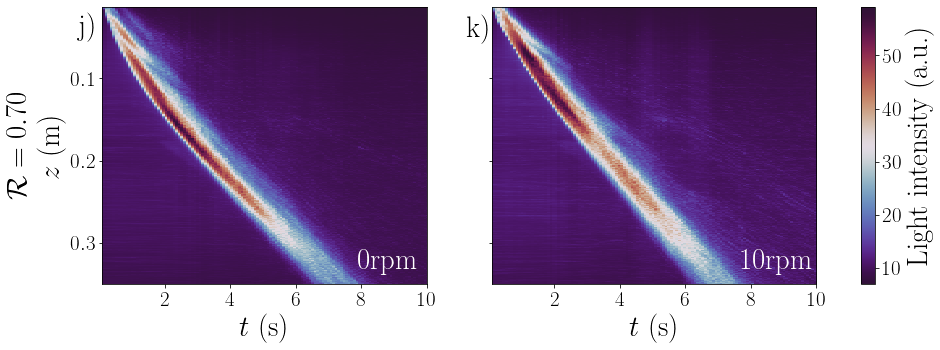}
        \label{subfig:profile224250}
    \end{subfigure}
    \begin{subfigure}[t]{.24\textwidth}
        % \caption{}
        \centering
        \includegraphics[height=4.96cm]{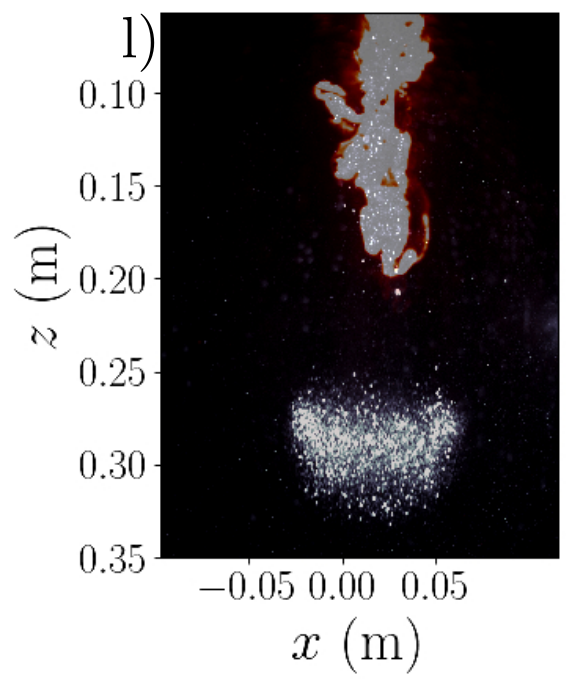}
        \label{subfig:rhoda224250}
    \end{subfigure}
    \caption{Left and middle columns respectively correspond to the Hovmöller diagrams of the horizontally-averaged light intensity of clouds falling respectively at 0rpm and 10rpm. Every row corresponds to a different Rouse number as indicated on the left-hand side. Right column: snapshots of particles falling with rhodamine at 10rpm.}%10rpm rhodamine 18/02, 19/02, 23/02, 01/03,  page 71 cahier manip
  \label{fig:Hovmoller}
\end{figure}

Further understanding can come from analysing how particles distribute and settle in columnar clouds. Figure \ref{fig:Hovmoller} combines snapshots and Hovmöller diagrams of the light intensity obtained for every image of an experiment after averaging the values of all pixels along every row. For very large Rouse numbers (see figures \ref{fig:Hovmoller}j-l), particles decouple from the fluid due to their inertia and fall as swarms with a constant velocity. For lower Rouse numbers (see figures \ref{fig:Hovmoller}a-i), compared to clouds falling in a still environment, the Hovmöller diagrams confirm that due to rotation, the cloud front velocity is approximately constant and detrainment is enhanced since the residence time of particles at any depth is visibly increased. On the Hovmöller diagrams, the fast concentrated frontal blob falling with constant velocity corresponds to the cloud. Conversely, the slow dilute region where particles settle on a much larger time scale on average corresponds to the stem of detrained particles (see respectively red and white regions in figures \ref{fig:Hovmoller}g-h). Furthermore, images with rhodamine confirm that the frontal blob corresponds to the turbulent cloud initially dyed with rhodamine, since this blob drags rhodamine down the tank (see figure \ref{fig:Hovmoller}i in particular). Thus, detrainment is enhanced by rotation; in addition, we observe that the frontal blob laterally shrinks as it falls in the tank. These observations might result from the interaction between the rotating ambient and the turbulent cloud at its interface, where the cloud seems to be gradually peeled off by a strong shear, thereby switching off particulate effects. Further investigation would require additional observation in a horizontal plane, as well as measurements of the fluid velocity field.

% %%%%%%%%%%%%%%%%%%%%%%%%%%%%%%%%%%%%%%%%%%%%%%%%
% %%%%%%%%%%%%%%%%%%%%%%%%%%%%%%%%%%%%%%%%%%%%%%%%

\section{\label{sec:Conclusion}Summary and conclusion}
Instantaneous releases of a buoyant material in still water behave differently if the material is salt water or heavy particles. Particle clouds initially behave as turbulent thermals, and the difference manifests through an increase of their entrainment capacity compared to salty clouds, with a maximum of enhancement for a finite inertia corresponding to $\mathcal{R}\simeq 0.3$ (figure \ref{fig:EntrainmentStill}). The origin of this enhancement likely stems from particulate effects which have already been observed in canonical turbulent flows, mainly preferential concentration and preferential sweeping. A second difference is due to a transition from the thermal regime to the swarm regime because particles decouple from turbulent eddies due to their inertia and eventually separate from them.
\par
Adding background rotation, the particulate enhancement of entrainment is inhibited (figure \ref{fig:entrainmentRotation}), likely due to the influence of the Coriolis force on the inflow of entrained fluid at the cloud interface. When the Coriolis force predominates on the dynamics of the whole cloud ($Ro\leq 1$), the latter transitions to a regime of vortical columnar flow (figure \ref{fig:RossbyTransition}). Together with the decoupling of particles from the flow (separation), this transition adds a new limitation on the duration of turbulence. The model of Morton \textit{et al.} \citep{mortonTurbulentGravitationalConvection1956} appears like an efficient framework to anticipate the different regimes experienced by the particle clouds throughout their fall.
\par
Polydisperse, bidisperse and monodisperse particle sets were used to assess the role of polydispersity in the clouds dynamics. It appears that polydispersity plays little part in turbulent clouds and vortical columnar flows. It essentially modifies the morphology of swarms during and after separation (section \ref{subsec:SeparationSwarm}).
\par
It is worth mentioning that the inhibition of the enhancement of entrainment due to background rotation is likely due to the fact that the clouds' inertia is not large enough for them to be completely insensitive to rotation. Increasing their initial inertia by releasing a larger mass excess $m_0$ could enable to observe a range of depths where rotation leaves them unaffected. Additional measurements for other (lower) values of $\Omega$ would also be beneficial to extract quantitative scalings with the rotation speed $\Omega$.
\par
Some open questions remain. According to the literature on particle-laden turbulent flows (e.g., \citep{balachandarTurbulentDispersedMultiphase2010,monchauxSettlingVelocityPreferential2017}) we expect inertial effects to be optimal when the ratio of the settling velocity $w_s$ over the \textit{local} flow velocity within the cloud is of order one -- in other words when a \textit{local} Rouse number is of order unity. For our smallest particles, this velocity ratio remains small because the clouds never fall deep enough to decelerate until this local Rouse number becomes of order unity. The entrainment rates of turbulent thermals and our clouds are constant before separation. Yet, if $\alpha$ varies on a length scale larger than our field of view (so that its variation could not be assessed), one can imagine that in a larger tank, these clouds would experience a further increase of $\alpha$ when the local Rouse number gets close to unity. This hypothesis cannot be ruled out without additional experiments at a larger scale.
\par
Further work on such particle clouds requires to quantify the flow produced by particles through PIV in a vertical laser sheet. Additional visualisations of the glass beads and of PIV particles in horizontal planes would prove enlightening to understand the structure of clouds in the presence of background rotation. It could clarify the mechanism through which the influence of rotation gradually contaminates an entire cloud from its periphery to its entirety, following similar work on plumes \cite{sutherlandPlumesRotatingFluid2021}. To this end, we would also benefit from experiments at a larger scale with a larger mass excess $m_0$ to better separate scales and dynamical regimes.

\section{Acknowledgements}
The authors acknowledge funding by the European Research Council under the European Union’s Horizon 2020 research and innovation program through Grant No. 681835-FLUDYCO-ERC-2015-CoG. The authors thank V. Dorel and L. Huguet for the setting of a first version of the experiment and metrology, as well as for their preliminary results that guided the present work.

%%%%%%%%%%%%%%%%%%%%%%%%%%%%%%%%%%%%%%%%%%%%%%%%
%%%%%%%%%%%%%%%%%%%%%%%%%%%%%%%%%%%%%%%%%%%%%%%%
\appendix

\section{\label{sec:ListExperiments}List of experiments}

Table \ref{tab:NumbersExperiments} lists the number of experiments processed for each couple $(\Omega, \mathcal{R})$. Numbers tend to be higher when particles are more sensitive to initial conditions, or when measurements are very sensitive to noise because of the considerable dilution of particles when the cloud falls. Numbers also vary because some experiments were not processed, because clouds occasionally went out of the laser sheet, or pieces of latex membrane fell with clouds in the laser sheet, introducing a bias in measurements.

\begin{table}[htb]
\renewcommand{\arraystretch}{1.1}
\begin{tabular}{c||c|c|c|c|c|c|c|c|c||c|c|c|c||c|c|c|c}
\diagbox[innerwidth=1cm]{$\Omega$}{$\mathcal{R}$} & \makecell{Salt\\ water} & \makecell{$6.00$\\ $\times 10^{-4}$} & \makecell{$7.57$\\ $\times 10^{-2}$ $\Circle$} & \makecell{$7.57$\\ $\times 10^{-2}$ $\CIRCLE$} & 0.308 & 0.696 & 1.48 & 2.73 & 3.70 & 0.168 & 0.406 & 0.814 & 4.13 & 0.210 & 0.255 & 0.304 & 0.354\\ 
\hline
0 rpm  & 6 & 5 & 16 & 8 & 5 & 6 & 5 & 4 & 10 & 8 & 5 & 6 & 6 & 6 & 3 & 2 & 4\\
5 rpm  & 12 & 0 & 19 & 7 & 15 & 10 & 10 & 5 & 5 & 9 & 14 & 9 & 5 & 4 & 4 & 4 & 4\\
10 rpm  & 13 & 0 & 21 & 6 & 15 & 9 & 11 & 10 & 5 & 10 & 5 & 5 & 5 & 5 & 7 & 5 & 5\\
20 rpm  & 17 & 0 & 18 & 3 & 12 & 15 & 15 & 5 & 9 & 5 & 6 & 5 & 5 & 6 & 5 & 5 & 5\\
\end{tabular}
\caption{Number of experiments which are processed for each couple $(\Omega, \mathcal{R})$. For $\mathcal{R} = 6.00\times 10^{-4}$, experiments were conducted only to determine the entrainment coefficient of clouds at 0rpm (see section \ref{subsec:OnePhaseThermals}).}
\label{tab:NumbersExperiments}
\end{table}

\section{\label{sec:ListNotations}List of notations}

Table \ref{tab:ListVariables} lists the notations used in this study for reference.\\

\tablefirsthead{\hline \textbf{Defined in} & \textbf{Variable} & \textbf{Description} \\}
\tablehead{\hline \textbf{Defined in} & \textbf{Variable} & \textbf{Description} \\}
\bottomcaption{List of variables, the section in which they are defined, and a short description.}
\xentrystretch{-0.15}
\begin{xtabular}{lcl}
\label{tab:ListVariables}
Section \ref{subsec:ExperimentalSetup} & $\rho_f$ & Density of ambient fluid\\
Section \ref{subsec:ExperimentalSetup} & $\nu$ & Viscosity of ambient fluid\\
Section \ref{subsec:ExperimentalSetup} & $\Omega$ & Angular velocity of the rotating table\\
Section \ref{subsec:ExperimentalSetup} & $D_\textrm{cyl}$ & Diameter of the cylinder\\
Section \ref{subsec:ExperimentalSetup} & $m_0$ & Total mass excess (1g)\\
Section \ref{subsec:ExperimentalSetup} & $\rho_p$ & Density of the glass beads\\
Section \ref{subsec:ExperimentalSetup} & $r_p$ & Mean radius of a particle set\\
Section \ref{subsec:ExperimentalSetup} & $H_0$ & Water height above the latex membrane\\
Section \ref{subsec:InitialCondRelease} & $l_M$ & Morton length\\
Section \ref{subsec:InitialCondRelease} & $g$ & Acceleration of gravity\\
Section \ref{subsec:InitialCondRelease} & $\rho_0$ & Initial cloud density\\
Section \ref{subsec:InitialCondRelease} & $U_\textrm{ref}$ & Reference velocity of clouds\\
Section \ref{subsec:ParticleDistributions} & $\sigma_p$ & Standard deviation of a distribution of particles' radii\\
Section \ref{subsec:ParticleDistributions} & $\mathcal{S}$ & Dimensionless number quantifying polydispersity\\
Section \ref{subsec:ParticleDistributions} & $p$ & A percentage to define bidisperse particle sets\\
Section \ref{subsec:DimlessNumbers} & $w_s$ & Terminal velocity of a single particle\\
Section \ref{subsec:DimlessNumbers} & $Re_p$ & Particulate Reynolds number\\
Section \ref{subsec:DimlessNumbers} & $w_s^\textrm{Stokes}$ & Stokes terminal velocity\\
Section \ref{subsec:DimlessNumbers} & $w_s^\textrm{Newton}$ & Newton's terminal velocity\\
Section \ref{subsec:DimlessNumbers} & $C_d$ & Drag coefficient of a spherical particle\\
Section \ref{subsec:DimlessNumbers} & $N_p$ & Total number of particles in a cloud\\
Section \ref{subsec:DimlessNumbers} & $\Pi$ & Dimensionless average radius of a particle set\\
Section \ref{subsec:DimlessNumbers} & $\mathcal{R}$ & Rouse number\\
\hline
Section \ref{subsec:OnePhaseThermals} & $r$ & Radius of a spherical thermal\\
Section \ref{subsec:OnePhaseThermals}  & $\rho$ & Density of a uniform spherical thermal\\
Section \ref{subsec:OnePhaseThermals}  & $z$ & Vertical position of the centre of mass of a uniform spherical thermal\\
Section \ref{subsec:OnePhaseThermals}  & $\dot{z}$ & Vertical velocity of the centre of mass of a thermal; by extension in\\
  &  & experiments, vertical velocity of the barycentre of particles within a cloud\\
Section \ref{subsec:OnePhaseThermals}  & $v_e$ & Entrainment velocity\\
Section \ref{subsec:OnePhaseThermals}  & $\alpha$ & Coefficient of entrainment\\
Section \ref{subsec:OnePhaseThermals}  & $C_D$ & Drag coefficient of a spherical thermal\\
Section \ref{subsec:ThermalRegime} & $z_f$ & Depth of a cloud's front\\
Section \ref{subsec:ThermalRegime} & $\dot{z}_f$ & Vertical velocity of a cloud's front\\
Section \ref{subsec:ThermalRegime} & $\alpha_\textrm{salt}$ & Entrainment coefficient of salt water thermals\\
Section \ref{subsec:SeparationSwarm} & $z_\textrm{sep}$ & Depth of separation between particles and eddies\\
Section \ref{subsec:SeparationSwarm} & $\sigma_x$ & Radius of a cloud computed as a standard deviation\\
Section \ref{subsec:SeparationSwarm} & $\sigma_z$ & Height of a cloud computed as a standard deviation\\
Section \ref{subsec:SeparationSwarm} & $\dot{z}_\textrm{f,s}$ & Constant vertical front velocity in the swarm or vortical columnar regimes\\
Section \ref{subsec:SeparationSwarm} & $\dot{z}_{p,e}$ & During separation: vertical velocity of particles still present inside eddies\\
Section \ref{subsec:SeparationSwarm} & $u(t)$ & During separation: velocity excess of swirling particles wrt separated particles\\
Section \ref{subsec:SeparationSwarm} & $\dot{z}_e$ & During separation:  vertical velocity of eddies\\
Section \ref{subsec:SeparationSwarm} & $\dot{\sigma}_x$ & Growth rate of the cloud radius\\
Section \ref{subsec:SeparationSwarm} & $\dot{\sigma}_z$ & Growth rate of the cloud height\\
Section \ref{subsec:SeparationSwarm} & $\mathcal{R}_\textrm{min}$ & Rouse number of the smallest particles within a cloud\\
Section \ref{subsec:SeparationSwarm} & $\mathcal{R}_\textrm{max}$ & Rouse number of the largest particles within a cloud\\
Section \ref{subsec:SeparationSwarm} & $\Delta w_s$ & Difference of terminal velocities between the largest and smallest particles\\
Section \ref{subsec:SeparationSwarm} & $j_p$ & During separation: volume flux of particles shed in the emerging swarm\\
Section \ref{subsec:SeparationSwarm} & $\phi$ & Particles' volume fraction\\
Section \ref{subsec:SwarmRegime} & $\dot{z}_\textrm{f,max}$ & Maximum cloud front velocity\\
\hline
Section \ref{subsec:OnePhaseThermals_rot} & $z_\textrm{f,col}$ & Depth of transition from a thermal to a vortical columnar flow\\
Section \ref{subsec:OnePhaseThermals_rot} & $Ro(z)$ & Cloud Rossby number\\
Section \ref{subsec:OnePhaseThermals_rot} & $Ro_\textrm{col}$ & Rossby number computed at the depth $z_\textrm{f,col}$\\
Section \ref{subsec:SwarmRotation} & $\sigma_{x,\infty}$ & (Possibly asymptotically) constant radius of a vortical columnar cloud\\
\hline
Appendix \ref{sec:cloudTracking} & $N_0$ & Otsu's method: number of pixels in class 0\\
Appendix \ref{sec:cloudTracking} & $N_1$ & Otsu's method: number of pixels in class 1\\
Appendix \ref{sec:cloudTracking} & $I_0$ & Otsu's method: average intensity in class 0\\
Appendix \ref{sec:cloudTracking} & $I_1$ & Otsu's method: average intensity in class 1\\
Appendix \ref{sec:cloudTracking} & $\delta^*$ & Cloud-tracking: interparticle distance where a cloud has low concentration\\
Appendix \ref{sec:cloudTracking} & $\Delta^*$ & Cloud-tracking: interparticle distance outside of clouds\\
Appendix \ref{sec:cloudTracking} & $I_\textrm{thr}$ & Cloud-tracking: threshold intensity to binarise images\\
Appendix \ref{sec:cloudTracking} & $\sigma$ & Cloud-tracking: size of the Gaussian kernel defining a hull around a cloud\\
Appendix \ref{sec:AppendixMeasurementsFromImages} & $N^*$ & A number of snapshots to compute standard deviations of light intensity\\
Appendix \ref{sec:AppendixMeasurementsFromImages} & $\Sigma_\textrm{sep}$ & Expanded uncertainty on the value of $z_\textrm{sep}$\\
\hline
\end{xtabular}

\section{\label{sec:cloudTracking}Measurements from the automatic cloud tracking}
\begin{figure}[htb]
    \centering
    \includegraphics[width=\textwidth]{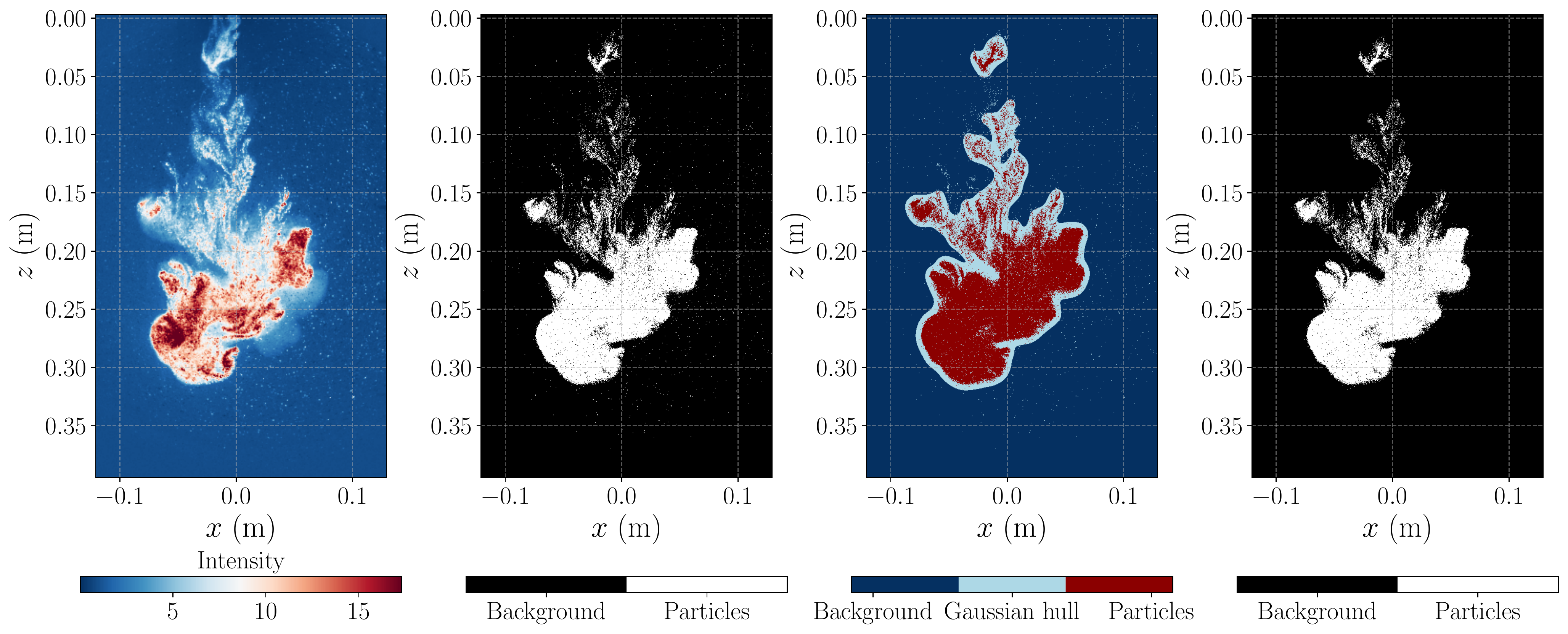}
    \caption{Illustration of the automatic cloud tracking. From left to right, the first image is the ratio of a raw image over a background image taken when no cloud is visible; the second image is a raw binarisation of the first image; the third image shows how to discriminate between the cloud (in the Gaussian hull) and old remaining particles; the last image corresponds to the final result i.e. the restriction of the second image to the Gaussian hull.}
    \label{fig:processingBeads}
\end{figure}

Several quantities are computed from an automatic processing of videos of falling glass beads (visualised by the camera with a green filter). This processing consists of an automatic detection of the particle cloud as follows. The signal received by the camera is never a flat field, even before cloud launching at $t=0$: this means that inhomogeneities are always present in the light intensity, even in the absence of particle cloud. To correct them, a background photograph is always saved at $t<0$. Then, all photographs of the cloud at $t>0$ are divided by this initial background image, so that pixels without particles should have a value of $1$, while pixels with particles should have larger values of relative light intensity. The aim, then, is to define a threshold on light intensity to binarise the image. This threshold should be time-dependent since the cloud keeps diluting so that its average light intensity keeps decreasing. An appropriate binarisation therefore depends on the histogram of each photograph, and Otsu's method \cite{otsuThresholdSelectionMethod1979} is adopted to process experiments. It consists in an optimisation algorithm which maximises the inter-class variance between the two resulting levels of intensity (denoted 0 or 1), the inter-class variance being proportional to $N_0N_1(I_0-I_1)^2$ with $N_{i\leq 1}$ the number of pixels in the level $i$, and $I_{i\leq 1}$ the average intensity in the level $i$. The second image in figure \ref{fig:processingBeads} shows the binarisation of the first image using Otsu's method to define the threshold.

A first limitation of this naive binarisation is that the tank is only emptied after several experiments, so that the background image usually contains some of the smallest particles from previous experiments which settle very slowly and behave as isolated tracers (see the background around the cloud in the second image in figure \ref{fig:processingBeads}). These isolated particles pollute the field of view, which makes it necessary to define a criterion to distinguish the falling cloud on one hand, and old remaining particles on the other hand. Otherwise, a naive binarisation either detects the cloud \textit{and} remaining particles (if the threshold is too low), or does not detect all the particles within the cloud (if the threshold is too large). Regions of high particle concentration are always very bright and easily detected. The main issue is to distinguish between regions of the cloud with low concentration, and detrimental particles from past experiments, since both of these appear in dark shades of gray.
\par
A solution to this problem lies in the interparticle distance. Small particles from past experiments are far away from one another, with a typical interparticle distance $\Delta^*$. Particles falling with the cloud in regions of low concentration have a typical interparticle distance $\delta^*<\Delta^*$ (first image in figure \ref{fig:processingBeads}). Then, the idea is to perform a naive binarisation with a threshold of intensity denoted $I_\textrm{thr}$ (second image in figure \ref{fig:processingBeads}), then blur the result with a Gaussian kernel of size $\sigma$ which verifies $\delta^* < \sigma < \Delta^*$. In doing so, regions of low concentration in the cloud will connect because $\delta^* < \sigma$, which means that blurred particles now overlap, resulting in a signal which is above the threshold $I_\textrm{thr}$. On the opposite, isolated particles from past experiments do not overlap after blurring the image since $\sigma < \Delta^*$, hence their signal is lost in noise below $I_\textrm{thr}$. Then, the resulting blurred image is binarised  with the same threshold $I_\textrm{thr}$ as before, creating a ``Gaussian hull" which only contains regions of both large intensity compared to $I_\textrm{thr}$, and whose interparticle distance is small compared to $\sigma$; see the third image in figure \ref{fig:processingBeads}. To finish with, the final result is the initial naive binarisation of faw images \textit{restricted} to the Gaussian hull only (fourth image in figure \ref{fig:processingBeads}). As observed in figure \ref{fig:processingBeads}, the Gaussian hull is able to capture some particles in regions of low concentration in the vicinity of the brightest regions, and gets rid of most particles in the background.
\par
Once finally binarised, images are used to compute several quantities: the cloud front position $z_f(t)$ is that of the lowermost white pixel; the coordinates $x(t)$ and $z(t)$ of the cloud centroid are computed as averages of the coordinates of all white pixels; the cloud vertical $\sigma_z(t)$ and lateral $\sigma_x(t)$ dimensions are computed as standard deviations of white pixels with respect to the centroid. Errorbars on these quantities correspond to a confidence interval of 95\% around the mean value, computed as 1.96 times the uncertainty obtained by averaging results of several realisations of an experiment, and from least squares regressions if any.

\section{\label{sec:AppendixMeasurementsFromImages}Measurements from raw images}

\subsection{\label{subsec:CoefficientEntrainment}Coefficient of entrainment}

\begin{figure}[htb]
    \centering
    \begin{subfigure}[t]{.195\textwidth}
        \caption{$\mathcal{R}=0$}
        \centering
        \includegraphics[height=3.3cm]{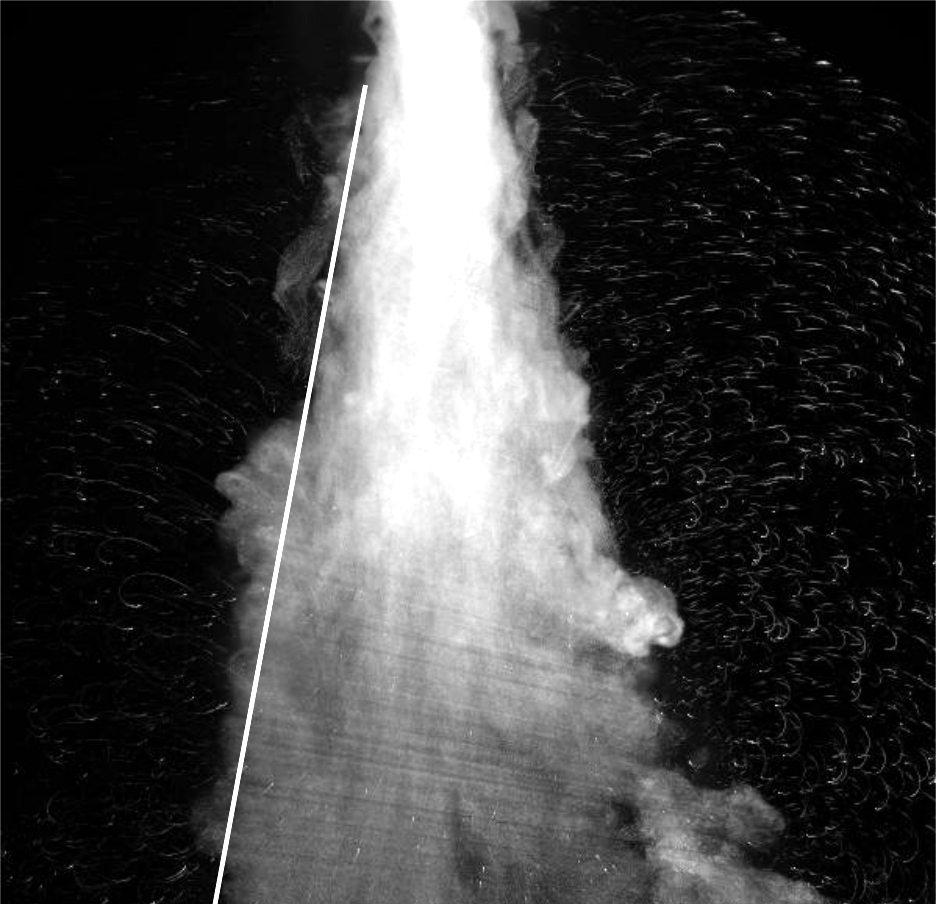}
        \label{subfig:illustrateAlpha_salt}
    \end{subfigure}
    \begin{subfigure}[t]{.195\textwidth}
        \caption{$\mathcal{R}=7.57\times 10^{-2}$}
        \centering
        \includegraphics[height=3.3cm]{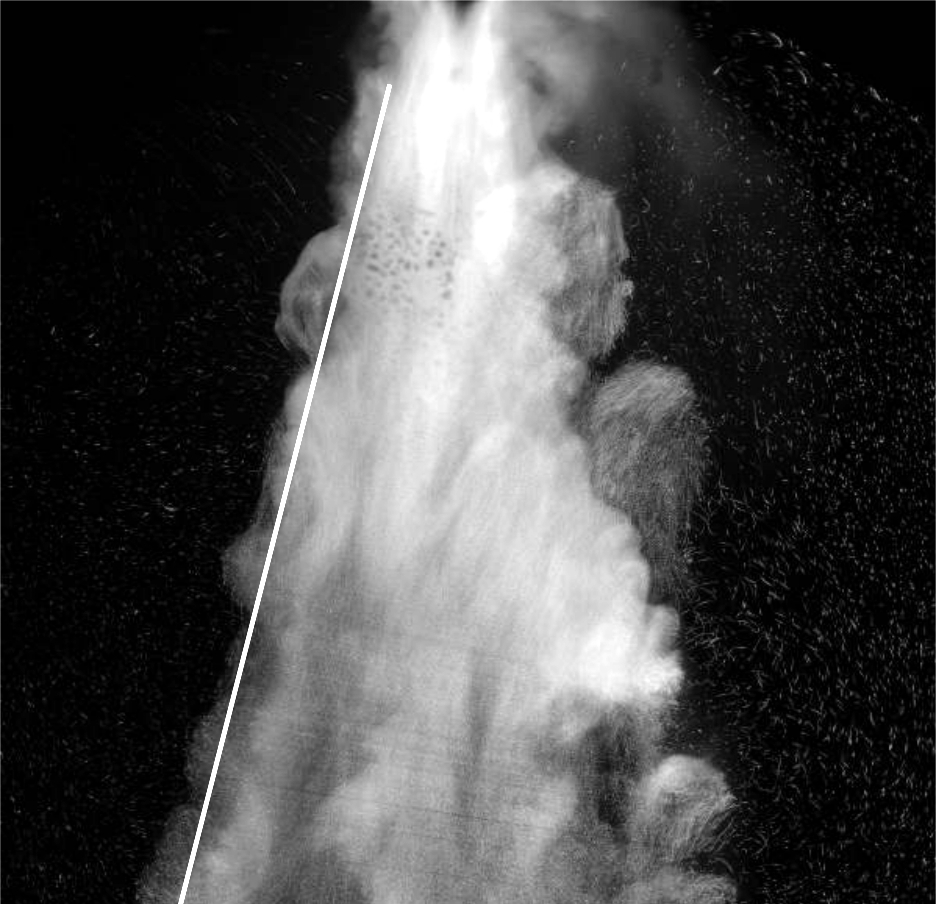}
        \label{subfig:illustrateAlpha_4070}
    \end{subfigure}
    \begin{subfigure}[t]{.195\textwidth}
        \caption{$\mathcal{R}=0.406$}
        \centering
        \includegraphics[height=3.3cm]{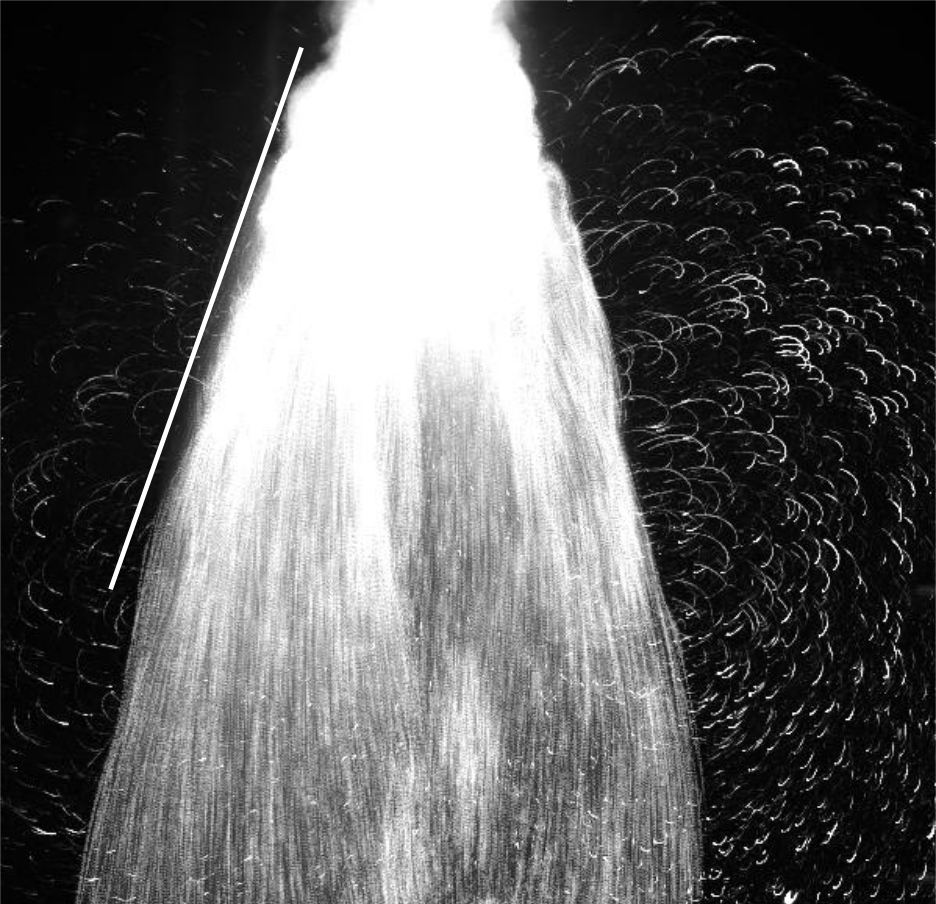}
        \label{subfig:illustrateAlpha_140150}
    \end{subfigure}
    \begin{subfigure}[t]{.195\textwidth}
        \caption{$\mathcal{R}=0.814$}
        \centering
        \includegraphics[height=3.3cm]{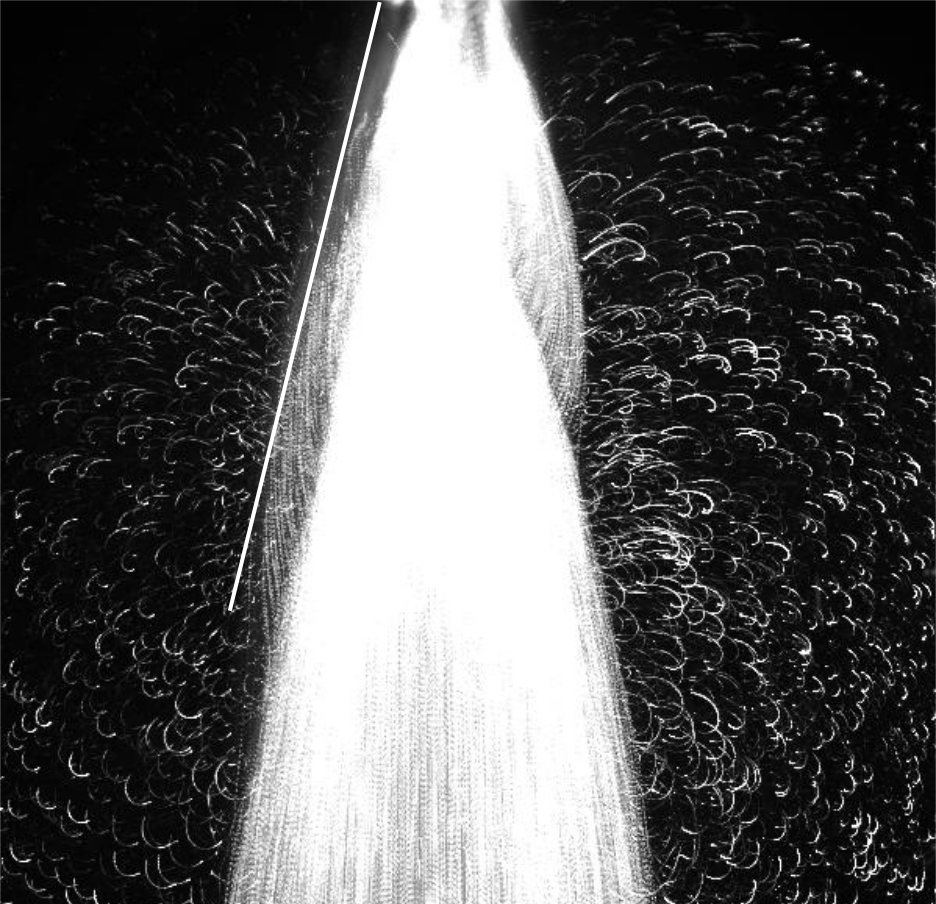}
        \label{subfig:illustrateAlpha_224250}
    \end{subfigure}
    \begin{subfigure}[t]{.195\textwidth}
        \caption{$\mathcal{R}=2.73$}
        \centering
        \includegraphics[height=3.3cm]{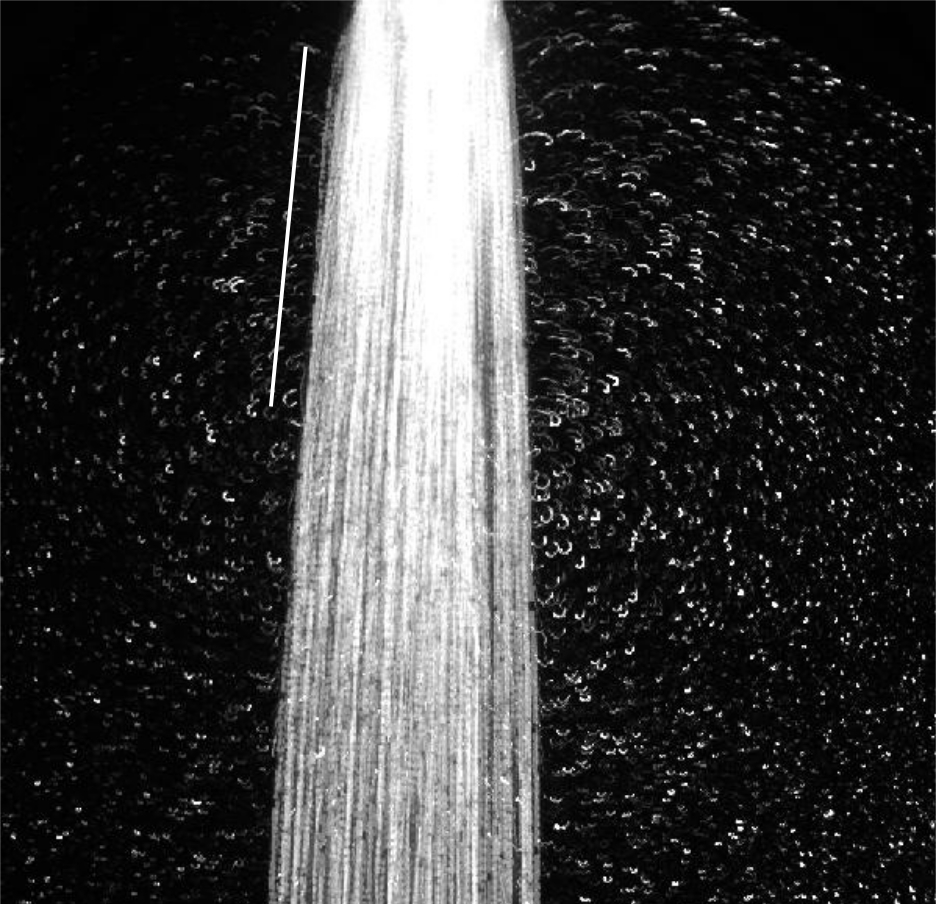}
        \label{subfig:illustrateAlpha_500750}
    \end{subfigure}
    \caption{For five Rouse numbers, images show a moving standard deviation of the light intensity during a typical experiment. Before separation, the angle of opening of clouds can be compared to the angle predicted from the average coefficient of entrainment $\alpha(\mathcal{R})$, shown by a thick solid white line.}
  \label{fig:illustrateDefinitionAlpha}
\end{figure}

The coefficient of entrainment $\alpha$ quantifies the growth rate of clouds, measured as the slope described by the edges of clouds as they fall in the water tank. The simplest and most robust way of quantifying this slope is by using all the photographs of a movie, as well as two integral images computed with the macro ZProject of ImageJ: the average and the pixel-by-pixel standard deviation of light intensity over the entire cloud fall (see figure \ref{fig:illustrateDefinitionAlpha}). With these three visualisations, the cloud edges are tracked, the slope $\alpha=dr/dz$ is determined by hand in each case, and we retain the mean value of the three methods for each realisation of an experiment. Finally we compute $\alpha$ and its error bar respectively as the mean and standard deviation of all the realisations of a given set ($\mathcal{R},\Omega$).

The slope $\alpha$ is determined over the distance where the cloud grows linearly, before separation and before transition to a vortical columnar flow. When a cloud never grows linearly (because particles are so large that the Rouse number $\mathcal{R}$ is above unity), an indicative value of $\alpha$ is measured: it corresponds to the initial growth rate of the cloud after rupturing the membrane, in the phase of acceleration of the cloud (see figure \ref{subfig:illustrateAlpha_500750}).

\subsection{\label{subsec:AppendixDepthSeparation}Depth of separation}

Experimentally, the depth of separation can be measured in two different ways. For the first one, the overlay of images from both cameras enables to compute the surface area of overlap between rhodamine and particles, both regions being defined from the cloud-tracking algorithm presented in appendix \ref{sec:cloudTracking}. Initially both regions are superimposed. Their overlap gradually shrinks in time or depth, evidencing an inflexion point which is considered as defining separation. Depending on experiments, the inflexion point may not be well defined. In that case the asymptotic states of complete overlap and nil overlap are detected, and separation is considered to happen at mid-time. This first method presents the advantage of being quantitative, but only a few experiments were performed with rhodamine. The second method is based on the trajectories of particles. Moving standard deviations are performed over a chosen number $N^*$ of snapshots (typically 30), which enables to observe the trajectory of individual particles over the $N^*$ snapshots at 50fps (see an example in figure \ref{subfig:OutOfEddy}). When particles separate from eddies, the patterns made by their trajectories transition from curved and randomly oriented (due to particles whirling inside eddies) to mostly straight and vertical. When half the visible particles have a vertical trajectory, separation is considered to occur. The difficulty to assess the proportion of particles having transitioned affects the determination of the depth of separation $z_{\textrm{sep}}$. This is accounted for by expanded uncertainties $\Sigma_{\textrm{sep}}$: they are measured in such a way that for $z$ lower than $z_{\textrm{sep}} - \Sigma_{\textrm{sep}}$ almost all particles swirl in eddies, and for $z$ larger than $z_{\textrm{sep}} + \Sigma_{\textrm{sep}}$ almost all particles have a vertical trajectory.
\par
The two methods to identify separation were both implemented for the same series of particle clouds of average Rouse number $\mathcal{R}=0.308$; the difference between their respective mean value of $z_{\textrm{sep}}$ corresponded to a low relative error of 2.1\%, which was considered satisfactory enough to validate the use of the second approach for all experiments.

\bibliography{apssamp}

\end{document}